\documentclass[aps,twocolumn,prb,superscriptaddress,amsmath,amssymb]{revtex4-2} 
\usepackage{dcolumn}
\usepackage{amsmath}
\usepackage{mathrsfs}
\usepackage{txfonts}
\usepackage{bm}
\usepackage[T1]{fontenc}
\usepackage{xspace}
\usepackage{ulem}
\usepackage{comment}
\usepackage{braket}
\usepackage{lipsum}
\usepackage{bbold}
\usepackage{mathtools}
\usepackage{graphicx}
\usepackage{hyperref}
\usepackage{color}
\usepackage{xcolor}
\usepackage[version=3]{mhchem}
\usepackage{placeins}
\hypersetup{
        colorlinks=true,
        citecolor=blue,
        urlcolor=blue,
        linkcolor=blue
}

\begin{document}
\let\emph\textit

\title{
    Temperature evolution of spin dynamics in extended Kitaev models within an Abrikosov fermion theory beyond mean field
}
\author{Daiki Sasamoto}
\email[sasamoto.daiki.r6@dc.tohoku.ac.jp]{}
\author{Joji Nasu}
\affiliation{
  Department of Physics, Graduate School of Science, Tohoku University, Sendai, Miyagi 980-8578, Japan
}

\date{\today}
\begin{abstract}
The $S=1/2$ Kitaev model provides an exactly solvable setting in which spin fractionalization produces itinerant Majorana fermions coupled to an emergent $\mathbb{Z}_2$ gauge field.
In realistic Kitaev magnets, however, non-Kitaev interactions break integrability and make their finite-temperature dynamics difficult to calculate.
We develop a finite-temperature parton mean-field theory based on the Abrikosov-fermion representation and incorporate the effects of exchange interactions through the random phase approximation to calculate the dynamical spin structure factor.
Compared with a Majorana-based formulation, the Abrikosov-fermion representation provides a natural way to incorporate the Heisenberg and $\Gamma$ interactions into the response calculation.
For the pure Kitaev model, the method reproduces a strong, nearly dispersionless peak at a finite low energy together with a broad continuum at higher energies.
It also reproduces the momentum dependence of the dynamical spin structure factor and its temperature evolution, including the downward shift of the low-energy spectral weight, in qualitative agreement with continuous-time quantum Monte Carlo calculations.
Within the present theory, the contraction of the continuum is traced to the temperature-induced reduction of the bond amplitudes, which narrows the parton bands and compresses the two-parton phase space.
For ferromagnetic Kitaev coupling, the energy of the low-energy peak is minimized at the $\Gamma$ point for a ferromagnetic Heisenberg interaction and maximized there for an antiferromagnetic interaction.
In the ferromagnetic Kitaev--$\Gamma$ model, the static response develops a finite-momentum softening, indicating an instability toward the period-three magnetic order reported in numerical studies.
For uniform-ferromagnetic states selected by a $[111]$ magnetic field, we find that a broad parton continuum coexists with magnonlike dispersive structures, producing spectral features consistent with a spinon-to-magnon crossover.
Taken together, these findings show that the present finite-temperature framework provides a versatile means of tracking characteristic changes in spin dynamics from the pure Kitaev limit into extended Kitaev models.
\end{abstract}
\maketitle


\section{Introduction}
\label{introduction}

Quantum spin liquids (QSLs) are insulating magnetic phases that avoid conventional long-range order down to zero temperature because strong quantum fluctuations and frustration suppress the formation of symmetry-breaking patterns~\cite{Anderson-1973,Balents-2010,Savary-Balents-2016,Zhou-Kanoda-2017,Broholm-Cava-2020}.
Unlike trivial paramagnets, QSLs can host long-range entanglement and emergent gauge structures, leading to fractionalized spin excitations and topological order beyond the Landau paradigm~\cite{Balents-2010,Savary-Balents-2016,Zhou-Kanoda-2017,Broholm-Cava-2020,Wen-2004}.
These features are of fundamental interest and also motivate the use of anyonic excitations and topological degeneracy as resources for robust quantum information processing~\cite{Wen-2004,Nayak-Simon-2008}.
A central example is the spin-$1/2$ Kitaev model on the honeycomb lattice.
Built from bond-dependent Ising-like interactions, it constitutes a rare two-dimensional example with an exactly solvable QSL ground state~\cite{Kitaev-2006}.
In the ground state, the spins fractionalize into itinerant Majorana fermions coupled to an emergent $\mathbb{Z}_2$ gauge field, accompanied by gapped flux (vison) excitations~\cite{Kitaev-2006,Hermanns-Kimchi-2018,Knolle-Moessner-2019}.
When time-reversal symmetry is broken by a weak magnetic field, the model can realize a chiral spin liquid hosting non-Abelian anyons and chiral Majorana edge modes, a long-standing motivation from the viewpoint of topological quantum computation~\cite{Nayak-Simon-2008,Kitaev-2006,Hermanns-Kimchi-2018,Knolle-Moessner-2019}.
This combination of exact solvability, fractionalized quasiparticles, and experimentally accessible dynamical responses makes the Kitaev model a central benchmark for QSL theory and for the development of analytical and numerical methods~\cite{Hermanns-Kimchi-2018,Knolle-Moessner-2019,Motome-Nasu-2020,Matsuda-Shibauchi-Kee-2025}.
Importantly, Kitaev interactions are not restricted to an abstract toy model.
In spin-orbit-coupled Mott insulators, spin and orbital degrees of freedom are entangled, and superexchange processes in edge-sharing octahedra can generate strong bond-directional interactions, providing a microscopic route to dominant Kitaev couplings~\cite{Jackeli-Khauliullin-2009}.
This mechanism has driven intensive searches for Kitaev candidate materials, most prominently in honeycomb iridates A$_2$IrO$_3$ (A=Li and Na) and the layered halide $\alpha$-RuCl$_3$~\cite{Singh-Gegenwart-2010,Plumb-2014,Rau-Lee-Kee-2014,Winter-2017,Takagi-2019}.
Although many candidates exhibit magnetic order at low temperatures, spectroscopic experiments have revealed broad scattering continua and unusual responses to temperature and magnetic field that have been discussed as signatures of proximate spin fractionalization~\cite{Chun-2015,Sandilands-2015,Banerjee-2016,Little-2017,Banerjee-2017,Banerjee-2018,Sears-2015,Kubota-2015,Wang-Reschke-2017}.
In the field-induced disordered regime of $\alpha$-RuCl$_3$, neutron, NMR, ESR, THz, and Raman experiments have further sharpened the debate over whether the observed continua and bound-state-like features should be interpreted in terms of fractionalized excitations or conventional magnetic quasiparticles~\cite{Baek-Do-Choi-2017,Banerjee-2018,Balz-Lampen-Kelley-2019,Ponomaryov-2020,Wulferding-2020,Sahasrabudhe-2024,Sarkis-Dixit-Rao-2026}.
Beyond scattering probes, transport and thermoelectric measurements, such as the thermal Hall effect, have emerged as complementary windows into fractionalized excitations and edge physics, while their interpretation in candidate materials remains under active debate~\cite{Kasahara-2018,Bruin-2022,Chern-Zhang-Kim-2021,Czajka-Gao-Hirschberger-Ong-2023,Hwang-Go-Seong-Shibauchi-Moon-2022,Motome-Nasu-2020,Trebst-Hickey-2022,Koyama-Nasu-2023,Koyama-Nasu-2024,Kato-Nasu-Sato-Okubo-Misawa-Motome-2025}.
More broadly, thermal Hall transport has been proposed as a probe of charge-neutral spinons in frustrated magnets beyond the Kitaev setting~\cite{Katsura-Nagaosa-Lee-2010,Samajdar-Chatterjee-2019,Mukherjee-2023,Sasamoto-2026-ThermalHall}.
Realistic Kitaev magnets are not captured by the pure Kitaev Hamiltonian alone.
Additional interactions, including Heisenberg interactions and off-diagonal symmetric couplings, as well as longer-range terms, are generally present and often sizable~\cite{Rau-Lee-Kee-2014,Winter-2017,Takagi-2019,Chaloupka-2013,Kimchi-2014,Katukuri-2014,Winter-2016,Yadav-2016}.
These non-Kitaev interactions break integrability and can stabilize conventional magnetic order.
At the same time, they provide a natural setting for studying the stability of Kitaev-like spin-liquid behavior against realistic perturbations~\cite{Hermanns-Kimchi-2018,Knolle-Moessner-2019,Chaloupka-2013,Kimchi-2014,Katukuri-2014,Winter-2016,Yadav-2016,Rousochatzakis-Perkins-Luo-Kee-2024,Zhang-Halasz-Zhu-Batista-2021}.
Once integrability is lost, however, the fractionalized degrees of freedom interact and the gauge sector is no longer static, making it substantially more difficult to compute dynamical correlations that can be compared directly with experiments~\cite{Hermanns-Kimchi-2018,Knolle-Moessner-2019,Chaloupka-2013,Rousochatzakis-Perkins-Luo-Kee-2024}.

In this context, dynamical response functions are especially important.
The dynamical spin structure factor $S(\bm{q},\omega)$ characterizes magnetic excitations and is closely related to the magnetic inelastic neutron scattering intensity, which includes polarization and magnetic form factors. Raman and terahertz (THz) spectroscopy and resonant inelastic X-ray scattering (RIXS) provide complementary probes with their own coupling operators and selection rules.  In suitable settings, broad continua rather than sharp magnon dispersions provide an important signature of fractionalization~\cite{Chun-2015,Sandilands-2015,Banerjee-2016,Little-2017,Knolle-2014,Halasz-Perkins-vanDenBrink-2016}.
At zero temperature in the pure Kitaev model, this continuum arises from Majorana-quasiparticle processes in the background of gauge fluxes created by local spin operators and provides a well-established example of two-dimensional fractionalized dynamics~\cite{Knolle-2014,Knolle-2015,Knolle-Perkins-2014}.
At finite temperature, even the exactly solvable limit becomes substantially more intricate because thermally excited $\mathbb{Z}_2$ fluxes reconstruct the Majorana spectrum and generate pronounced crossover behavior in the dynamical response~\cite{Nasu-Udagawa-Motome-2014,Nasu-Udagawa-Motome-2015,Motome-Nasu-2020}.
For the integrable Kitaev model, the temperature evolution of $S(\bm{q},\omega)$ has been studied using methods that explicitly sample the gauge sector while treating Majorana fermions, such as Majorana-based cluster dynamical mean-field (CDMFT) theory and continuous-time quantum Monte Carlo (CTQMC)~\cite{Yoshitake-Nasu-Motome-2016,Yoshitake-Nasu-2017}.
By contrast, perturbative and parton-based studies of integrability-breaking interactions have primarily addressed zero temperature or low-energy limits, clarifying long-range dynamical correlations, power-law response, modified spin structure factors, and bound-state formation near magnetic instabilities~\cite{Tikhonov-Feigelman-Kitaev-2011,Song-You-Balents-2016,Gohlke-Verresen-Moessner-Pollmann-2017,Winter-Riedl-2017,Knolle-Bhattacharjee-Moessner-2018,Zhang-Halasz-Zhu-Batista-2021,Cookmeyer-Moore-2023}.

Parton mean-field theory and its extensions beyond mean field provide a practical framework for studying dynamical spin correlations in frustrated magnets~\cite{Ghioldi-Gonzalez-2018,Sasamoto-Nasu-2026-Kagome}.
For the triangular-lattice $J_1$--$J_2$ Heisenberg model, fermionic-parton mean-field theory combined with RPA captures the dynamical response and magnetic instabilities of a U(1) Dirac QSL in agreement with numerical studies~\cite{Willsher-Knolle-2025}.
In Kitaev systems, RPA treatments built on parton mean-field states reproduce the characteristic dynamical response of the exactly solvable limit and capture magnetic instabilities induced by non-Kitaev interactions or a magnetic field~\cite{Rao-Moessner-Knolle-2025,Rao-Moessner-Knolle-2026}.
The temperature evolution of $S(\bm{q},\omega)$ in the pure Kitaev model also provides characteristic signatures of fractionalization~\cite{Yoshitake-Nasu-Motome-2016,Yoshitake-Nasu-2017}.
Extending these response theories to finite temperature in models with non-Kitaev interactions is therefore important for interpreting experiments on candidate materials.

In this paper, we develop a finite-temperature parton mean-field theory based on the Abrikosov-fermion representation and incorporate interaction effects through RPA to calculate the dynamical spin structure factor of extended Kitaev models.  This formulation provides a natural way to include the Heisenberg and $\Gamma$ interactions in the response calculation, allowing liquid and selected magnetically ordered saddles to be treated within a common framework.
We first benchmark the method against previous calculations for the pure Kitaev model.
The theory reproduces a strong, nearly dispersionless low-energy peak with a nonzero gap, together with a broad continuum at higher energies.  It also captures the characteristic momentum dependence of the dynamical spin structure factor and its temperature evolution, including the downward shift of the low-energy spectral weight, in qualitative agreement with previous CDMFT+CTQMC calculations~\cite{Yoshitake-Nasu-Motome-2016}.  This agreement supports the ability of the present approach to capture the characteristic spin dynamics.  Within this mean-field description, we identify the temperature-induced reduction of the intersite bond amplitudes as the mechanism that narrows the parton bands, compresses the two-parton phase space, and thereby contracts the continuum toward lower energies.
For ferromagnetic Kitaev coupling, the energy of the low-energy peak is minimized at the $\Gamma$ point for a ferromagnetic Heisenberg interaction and maximized there for an antiferromagnetic interaction, in qualitative agreement with exact diagonalization.
In the ferromagnetic Kitaev--$\Gamma$ model studied here, the static response softens at a finite wave vector, identifying the same period-three magnetic tendency found in independent numerical studies.
Most notably, comparisons with linear spin-wave theory for uniform-ferromagnetic states selected by a magnetic field in both the Kitaev--Heisenberg and Kitaev--$\Gamma$ models reveal that a broad two-parton continuum can coexist with pronounced magnonlike dispersive features.
This coexistence is consistent with a spinon-to-magnon crossover at the level of the dynamical response. Together with the agreement of the magnetic-instability wave vector with independent numerical studies, these results motivate extending the approach to additional anisotropic and longer-range interactions.

The remainder of this paper is organized as follows.  Section~\ref{sec:Model} introduces the extended Kitaev Hamiltonian and the parameter regimes studied here.  Section~\ref{sec:Method} develops the Majorana- and Abrikosov-fermion mean-field formulations, their exact correspondence at the quadratic level, and the finite-temperature RPA response theory.  Section~\ref{sec:results} presents the pure-Kitaev benchmark and the Kitaev--Heisenberg and Kitaev--$\Gamma$ results, including finite-momentum softening and the response of field-induced uniform-ferromagnetic states.  Section~\ref{sec:Discussion} discusses the thermal contraction of the continuum, local-moment normalization, comparisons with exact diagonalization, and the relation between two-parton and magnonlike responses.  Section~\ref{sec:Summary} summarizes the main conclusions and future directions.  Technical derivations, numerical details, and complementary benchmark calculations are collected in the Appendices.

\section{Model}
\label{sec:Model}

\begin{figure}[t]
  \centering
      \includegraphics[width=\columnwidth,clip]{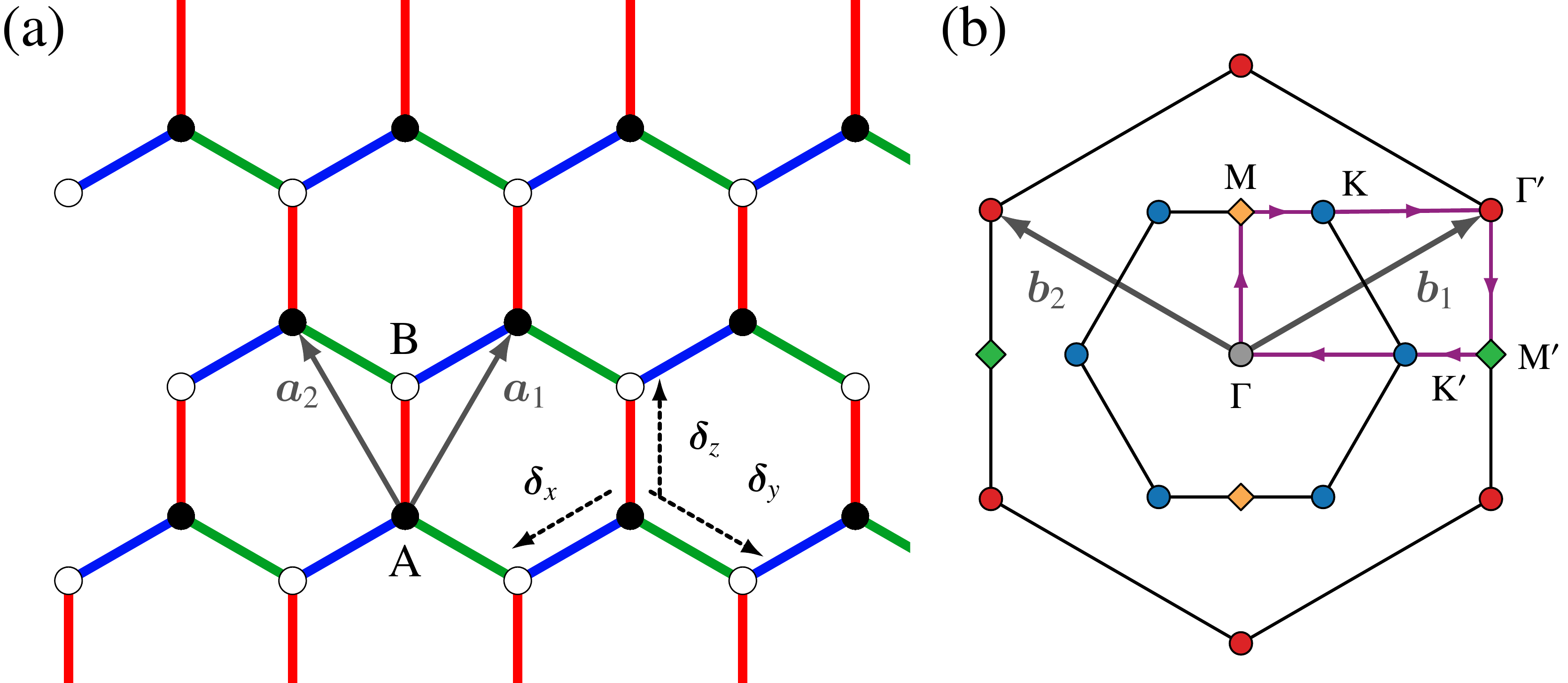}
      \caption{
        (a)
        Honeycomb lattice on which the $S=1/2$ extended Kitaev model is defined.
        The blue, green, and red lines denote the $x$, $y$, and $z$ bonds, respectively.
        The gray arrows indicate the primitive translation vectors $\bm{a}_{1}$ and $\bm{a}_{2}$.
        The black (white) circles represent the A (B) sublattice.
        (b)
        First and extended Brillouin zones of the original honeycomb lattice.
        Filled symbols denote the high-symmetry points.
        The gray arrows indicate the primitive reciprocal lattice vectors $\bm{b}_{1}$ and $\bm{b}_{2}$ corresponding to $\bm{a}_{1}$ and $\bm{a}_{2}$ in real space.
      }
      \label{fig:honeycomb_lattice.pdf}
\end{figure}

\begin{figure}[t]
  \centering
      \includegraphics[width=\columnwidth,clip]{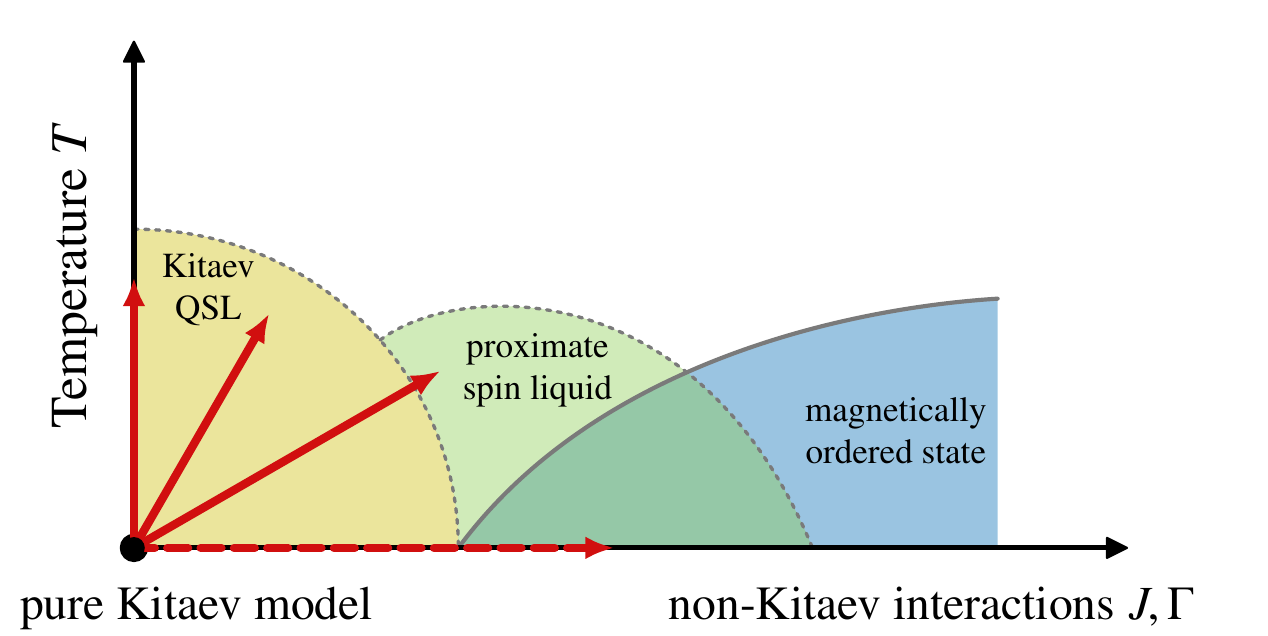}
      \caption{
        Schematic view of the parameter regime considered here.
        The yellow region comprises the low-temperature Kitaev QSL and its connected Kitaev-dominated thermal regime. The blue and unshaded regions denote magnetic order and a conventional paramagnet, respectively. The translucent green region denotes a proximate spin-liquid regime in which Kitaev-like dynamical correlations persist. It can overlap the ordered regime because it is a crossover rather than a distinct thermodynamic phase. Dashed gray curves indicate crossovers, while the solid gray curve marks the magnetic ordering boundary. The red dashed arrow denotes the zero-temperature direction studied in Refs.~\cite{Rao-Moessner-Knolle-2025,Rao-Moessner-Knolle-2026}. The red solid arrows denote the finite-temperature directions investigated in this work.
      }
      \label{fig:schematic_phase_diagram.pdf}
\end{figure}

We consider the extended Kitaev Hamiltonian
\begin{align}
    \label{eq:KJG-Hamiltonian}
    \mathcal{H}
    &=K \sum_{\gamma=x,y,z} \sum_{\langle i,j\rangle_{\gamma}}S_{i}^{\gamma}S_{j}^{\gamma}
    +J\sum_{\langle i,j\rangle}\bm{S}_{i}\cdot\bm{S}_{j}\notag\\
    &\quad
    +\Gamma\sum_{\gamma=x,y,z}\sum_{\langle i,j\rangle_{\gamma}}
    \left(S_{i}^{\alpha_{\gamma}}S_{j}^{\beta_{\gamma}}
    +S_{i}^{\beta_{\gamma}}S_{j}^{\alpha_{\gamma}}\right)
    -\sum_{i}\bm{h}\cdot\bm{S}_{i},
\end{align}
where $S_{i}^{\alpha}$ $(\alpha=x,y,z)$ is an $S=1/2$ spin operator at site $i$, and $\langle i,j\rangle_{\gamma}$ denotes a nearest-neighbor $\gamma$ bond of the honeycomb lattice [Fig.~\ref{fig:honeycomb_lattice.pdf}(a)].
The coupling $K$ is AFM for $K>0$ and FM for $K<0$.
In the $\Gamma$ term, the spin components
$(\alpha_{\gamma},\beta_{\gamma})$ are defined as
$(\alpha_{x},\beta_{x})=(y,z)$,
$(\alpha_{y},\beta_{y})=(z,x)$, and
$(\alpha_{z},\beta_{z})=(x,y)$.
The coupling $J$ denotes the isotropic Heisenberg interaction, and $\Gamma$ denotes the symmetry-allowed off-diagonal interaction.
The last term is the Zeeman coupling to an external magnetic field $\bm{h}$, measured in the same energy units as the exchange couplings.
To specify its orientation, we introduce
$\bm{h}=h\bm{e}_{\nu}$ with
$\bm{e}_{b}=(1,-1,0)/\sqrt{2}$ and
$\bm{e}_{c}=(1,1,1)/\sqrt{3}$.
The response of the Kitaev model depends on the field orientation.
In the pure Kitaev limit, a weak field with nonzero $x$, $y$, and $z$ components generates a chiral Majorana mass proportional to $h_xh_yh_z$ and opens a topological gap~\cite{Kitaev-2006}.
We therefore focus on the $c$-axis direction and use $\bm{h}=h\bm{e}_{c}$ for all finite-field results presented in this work, while the zero-field calculations are obtained by setting $\bm{h}=\bm{0}$.
Throughout this paper, energies are measured in units of $\left|K\right|$, and the primitive translation vectors of the honeycomb lattice have unit length.
With the coordinate axes shown in Fig.~\ref{fig:honeycomb_lattice.pdf}, we use $\bm a_{1}=(1/2,\sqrt{3}/2)$ and $\bm a_{2}=(-1/2,\sqrt{3}/2)$.  The corresponding reciprocal vectors are $\bm b_{1}=(2\pi,2\pi/\sqrt{3})$ and $\bm b_{2}=(-2\pi,2\pi/\sqrt{3})$.
A site $i=(l,s)$ has position $\bm r_{l,s}=\bm R_l+\bm d_s$, where $\bm R_l$ is a Bravais lattice vector and $s=\mathrm A,\mathrm B$. We choose $\bm d_{\mathrm A}=\bm 0$ and $\bm d_{\mathrm B}=(0,1/\sqrt{3})$. The directed nearest-neighbor vectors from A to B are $\bm\delta_x=\bm d_{\mathrm B}-\bm a_1$, $\bm\delta_y=\bm d_{\mathrm B}-\bm a_2$, and $\bm\delta_z=\bm d_{\mathrm B}$.
Throughout this paper, bond operators and mean-field channels on a nearest-neighbor pair $(i,j)$ are defined with $i\in\mathrm{A}$ and $j\in\mathrm{B}$.

The physical regime targeted here is illustrated schematically in Fig.~\ref{fig:schematic_phase_diagram.pdf}.
The dashed red arrow indicates the introduction of non-Kitaev terms to the pure Kitaev model at zero temperature.
Previous Majorana-parton response theories have already addressed this direction in generic Kitaev QSLs and in magnetically ordered phases proximate to them~\cite{Rao-Moessner-Knolle-2025,Rao-Moessner-Knolle-2026}.
In particular, Ref.~\cite{Rao-Moessner-Knolle-2026} formulated a zero-temperature response theory combining Majorana mean field and RPA for such ordered phases, including zigzag-ordered phases and an application to a material-motivated model for $\alpha$-RuCl$_3$.
The solid red directions instead enter the finite-temperature plane, either by heating the pure Kitaev model or by increasing temperature and non-Kitaev interactions together.
The dynamical response of generic extended Kitaev models along these directions remains largely unexplored and is the main subject of the present finite-temperature parton-RPA calculation.
At the pure Kitaev point, the zero-temperature ground state is an exactly solvable QSL, while at finite temperature thermally excited fluxes modify the dynamical response.
Upon increasing non-Kitaev interactions, the zero-temperature state may enter a magnetically ordered phase.
Heating such an ordered state leads to a paramagnet rather than to a thermodynamic Kitaev-QSL phase. Close to the Kitaev limit, however, the paramagnetic response can retain Kitaev-like fractionalized features over a finite-temperature crossover regime.
Experiments often probe dynamical correlations at finite temperature and finite energy, where remnants of fractionalized Kitaev excitations can persist even outside the ideal spin-liquid ground-state phase.
Our goal is to describe the evolution of $S(\bm{q},\omega)$ in this experimentally relevant proximate regime, rather than only at the pure Kitaev point.

Microscopically, such non-Kitaev terms arise from direct electron hopping between neighboring transition-metal ions and from departures from the ideal edge-sharing octahedral geometry, and are therefore generally unavoidable in spin-orbit-entangled honeycomb magnets~\cite{Jackeli-Khauliullin-2009,Chaloupka-2010,Rau-Lee-Kee-2014}.
Descriptions of individual compounds often require longer-range interactions and/or a small $\Gamma^{\prime}$ term~\cite{Eichstaedt-2019,Kaib-2021,Suzuki-2021,Li-2021,Laurell-2020,Maksimov-Chernyshev-2020,Moller-Maksimov-Jiang-White-Valenti-Chernyshev-2025}, while a recurring outcome of microscopic studies is that the nearest-neighbor sector of the leading $d^5$ candidate materials is governed by a dominant ferromagnetic Kitaev interaction together with a positive off-diagonal interaction~\cite{Katukuri-2014,Yamaji-2014,Eichstaedt-2019,Kaib-2021,Suzuki-2021,Li-2021,Laurell-2020}.

Among the layered honeycomb compounds, $\alpha$-RuCl$_3$ is currently the most intensively studied Kitaev material.
First-principles, RIXS, and many-body analyses consistently place it in the regime of dominant FM Kitaev interaction and $\Gamma>0$, whereas the nearest-neighbor Heisenberg interaction is comparatively small and its precise sign remains under discussion~\cite{Eichstaedt-2019,Kaib-2021,Suzuki-2021,Li-2021,Laurell-2020,Maksimov-Chernyshev-2020,Moller-Maksimov-Jiang-White-Valenti-Chernyshev-2025}.
A particularly instructive comparison is provided by Na$_{2}$IrO$_{3}$. A recent microscopic study of $d^5$ Kitaev materials found that Na$_{2}$IrO$_{3}$ and $\alpha$-RuCl$_{3}$ share the same dominant sign structure, namely FM Kitaev interaction and positive off-diagonal interaction, while the sign of $J$ is opposite in the two compounds, being AFM in Na$_{2}$IrO$_{3}$ and FM in $\alpha$-RuCl$_3$~\cite{Liu-2022}.
Experimental high-field data on Na$_{2}$IrO$_{3}$ likewise point to strong FM Kitaev correlations~\cite{Das-2019}.
Motivated by this material trend, we use the nearest-neighbor $K$-$J$-$\Gamma$ Hamiltonian as a common minimal model.  To isolate the effects of the two leading non-Kitaev interactions, the dynamical calculations examine separately the Kitaev--Heisenberg model ($\Gamma=0$) for both signs of $J$ and the Kitaev--$\Gamma$ model ($J=0$) for positive $\Gamma$.
Beyond these $d^5$ materials, first-principles studies have proposed honeycomb $A_2$PrO$_3$ ($A=\mathrm{Li},\mathrm{Na}$), with Pr$^{4+}$ ions in the $4f^1$ configuration, as candidates with dominant AFM Kitaev interactions~\cite{Jang-Sano-Kato-Motome-2019,Jang-Sano-Kato-Motome-2020}.
In real materials, however, the dominant Kitaev interaction is typically accompanied by residual non-Kitaev couplings strong enough to induce long-range order at the lowest temperatures.
For this reason, the experimentally relevant question is often not the realization of an ideal Kitaev spin liquid, but rather the proximity to it.
Here, the term proximate Kitaev spin liquid denotes a regime adjacent to the Kitaev spin liquid in parameter space, where the ground state may already be magnetically ordered but thermodynamic and dynamical responses over a broad temperature or energy window retain signatures of fractionalization~\cite{Catuneanu-2018,Gohlke-2018,Gohlke-Verresen-Moessner-Pollmann-2017,Wang-2019,Revelli-2020}. In this setting, the relevant fractionalized degrees of freedom are itinerant Majorana fermions and $\mathbb{Z}_{2}$ gauge fluxes rather than conventional magnons~\cite{Kitaev-2006,Nasu-Knolle-2016}.
Numerical studies of extended $K$-$J$-$\Gamma$ models indeed found proximate spin-liquid phases near the FM Kitaev limit and showed that Kitaev-like signatures can persist well into neighboring ordered regimes~\cite{Catuneanu-2018,Gohlke-2018,Gohlke-Verresen-Moessner-Pollmann-2017,Wang-2019}.
This viewpoint is strongly motivated by experiments on $\alpha$-RuCl$_{3}$.
Despite its zigzag ordered ground state, Raman and neutron scattering have revealed a broad magnetic continuum extending beyond the sharp magnon sector~\cite{Sandilands-2015,Banerjee-2016,Banerjee-2017,Do-2017,Winter-Riedl-2017}.
These features are widely discussed as signatures of fractionalized excitations or, more generally, as evidence that the material lies close to a Kitaev spin-liquid regime.
Consistently, local-probe and thermodynamic measurements have reported signatures compatible with the characteristic fractional excitation structure of the Kitaev model~\cite{Jansa-2018,Widmann-2019}, while theoretical analysis of Raman response identified a fermionic signal expected from Majorana fractionalization~\cite{Nasu-Knolle-2016}.
A direct comparison between Raman susceptibility measurements and quantum Monte Carlo calculations further showed that this fermionic response survives even when non-Kitaev interactions become relevant, thereby providing a concrete justification for a proximate-spin-liquid description of an ordered material~\cite{Wang-Osterhoudt-2020}.
Related RIXS experiments on honeycomb iridates identified a proximate spin-liquid regime above the magnon scale even inside magnetically ordered phases~\cite{Revelli-2020}, whereas H$_{3}$LiIr$_{2}$O$_{6}$ shows no long-range order and exhibits a broad momentum-independent continuum consistent with dominant FM Kitaev interactions in a disordered spin-liquid regime~\cite{Li-Winter-2018,Knolle-Moessner-Perkins-2019,delaTorre-2023}.
These observations motivate the study of parameter regions that are magnetically ordered at the lowest temperatures but remain sufficiently close to the FM Kitaev limit to exhibit Majorana-like excitation continua.

\section{Method}
\label{sec:Method}

Our method builds on an RPA response theory formulated for interacting Majorana fermions around a Majorana mean-field saddle~\cite{Rao-Moessner-Knolle-2025}.
That approach was shown to capture the characteristic dynamical response at the pure Kitaev point, together with the collective modes and magnetic instabilities generated by non-Kitaev interactions and magnetic fields~\cite{Rao-Moessner-Knolle-2025}.
More recently, the same Majorana-parton response approach has been applied to magnetically ordered phases proximate to Kitaev spin liquids, where the magnetic order is incorporated at the saddle-point level and the spin dynamics is subsequently evaluated in an RPA-type approximation~\cite{Rao-Moessner-Knolle-2026}.
Related self-consistent parton mean-field calculations combined with RPA for a $J_{3}$-perturbed Kitaev spin liquid have likewise revealed paramagnonlike collective modes coexisting with a Majorana continuum and their softening toward magnetic order~\cite{Chen-Wang-2026}.
Complementary Schwinger-boson perturbation theory for spin-$S$ Kitaev--Heisenberg magnets has likewise used an RPA construction around a spin-liquid saddle to track magnetic instabilities and two-spinon continua near the Kitaev limits~\cite{Sasamoto-Nasu-2026-KH}.
These results show that an RPA resummation built on fractionalized parton quasiparticles provides a useful connection between the exactly solvable Kitaev limit, nearby nonintegrable spin-liquid regimes, and proximate ordered regimes.
Motivated by these developments, we formulate the corresponding Abrikosov-fermion approach, in which the susceptibility and vertex are explicit matrices and finite-temperature mean-field solutions can be incorporated directly.

\subsection{Majorana fermion mean-field theory for the extended Kitaev model}
\label{sec:Majorana fermion mean-field theory for extended Kitaev model}

Related Majorana representations and their connections to alternative formulations of the Kitaev model are discussed in Ref.~\cite{Fu-Knolle-Perkins-2018}.
We represent the spin operators $S_{i}^{\alpha}\:\left(\alpha=x,y,z\right)$ on lattice site $i$ as
\begin{align}
    \label{eq:spin-Majorana-representation}
    S_{i}^{\alpha}=ic_{i}b_{i}^{\alpha}.
\end{align}
The four Majorana fermions satisfy
\begin{align}
    \label{eq:commutation-relation-Majorana}
    \{c_i,c_j\}&=\delta_{ij},
    &
    \{b_i^\alpha,b_j^\beta\}&=\delta_{ij}\delta_{\alpha\beta},
    &
    \{c_i,b_j^\alpha\}&=0,
\end{align}
where $\alpha,\beta=x,y,z$.
Each operator is Hermitian and squares to $1/2$.
Since this rewriting of spin operators in terms of Majorana fermions enlarges the Hilbert space, one must impose the on-site constraint
\begin{align}
    \label{eq:onsite-consraint}
    D_{i}=4c_{i}b_{i}^{x}b_{i}^{y}b_{i}^{z}=1.
\end{align}

Rewriting the pure Kitaev model in terms of Majorana fermions, one obtains
\begin{align}
    \mathcal{H}_{\text{K}}&=K\sum_{\langle i,j\rangle_{\gamma}}\left(ic_{i}b_{i}^{\gamma}\right)\left(ic_{j}b_{j}^{\gamma}\right)\notag\\
    &=-iK\sum_{\langle i,j\rangle_{\gamma}}u_{ij}^{\gamma}c_{i}c_{j},
    \label{eq:Kitaev-Hamiltonian-Majorana}\\
    u_{ij}^{\gamma}&=ib_{i}^{\gamma}b_{j}^{\gamma}.
    \label{eq:Kitaev-bond-variable}
\end{align}
The bond variable $u_{ij}^{\gamma}$ defined in Eq.~\eqref{eq:Kitaev-bond-variable} obeys $[u_{ij}^{\gamma},\mathcal{H}_{\text{K}}]=0$ and is therefore a conserved link operator.  It is nevertheless gauge dependent.  For the two endpoints of the link, $\{D_i,u_{ij}^{\gamma}\}=\{D_j,u_{ij}^{\gamma}\}=0$, whereas $[D_{\ell},u_{ij}^{\gamma}]=0$ for $\ell\neq i,j$.  Equivalently, the local $\mathbb{Z}_2$ gauge transformation generated by $D_i$ reverses the signs of all four Majorana operators at site $i$ and hence of every link variable incident on $i$, while leaving $\mathcal{H}_{\text{K}}$ invariant.
The gauge-invariant $\mathbb{Z}_{2}$ flux operator on each hexagonal plaquette $p$ is defined in terms of spins as
\begin{align}
    \label{eq:definition-Wp}
    W_p=\prod_{i\in p}(2S_i^{\alpha_i})=\prod_{\langle i,j\rangle_\gamma\in\partial p}(2u_{ij}^\gamma).
\end{align}
The first product runs over the six sites of $p$, with $\alpha_i$ denoting the type of the bond at site $i$ that is not on the plaquette boundary $\partial p$. The second equality holds in the physical subspace $D_i=1$, with all boundary bonds oriented from A to B as in Fig.~\ref{fig:honeycomb_lattice.pdf}(a). Under a local gauge transformation, the signs of the two boundary links incident on a site change together, so $W_p$ is gauge invariant and commutes with every $D_i$. The uniform gauge $u_{\mathrm{AB}}^\gamma=+1/2$ gives $W_p=+1$ on every plaquette, corresponding to the flux-free ground-state sector.

To express the pure Kitaev model as a fermionic mean-field theory, we introduce the real mean-field channels
\begin{align}
    C_{\gamma}
    &=\langle ic_{i}c_{j}\rangle
    \label{eq:mean-field-channel-Kitaev-Majorana-C}\\
    Q^{\alpha\beta}_{\gamma}
    &=\langle ib_{i}^{\alpha}b_{j}^{\beta}\rangle
    \label{eq:mean-field-channel-Kitaev-Majorana-Q}
\end{align}
and rewrite the Hamiltonian as
\begin{align}
    \label{eq:Kitaev-Hamiltonian-Majorana-mean-field}
    \mathcal{H}^{\text{MF}}=-iK\sum_{\langle i,j\rangle_{\gamma}}\left(Q_{\gamma}^{\gamma\gamma}c_{i}c_{j}+C_{\gamma}b_{i}^{\gamma}b_{j}^{\gamma}+i C_{\gamma}Q_{\gamma}^{\gamma\gamma}\right)
\end{align}
At zero temperature in the flux-free sector, this saddle-point decoupling reproduces the exact ground-state energy and nearest-neighbor static spin correlation.  This statement does not extend to arbitrary finite-temperature static quantities, for which the exact treatment requires thermal averaging over flux sectors.
The subscript $\gamma$ on $C_{\gamma}$ and $Q^{\alpha\beta}_{\gamma}$ specifies the bond type. We orient every bond from site $i$ on the A sublattice to site $j$ on the B sublattice.
The zero-temperature flux-free saddle gives
\begin{align}
    C_{\gamma}
    &\simeq\mp0.2624,\label{eq:mean-field-value-of-Kitaev-model-C}\\
    Q^{\gamma\gamma}_{\gamma}
    &=\frac{1}{2}.\label{eq:mean-field-value-of-Kitaev-model-Q}
\end{align}
At the nonmagnetic pure-Kitaev saddle, the onsite mixed-Majorana covariances vanish and Wick factorization gives $\langle D_i\rangle_0=0$, rather than $1$, reflecting the unprojected character of the Gaussian saddle.
The $-$ sign of $C_{\gamma}$ corresponds to the FM Kitaev model ($K<0$), whereas the $+$ sign corresponds to the AFM Kitaev model ($K>0$).
From these values, one obtains the ground-state static spin correlation on a $\gamma$ bond,
$\langle S_{i}^{\gamma}S_{j}^{\gamma}\rangle$, as
\begin{align}
    \label{eq:exact-spin-correlator}
    \langle S_{i}^{\gamma}S_{j}^{\gamma}\rangle=-C_{\gamma}Q_{\gamma}^{\gamma\gamma}\simeq\pm 0.1312.
\end{align}

We now extend this Majorana-fermion mean-field construction away from the pure Kitaev limit. Rewriting Eq.~\eqref{eq:KJG-Hamiltonian} in the Majorana representation gives
\begin{align}
    \label{eq:KJG-Hamiltonian-Majorana}
    \mathcal{H}
    &=K\sum_{\gamma=x,y,z}\sum_{\langle i,j\rangle_{\gamma}}\left(ic_{i}b_{i}^{\gamma}\right)\left(ic_{j}b_{j}^{\gamma}\right)\notag\\
    &+J\sum_{\langle i,j\rangle}\sum_{\alpha=x,y,z}\left(ic_{i}b_{i}^{\alpha}\right)\left(ic_{j}b_{j}^{\alpha}\right)\notag\\
    &+\Gamma\sum_{\gamma=x,y,z}\sum_{\langle i,j\rangle_{\gamma}}
    \left[
    \left(ic_{i}b_{i}^{\alpha_{\gamma}}\right)
    \left(ic_{j}b_{j}^{\beta_{\gamma}}\right)
    +\left(ic_{i}b_{i}^{\beta_{\gamma}}\right)
    \left(ic_{j}b_{j}^{\alpha_{\gamma}}\right)
    \right]\notag\\
    &-\sum_{i}\sum_{\alpha=x,y,z}h^{\alpha}\left(ic_{i}b_{i}^{\alpha}\right).
\end{align}
For a general bond, the spin bilinear is rewritten in the Majorana representation and decoupled as
\begin{align}
    \label{eq:general-majorana-decoupling}
    S_i^\alpha S_j^\beta
    &=
    (ic_i b_i^\alpha)(ic_j b_j^\beta)
    =
    -\left(ic_i c_j\right)\left(ib_i^\alpha b_j^\beta\right)
    \notag\\
    &\simeq
    -\left[
    Q_{ij}^{\alpha\beta}\left(ic_i c_j\right)
    +C_{ij}\left(ib_i^\alpha b_j^\beta\right)
    -C_{ij}Q_{ij}^{\alpha\beta}
    \right]\notag\\
    &+m_i^\alpha\left(ic_j b_j^\beta\right)
    +m_j^\beta\left(ic_i b_i^\alpha\right)
    -m_i^\alpha m_j^\beta ,
\end{align}
where
\begin{align}
    \label{eq:general-majorana-channels}
    C_{ij}
    &=
    \langle ic_i c_j\rangle,
    &
    Q_{ij}^{\alpha\beta}
    &=
    \langle ib_i^\alpha b_j^\beta\rangle,
    &
    m_i^\alpha
    &=
    \langle S_i^\alpha\rangle
    =
    \langle ic_i b_i^\alpha\rangle .
\end{align}
In Eq.~\eqref{eq:general-majorana-decoupling}, the terms involving the intersite fields $C_{ij}$ and $Q_{ij}^{\alpha\beta}$ describe the liquid-channel decoupling and yield Eq.~\eqref{eq:Kitaev-Hamiltonian-Majorana-mean-field} in the pure Kitaev limit.
The terms involving the onsite moments $m_i^\alpha$ and $m_j^\beta$ describe the magnetic-channel decoupling. They vanish at a liquid saddle but must be retained when the saddle point carries finite magnetic moments.
Thus, in the liquid-saddle calculations we impose $m_i^\alpha=0$ and use translationally invariant bond channels $C_{\gamma}$ and $Q_{\gamma}^{\alpha\beta}$, while the uniform-ferromagnetic calculations also retain the onsite magnetic channel $m_i^\alpha$.

To impose the constraint in Eq.~\eqref{eq:onsite-consraint} on average, we introduce the Lagrange multipliers $\lambda^{\alpha}$ and add
\begin{align}
    \label{eq:constraint-average}
    \mathcal{H}^{\text{c}}=\sum_{i,\alpha}i\lambda^{\alpha}\left(c_{i}b_{i}^{\alpha}+\frac{1}{2}\varepsilon_{\alpha\beta\gamma}b_{i}^{\beta}b_{i}^{\gamma}\right)
\end{align}
to the mean-field Hamiltonian. The parameters $\lambda^{\alpha}$ are determined self-consistently together with the mean-field channels.
Equation~\eqref{eq:constraint-average} imposes the three bilinear forms of the local constraint at the expectation-value level and defines the average-constraint convention used below.
We write the quadratic mean-field Hamiltonian in matrix form to obtain the Majorana fermion spectrum and solve the self-consistency equations.
The following expressions use the primitive two-sublattice unit cell for both the liquid and uniform-ferromagnetic calculations.
We write $i=(l,s)$, where $l=1,2,\ldots,N/2$ labels the unit cell and $s=\text{A},\text{B}$ labels the sublattice, and define the eight-component Majorana spinor
\begin{align}
    \label{eq:definition-majorana-spinor-real-space}
    \mathcal{B}_{l}^{T}
    =
    \Bigl(
    c_{(l,\text{A})}\:
    b_{(l,\text{A})}^{x}\:
    b_{(l,\text{A})}^{y}\:
    b_{(l,\text{A})}^{z}\:
    c_{(l,\text{B})}\:
    b_{(l,\text{B})}^{x}\:
    b_{(l,\text{B})}^{y}\:
    b_{(l,\text{B})}^{z}
    \Bigr).
\end{align}
Then the mean-field Hamiltonian can be written as
\begin{align}
    \label{eq:BdG-Majorana-real-space}
    \mathcal{H}^{\text{MF}}
    =
    \frac{i}{2}\sum_{l,l^{\prime}}
    \mathcal{B}_{l}^{T}
    \widetilde{H}_{ll^{\prime}}
    \mathcal{B}_{l^{\prime}},
\end{align}
Here $\widetilde{H}_{ll^{\prime}}$ is the $8\times 8$ real-space kernel determined by the self-consistent mean-field parameters $C_{\gamma}$, $Q_{\gamma}^{\alpha\beta}$, $\lambda^{\alpha}$, and, when retained, $m_i^\alpha$.
Because the Majorana fermions are real fields, the kernel blocks satisfy
\begin{align}
    \label{eq:definition-majorana-matrix-real-space}
    \widetilde{H}_{ll^{\prime}}^{*}
    =
    \widetilde{H}_{ll^{\prime}},
    \qquad
    \widetilde{H}_{ll^{\prime}}^{T}
    =
    -\widetilde{H}_{l^{\prime}l}.
\end{align}

We use the actual site positions in all spatial Fourier transforms, consistently with the response convention of Ref.~\cite{Sasamoto-Nasu-2026-KH}. For each component $a=1,\ldots,8$ of $\mathcal B_l$, $\bm r_{l,a}$ denotes the position of its associated site. Thus, it equals $\bm r_{l,\mathrm A}$ for $a=1,\ldots,4$ and $\bm r_{l,\mathrm B}$ for $a=5,\ldots,8$. The Fourier transform is
\begin{align}
    \label{eq:fourier-transformation-majorana}
    \mathcal{B}_{\bm{k},a}
    =
    \sqrt{\frac{2}{N}}\sum_{l}\mathcal{B}_{l,a}e^{-i\bm{k}\cdot\bm{r}_{l,a}},
\end{align}
and the reality condition of Majorana fermions implies
\begin{align}
    \label{eq:reality-condition-majorana-momentum-space}
    \mathcal{B}_{\bm{k}}^{\dagger}
    =
    \mathcal{B}_{-\bm{k}}^{T}.
\end{align}
Then the Hamiltonian becomes
\begin{align}
    \label{eq:BdG-Majorana-momentum-space}
    \mathcal{H}^{\text{MF}}
    =
    \frac{i}{2}\sum_{\bm{k}}^{\text{BZ}}
    \mathcal{B}_{-\bm{k}}^{T}
    \widetilde{H}_{\bm{k}}
    \mathcal{B}_{\bm{k}},
\end{align}
where the sum over $\bm{k}$ is taken over the first Brillouin zone, and
\begin{align}
    \label{eq:definition-majorana-matrix-momentum-space}
    [\widetilde{H}_{\bm{k}}]_{ab}
    =
    \sum_{l^{\prime}}
    [\widetilde{H}_{ll^{\prime}}]_{ab}
    e^{-i\bm{k}\cdot\left(\bm{r}_{l,a}-\bm{r}_{l^{\prime},b}\right)}.
\end{align}
The matrix $\widetilde{H}_{\bm{k}}$ is independent of $l$ because $\widetilde{H}_{ll^{\prime}}$ depends only on the relative coordinate $\bm{R}_{l}-\bm{R}_{l^{\prime}}$, where $\bm R_l$ denotes the position vector of unit cell $l$.
To make the parallel with the Abrikosov fermion formulation explicit, we introduce the Hermitian matrix
\begin{align}
    \label{eq:definition-hermitian-majorana-matrix}
    H_{\bm{k}}^{\text{M}}
    =
    i\widetilde{H}_{\bm{k}},
\end{align}
so that Eq.~\eqref{eq:BdG-Majorana-momentum-space} can be rewritten as
\begin{align}
    \label{eq:BdG-Majorana-momentum-space-hermitian}
    \mathcal{H}^{\text{MF}}
    =
    \frac{1}{2}\sum_{\bm{k}}^{\text{BZ}}
    \mathcal{B}_{\bm{k}}^{\dagger}
    H_{\bm{k}}^{\text{M}}
    \mathcal{B}_{\bm{k}}.
\end{align}
The matrix $H_{\bm{k}}^{\text{M}}$ satisfies
\begin{align}
    \label{eq:symmetry-hermitian-majorana-matrix}
    \left(H_{\bm{k}}^{\text{M}}\right)^{\dagger}
    =
    H_{\bm{k}}^{\text{M}},
    \qquad
    \left(H_{\bm{k}}^{\text{M}}\right)^{*}
    =
    -H_{-\bm{k}}^{\text{M}}.
\end{align}

Since $\mathcal{B}_{\bm{k}}$ obeys the Majorana reality condition, the diagonalization differs slightly from the conventional BdG problem for complex fermions.
We therefore introduce the particle-hole-paired quasiparticle basis
\begin{align}
    \label{eq:definition-majorana-quasiparticle-basis}
    \Xi_{\bm{k}}^{T}
    =
    \left(
    a_{\bm{k},1}\:
    \cdots\:
    a_{\bm{k},4}\:
    a_{-\bm{k},1}^{\dagger}\:
    \cdots\:
    a_{-\bm{k},4}^{\dagger}
    \right),
\end{align}
and the transformation
\begin{align}
    \label{eq:Majorana-Bogoliubov-transformation}
    \mathcal{B}_{\bm{k}}
    =
    U_{\bm{k}}
    \Xi_{\bm{k}},
\end{align}
where
\begin{align}
    \label{eq:definition-majorana-unitary-matrix}
    U_{\bm{k}}
    =
    \left(
    \bm{u}_{\bm{k},1}\:
    \cdots\:
    \bm{u}_{\bm{k},4}\:
    \bm{u}_{-\bm{k},1}^{*}\:
    \cdots\:
    \bm{u}_{-\bm{k},4}^{*}
    \right)
\end{align}
is an $8\times 8$ unitary matrix satisfying
\begin{align}
    \label{eq:unitarity-majorana-unitary-matrix}
    U_{\bm{k}}^{\dagger}U_{\bm{k}}
    =
    U_{\bm{k}}U_{\bm{k}}^{\dagger}
    =
    \bm{1}_{8\times 8}.
\end{align}
The column vectors $\bm{u}_{\bm{k},\eta}$ are determined by the eigenvalue equations
\begin{align}
    \label{eq:eigenvalue-equation-majorana-particle}
    H_{\bm{k}}^{\text{M}}
    \bm{u}_{\bm{k},\eta}
    =
    \varepsilon_{\bm{k},\eta}
    \bm{u}_{\bm{k},\eta},
    \qquad
    \varepsilon_{\bm{k},\eta}>0,
\end{align}
and
\begin{align}
    \label{eq:eigenvalue-equation-majorana-hole}
    H_{\bm{k}}^{\text{M}}
    \bm{u}_{-\bm{k},\eta}^{*}
    =
    -\varepsilon_{-\bm{k},\eta}
    \bm{u}_{-\bm{k},\eta}^{*},
\end{align}
which show that the spectrum appears in particle-hole pairs as a consequence of the Majorana reality condition.
The diagonal energy matrix is therefore
\begin{align}
    \label{eq:definition-majorana-energy-matrix}
    E_{\bm{k}}^{\text{M}}
    =
    U_{\bm{k}}^{\dagger}
    H_{\bm{k}}^{\text{M}}
    U_{\bm{k}}
    =
    \text{diag}\{
    \varepsilon_{\bm{k},1},\ldots,\varepsilon_{\bm{k},4},
    -\varepsilon_{-\bm{k},1},\ldots,-\varepsilon_{-\bm{k},4}
    \}.
\end{align}
Using this transformation, the mean-field Hamiltonian can be written in diagonal form as
\begin{align}
    \label{eq:BdG-Majorana-diagonal}
    \mathcal{H}^{\text{MF}}
    =
    \frac{1}{2}\sum_{\bm{k}}^{\text{BZ}}
    \Xi_{\bm{k}}^{\dagger}
    E_{\bm{k}}^{\text{M}}
    \Xi_{\bm{k}}
    =
    \sum_{\bm{k}}^{\text{BZ}}\sum_{\eta=1}^{4}
    \varepsilon_{\bm{k},\eta}
    \left(
    a_{\bm{k},\eta}^{\dagger}a_{\bm{k},\eta}
    -\frac{1}{2}
    \right).
\end{align}
The thermal occupation is given by the Fermi-Dirac distribution
\begin{align}
    \label{eq:distribution-function-majorana}
    g(\varepsilon_{\bm{k},\eta})
    =
    \langle a_{\bm{k},\eta}^{\dagger}a_{\bm{k},\eta}\rangle
    =
    \frac{1}{e^{\varepsilon_{\bm{k},\eta}/T}+1}.
\end{align}
Throughout this paper, we set the Boltzmann constant $k_{\mathrm{B}}$ to unity.
The Majorana bilinears entering the self-consistency equations can then be evaluated in the diagonal basis.
For $a,b=1,\ldots,8$, let $u_{a\eta}(\bm{k})$ denote the $a$-th component of $\bm{u}_{\bm{k},\eta}$.
Then
\begin{align}
    \label{eq:bilinear-expectation-value-majorana}
    i\langle \mathcal{B}_{l,a}\mathcal{B}_{l^{\prime},b}\rangle
    &=
    i\frac{2}{N}\sum_{\bm{k}}^{\text{BZ}}\sum_{\eta=1}^{4}
    \Bigl[
    u_{a\eta}(\bm{k})u_{b\eta}^{*}(\bm{k})
    \left\{1-g(\varepsilon_{\bm{k},\eta})\right\}\notag\\
    &+u_{a\eta}^{*}(-\bm{k})u_{b\eta}(-\bm{k})
    g(\varepsilon_{-\bm{k},\eta})
    \Bigr]
    e^{i\bm{k}\cdot\left(\bm{r}_{l,a}-\bm{r}_{l^{\prime},b}\right)}.
\end{align}
From Eq.~\eqref{eq:bilinear-expectation-value-majorana}, the mean-field channels $C_{\gamma}$, $Q_{\gamma}^{\alpha\beta}$, and the Lagrange multipliers $\lambda^{\alpha}$ are determined self-consistently.

\subsection{Abrikosov fermion mean-field theory for the extended Kitaev model}
\label{sec:Abrikosov fermion mean-field theory for extended Kitaev model}

Related SU(2) slave-fermion formulations have also been used to describe the exactly solvable Kitaev model and its superconducting spinon structure~\cite{Burnell-Nayak-2011}.
For a spin $S=1/2$ at lattice site $i$, each spin component $S_{i}^{\alpha} \left(\alpha=x,y,z\right)$ is represented in terms of a two-component fermion
$\bm{f}_{i}=\left(f_{i\uparrow},f_{i\downarrow}\right)^T$ as
\begin{align}
    \label{eq:Abrikosov fermion representation}
    S_{i}^{\alpha}
    =\frac{1}{2}\sum_{\mu,\nu=\uparrow,\downarrow}f_{i\mu}^{\dagger}\sigma_{\mu\nu}^{\alpha}f_{i\nu},
\end{align}
where $\bm{\sigma}=(\sigma^{x},\sigma^{y},\sigma^{z})$ denotes a set of Pauli matrices.
The operators $f_{i\uparrow}$ and $f_{i\downarrow}$ obey the canonical anticommutation relations $\{f_{i\mu},f_{j\nu}^{\dagger}\}=\delta_{ij}\delta_{\mu\nu}$, together with $\{f_{i\mu},f_{j\nu}\}=\{f_{i\mu}^{\dagger},f_{j\nu}^{\dagger}\}=0$.
Because this fermionic representation enlarges the Hilbert space, the physical subspace is specified by the three local SU(2) constraints
\begin{subequations}
\label{eq:Abrikosov-local-constraints}
\begin{align}
    n_i-1
    &=\sum_{\mu}f_{i\mu}^{\dagger}f_{i\mu}-1=0,
    \label{eq:Abrikosov-number-constraint}\\
    f_{i\downarrow}f_{i\uparrow}
    &=0,
    \qquad
    f_{i\uparrow}^{\dagger}f_{i\downarrow}^{\dagger}
    =0.
    \label{eq:Abrikosov-pair-constraints}
\end{align}
\end{subequations}
The first condition enforces single occupancy and projects onto a spin-$1/2$ degree of freedom with $\bm S_i^2=3/4$. The other two are its particle-hole partners. At the mean-field level, the three constraints are imposed on average by three real uniform Lagrange multipliers, which enter through
\begin{align}
    \mathcal H_{\mathrm{AF}}^{\mathrm c}
    =\sum_i\Bigl[
    &\lambda^z(n_i-1)\notag\\
    &-(\lambda^x+i\lambda^y)
    f_{i\uparrow}^{\dagger}f_{i\downarrow}^{\dagger}
    -(\lambda^x-i\lambda^y)
    f_{i\downarrow}f_{i\uparrow}\Bigr].
    \label{eq:Abrikosov-constraint-term}
\end{align}
Under the local canonical transformation introduced below, Eq.~\eqref{eq:Abrikosov-constraint-term} is exactly the Abrikosov-fermion form of the Majorana constraint term in Eq.~\eqref{eq:constraint-average}. Thus, the same $\lambda^{x,y,z}$ are determined self-consistently in the two representations.

Rather than introducing an independent Abrikosov-fermion decoupling, we construct a quadratic Hamiltonian that is exactly equivalent to the Majorana mean-field Hamiltonian through the local canonical transformation
\begin{align}
    f_{i\uparrow}
    &=\frac{1}{\sqrt{2}}\left(c_{i}+ib_{i}^{z}\right),
    \label{eq:MF-AF-relation-up}\\
    f_{i\downarrow}
    &=\frac{i}{\sqrt{2}}\left(b_{i}^{x}+ib_{i}^{y}\right).
    \label{eq:MF-AF-relation-down}
\end{align}
The explicit relations between the Majorana bilinears and the Abrikosov-fermion bond operators are given in Appendix~\ref{app:Connection between Majorana-fermion and Abrikosov-fermion formulations}.
Using these relations, the quadratic Majorana mean-field Hamiltonian in Eq.~\eqref{eq:BdG-Majorana-real-space} is exactly mapped, through the local canonical transformation, onto the equivalent Abrikosov-fermion BdG form
\begin{align}
    \label{eq:BdG-Abrikosov-fermion-real-space}
    \mathcal{H}^{\text{MF}}=\frac{1}{2}\sum_{l,l^{\prime}}\mathcal{F}_{l}^{\dagger}H_{ll^{\prime}}\mathcal{F}_{l^{\prime}},
\end{align}
where $i=(l,s)$, $l=1,2,\ldots,N/2$, and $s=\text{A},\text{B}$.
Here $H_{ll^{\prime}}$ is the $8\times 8$ real-space BdG kernel, whose blocks obey $H_{ll^{\prime}}^{\dagger}=H_{l^{\prime}l}$, and $\mathcal{F}_{l}^{\dagger}$ is the eight-component operator vector defined below.
\begin{align}
    \label{eq:definition-Fl}
    \mathcal{F}_{l}^{\dagger}=\Bigl(f_{(l,\text{A}),\uparrow}^{\dagger}\:f_{(l,\text{B}),\uparrow}^{\dagger}\:f_{(l,\text{A}),\downarrow}^{\dagger}\:f_{(l,\text{B}),\downarrow}^{\dagger}\notag\\
    f_{(l,\text{A}),\uparrow}\:f_{(l,\text{B}),\uparrow}\:f_{(l,\text{A}),\downarrow}\:f_{(l,\text{B}),\downarrow}\Bigr).
\end{align}
By introducing the Fourier transform of the operator vector $\mathcal{F}_{l}$, the Hamiltonian can be written as
\begin{align}
    \label{eq:BdG-Abrikosov-fermion-momentum-space}
    \mathcal{H}^{\text{MF}}=\frac{1}{2}\sum_{\bm{k}}^{\text{BZ}}\mathcal{F}_{\bm{k}}^{\dagger}H_{\bm{k}}\mathcal{F}_{\bm{k}}, 
\end{align}
where the sum over $\bm{k}$ is taken over the first Brillouin zone. The corresponding momentum-space operator vector $\mathcal{F}_{\bm{k}}^{\dagger}$ is defined by
\begin{align}
    \label{eq:definition-Fk}
    \mathcal{F}_{\bm{k}}^{\dagger}=\Bigl(f_{\bm{k},\text{A},\uparrow}^{\dagger}\:f_{\bm{k},\text{B},\uparrow}^{\dagger}\:f_{\bm{k},\text{A},\downarrow}^{\dagger}\:f_{\bm{k},\text{B},\downarrow}^{\dagger}\notag\\
    f_{-\bm{k},\text{A},\uparrow}\:f_{-\bm{k},\text{B},\uparrow}\:f_{-\bm{k},\text{A},\downarrow}\:f_{-\bm{k},\text{B},\downarrow}\Bigr).
\end{align}
Using the same site-position convention, we define
\begin{align}
    \label{eq:fourier-transformation}
    f_{\bm{k},s,\mu}^{\dagger}=\sqrt{\frac{2}{N}}\sum_{l}f_{(l,s),\mu}^{\dagger}e^{i\bm{k}\cdot\bm{r}_{l,s}},
\end{align}
where $s=\mathrm A,\mathrm B$ and $\mu=\uparrow,\downarrow$. The $8\times8$ BdG matrix is
\begin{align}
    \label{eq:definition-K-momentum-space}
    [H_{\bm{k}}]_{mn}=\sum_{l^{\prime}}[H_{ll^{\prime}}]_{mn}e^{-i\bm{k}\cdot\left(\bm{r}_{l,m}-\bm{r}_{l^{\prime},n}\right)}.
\end{align}
Here $\bm r_{l,m}$ is the position of the site associated with Nambu component $m=1,\ldots,8$ in Eq.~\eqref{eq:definition-Fl}. The particle and hole components at a given site carry the same Fourier phase in $\mathcal F_{\bm k}$.
The matrix $H_{\bm{k}}$ is independent of $l$ because $H_{ll^{\prime}}$ depends only on the relative coordinate $\bm{R}_{l}-\bm{R}_{l^{\prime}}$.
We diagonalize it by the Bogoliubov transformation $E_{\bm{k}}=T_{\bm{k}}^{\dagger}H_{\bm{k}}T_{\bm{k}}$, where the $8\times 8$ unitary matrix $T_{\bm{k}}$ satisfies $T_{\bm{k}}^{\dagger}T_{\bm{k}}=T_{\bm{k}}T_{\bm{k}}^{\dagger}=\bm{1}_{8\times 8}$.
The diagonal energy matrix is
$E_{\bm{k}}=\text{diag}\{\varepsilon_{\bm{k},1},\ldots,\varepsilon_{\bm{k},4},-\varepsilon_{-\bm{k},1},\ldots,-\varepsilon_{-\bm{k},4}\}\equiv\text{diag}\{\varepsilon_{\bm{k},\eta}\}_{\eta=1}^{8}$.
For $\eta=5,\ldots,8$, the notation in the last expression is defined by $\varepsilon_{\bm{k},\eta}\equiv-\varepsilon_{-\bm{k},\eta-4}$.
Using this Bogoliubov transformation, the mean-field Hamiltonian can be written in diagonal form as
\begin{align}
    \label{eq:BdG-Abrikosov-fermion-diagonal}
    \mathcal{H}^{\text{MF}}=\frac{1}{2}\sum_{\bm{k}}^{\text{BZ}}\Gamma_{\bm{k}}^{\dagger}E_{\bm{k}}\Gamma_{\bm{k}},
\end{align}
where the basis that diagonalizes the Hamiltonian is defined as
$\Gamma_{\bm{k}}=T_{\bm{k}}^{-1}\mathcal{F}_{\bm{k}}$. Its components are
\begin{align}
    \label{eq:diagonalized-basis}
    \Gamma_{\bm{k}}=\left(\gamma_{\bm{k},1},\ldots,\gamma_{\bm{k},4},\gamma_{-\bm{k},1}^{\dagger},\ldots,\gamma_{-\bm{k},4}^{\dagger}\right)^{T}.
\end{align}
For the positive-energy branches $\eta=1,\ldots,4$, the thermal occupation is $\langle\gamma_{\bm{k},\eta}^{\dagger}\gamma_{\bm{k},\eta}\rangle=g(\varepsilon_{\bm{k},\eta})$, where $g$ is defined in Eq.~\eqref{eq:distribution-function-majorana}.

\subsection{Calculation of the spin structure factor}
\label{sec:Calculation of the spin structure factor}

We calculate the dynamical spin structure factor, which is directly accessible in inelastic neutron-scattering experiments, in the Abrikosov fermion representation.
We define the fluctuation about the static moment by $\delta S_i^\alpha=S_i^\alpha-m_i^\alpha$, where $m_i^\alpha=\langle S_i^\alpha\rangle$.
The connected dynamical spin structure factor is defined directly in terms of these fluctuations as
\begin{align}
    \label{eq:definition-dynamical-spin-structure-factor}
    S^{\alpha\alpha^{\prime}}(\bm{q},\omega)=\int_{-\infty}^{\infty}\frac{dt}{2\pi}e^{i\omega t}\frac{1}{N}\sum_{i,j}\langle \delta S_{i}^{\alpha}(t)\delta S_{j}^{\alpha^{\prime}}(0)\rangle e^{-i\bm{q}\cdot\left(\bm{r}_{i}-\bm{r}_{j}\right)},
\end{align}
Here, $\bm r_i=\bm r_{l,s}$ for $i=(l,s)$.
Thus, the elastic Bragg contribution from the static moments is excluded from the outset.
We focus on the inelastic part, excluding any strictly zero-energy contribution.
In practice, we evaluate the corresponding Matsubara susceptibility and perform the analytic continuation at the final stage.
We introduce the bare imaginary-time correlation function
\begin{align}
    \label{eq:definition-imaginary-time-spin-correlation}
    \chi_{ij}^{\alpha\alpha^{\prime}(0)}(\tau)
    =
    \left\langle \mathcal{T}_{\tau}\delta S_{i}^{\alpha}(\tau)\delta S_{j}^{\alpha^{\prime}}(0)\right\rangle_{0},
    \qquad
    0\leq \tau<\beta ,
\end{align}
where \(\mathcal{T}_{\tau}\) is the imaginary-time-ordering operator and
\(\langle\cdots\rangle_{0}\) denotes the thermal average with respect to
\(\mathcal{H}^{\text{MF}}\).
The imaginary-time spin operator is given by
\begin{align}
    \label{eq:imaginary-time-spin}
    S_{i}^{\alpha}(\tau)=\frac{1}{2}\sum_{\mu,\nu}\bar{f}_{i\mu}(\tau)\sigma^{\alpha}_{\mu\nu}f_{i\nu}(\tau),
\end{align}
with
\begin{align}
    \label{eq:imaginary-time-fermion}
    \bar{f}_{i\mu}(\tau)
    &=e^{\tau\mathcal{H}^{\text{MF}}}f_{i\mu}^{\dagger}e^{-\tau\mathcal{H}^{\text{MF}}},\\
    f_{i\nu}(\tau)
    &=e^{\tau\mathcal{H}^{\text{MF}}}f_{i\nu}e^{-\tau\mathcal{H}^{\text{MF}}}.
\end{align}
Here the bar on \(\bar{f}\) denotes the imaginary-time-evolved creation operator and should not be interpreted as complex conjugation.
The corresponding bare Matsubara susceptibility is
\begin{align}
    \label{eq:definition-matsubara-susceptibility}
    \chi^{\alpha\alpha^{\prime}(0)}(\bm{q},i\omega_{n})
    =
    \frac{1}{N}\sum_{i,j}e^{-i\bm{q}\cdot(\bm{r}_{i}-\bm{r}_{j})}
    \int_{0}^{\beta}d\tau\,e^{i\omega_{n}\tau}
    \chi_{ij}^{\alpha\alpha^{\prime}(0)}(\tau),
\end{align}
where \(\omega_{n}=2\pi nT\) is a bosonic Matsubara frequency.
Because the mean-field Hamiltonian is quadratic, this fluctuation susceptibility can be evaluated by Wick's theorem in imaginary time.
In the Nambu operator basis \(\mathcal{F}_{\bm{k}}\), we define the fermionic Matsubara Green's function
\begin{align}
    \label{eq:definition-nambu-green-matsubara}
    \mathcal{G}_{\bm{k},mm^{\prime}}(i\nu_{\ell})
    \equiv
    -\int_{0}^{\beta}d\tau\,e^{i\nu_{\ell}\tau}
    \left\langle \mathcal{T}_{\tau}\mathcal{F}_{\bm{k},m}(\tau)\mathcal{F}_{\bm{k},m^{\prime}}^{\dagger}(0)\right\rangle_{0},
\end{align}
where \(\nu_{\ell}=(2\ell+1)\pi T\) is a fermionic Matsubara frequency.
Using \(\mathcal{F}_{\bm{k}}=T_{\bm{k}}\Gamma_{\bm{k}}\), one obtains
\begin{align}
    \label{eq:nambu-green-bdg-matsubara}
    \mathcal{G}_{\bm{k},mm^{\prime}}(i\nu_{\ell})
    =
    \sum_{\eta=1}^{8}
    \left(T_{\bm{k}}\right)_{m\eta}
    \frac{1}{i\nu_{\ell}-\varepsilon_{\bm{k},\eta}}
    \left(T_{\bm{k}}^{\dagger}\right)_{\eta m^{\prime}}.
\end{align}
After summing over the internal fermionic Matsubara frequency, the Wick bubble at $i\omega_n\ne0$ can be written in the generic pole form
\begin{align}
    \label{eq:chi0-pole-form}
    \chi^{\alpha\alpha^{\prime}(0)}(\bm{q},i\omega_{n})
    =
    \frac{1}{N}\sum_{\bm{k}}^{\mathrm{BZ}}\sum_{\eta,\eta^{\prime}=1}^{8}
    \mathcal{W}_{\eta\eta^{\prime}}^{\alpha\alpha^{\prime}}(\bm{k},\bm{q})
    \frac{g(\varepsilon_{\bm{k},\eta})-g(\varepsilon_{\bm{k}+\bm{q},\eta^{\prime}})}
    {i\omega_{n}+\varepsilon_{\bm{k},\eta}-\varepsilon_{\bm{k}+\bm{q},\eta^{\prime}}},
\end{align}
Here, $\eta$ and $\eta^{\prime}$ run independently over the eight signed BdG branches.
The weight \(\mathcal{W}_{\eta\eta^{\prime}}^{\alpha\alpha^{\prime}}(\bm{k},\bm{q})\) contains the spin matrices, sublattice matrix elements, BdG transformation matrices, and the signs associated with the normal and anomalous Wick contractions.
Equation~\eqref{eq:chi0-pole-form} is the Matsubara representation of the bare bubble used in the numerical calculation.
The retarded bare susceptibility is obtained only after the analytic continuation
\begin{align}
    \label{eq:analytic-continuation-chi0}
    \chi^{\alpha\alpha^{\prime}(0),R}(\bm{q},\omega)
    =
    \chi^{\alpha\alpha^{\prime}(0)}(\bm{q},i\omega_{n}\to \omega+i\delta).
\end{align}

\subsection{Random phase approximation}
\label{sec:Random phase approximation}

\begin{figure}[t]
  \centering
      \includegraphics[width=\columnwidth,clip]{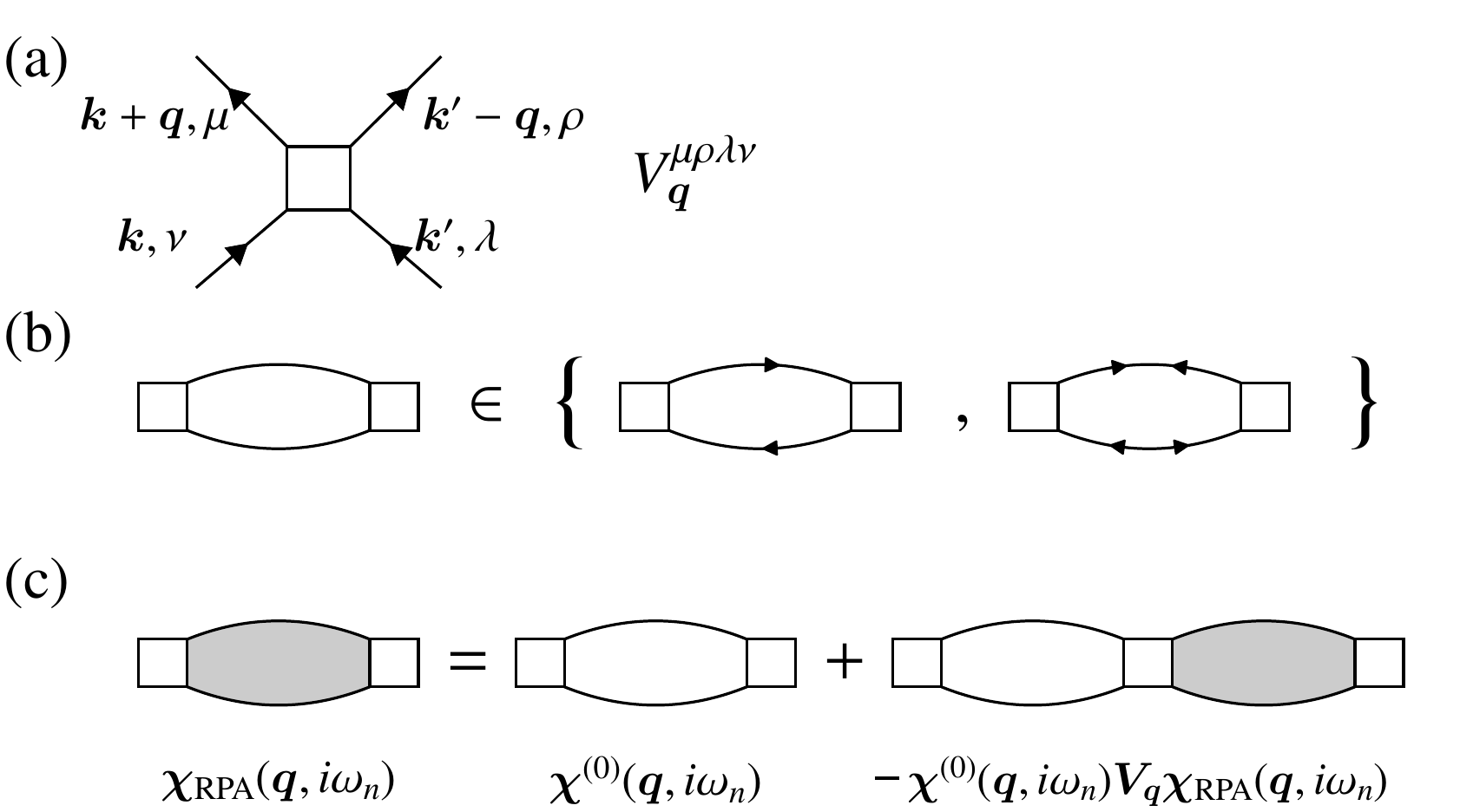}
      \caption{
      (a) Schematic representation of the four-point vertex $V_{\bm{q}}^{\mu\rho\lambda\nu}$ defined in Eq.~\eqref{eq:definition-vertex}.
      (b) Bubble diagrams contributing to the bare Matsubara Wick susceptibility $\bm{\chi}^{(0)}(\bm{q},i\omega_{n})$.
      In addition to the conventional particle-number-conserving bubble, anomalous bubble contributions also appear, reflecting the particle-number-nonconserving nature of the Abrikosov fermion mean-field theory.
      (c) Feynman-diagram representation of the Dyson equation for the RPA susceptibility.
      The shaded bubble denotes the RPA-dressed Matsubara susceptibility $\bm{\chi}_{\mathrm{RPA}}(\bm{q},i\omega_{n})$, obtained by connecting the bare matrices $\bm{\chi}^{(0)}$ constructed from the Wick bubbles in (b) through the four-point vertex shown in (a).
      }
      \label{fig:RPA_diagram.pdf}
\end{figure}

We now outline the RPA used to incorporate interaction effects beyond the Abrikosov fermion mean-field theory.
Details of the RPA formulation are given in Appendices~\ref{app:local-moment-normalization} and~\ref{app:Detailed calculation of the random-phase approximation}.
For comparison, the Majorana-fermion response theory and the differences between the unprojected spin responses in the two representations are discussed in Appendix~\ref{app:majorana-perturbation-theory}.
The starting point is the bare Matsubara Wick susceptibility of the spin fluctuations defined in Eq.~\eqref{eq:definition-matsubara-susceptibility}.
It is represented by the bubble diagrams in Fig.~\ref{fig:RPA_diagram.pdf}(b), including the anomalous contribution shown on the right that arises from the particle-number-nonconserving mean-field Hamiltonian.
In the actual RPA calculation, we do not project the bare bubble onto the final Cartesian spin components before the resummation.
Instead, we retain both channel and sublattice indices.  The rows and columns of the susceptibility matrices are labeled by $(a,u)$, where $a\in\{+,-,\uparrow,\downarrow\}$ denotes the channel and $u\in\{\mathrm{A},\mathrm{B}\}$ the sublattice.
The four channel operators are
$S_{i}^{+}=f_{i\uparrow}^{\dagger}f_{i\downarrow}$,
$S_{i}^{-}=f_{i\downarrow}^{\dagger}f_{i\uparrow}$,
$S_{i}^{\uparrow}=f_{i\uparrow}^{\dagger}f_{i\uparrow}$, and
$S_{i}^{\downarrow}=f_{i\downarrow}^{\dagger}f_{i\downarrow}$.
The bare and RPA susceptibilities of the channel fluctuations $\delta S_i^a=S_i^a-\langle S_i^a\rangle_0$ are represented by $8\times8$ matrices with elements
\begin{align}
    \label{eq:channel-susceptibility-matrices}
    \left[\bm\chi^{(0)}(\bm q,i\omega_n)\right]_{(a,u),(b,v)}
    &=\chi^{(0)}_{(a,u),(b,v)}(\bm q,i\omega_n),\notag\\
    \left[\bm\chi_{\mathrm{RPA}}(\bm q,i\omega_n)\right]_{(a,u),(b,v)}
    &=\chi_{\mathrm{RPA};(a,u),(b,v)}(\bm q,i\omega_n).
\end{align}
Before the RPA resummation, we rescale the bare susceptibility to satisfy the local-moment sum rule, as detailed in Appendix~\ref{app:local-moment-normalization}.  The explicit Wick-contraction formula for its elements is given in Appendix~\ref{app:Detailed calculation of the random-phase approximation}, and $\bm\chi_{\mathrm{RPA}}$ is determined by the Dyson equation below.
Following electronic mean-field treatments combined with RPA~\cite{Baym-Kadanoff-1961,Schrieffer-Wen-Zhang-1989,Yonemitsu-Batistic-Bishop-1991,Yonemitsu-Bishop-1992,Lorenzana-Grynberg-Yu-Yonemitsu-Bishop-1993,Belitz-Kirkpatrick-1997,Bishop-McMullen-2000,Bunemann-Gebhard-2001}, we use the exchange interaction as the RPA vertex for fluctuations about the mean-field saddle.  The mean-field propagators determine $\bm\chi^{(0)}$, and repeated scattering through this vertex is then resummed.  We write the interaction as
\begin{align}
    \label{eq:Hamiltonian-int}
    \mathcal{H}_{\mathrm{int}}=\frac{1}{2}\sum_{i,j}\sum_{\alpha,\alpha^{\prime}}J_{ij}^{\alpha\alpha^{\prime}}S_{i}^{\alpha}S_{j}^{\alpha^{\prime}},
\end{align}
where the sum over $(i,j)$ is unrestricted and the prefactor $1/2$ avoids double counting bonds.
For the present model, Eq.~\eqref{eq:Hamiltonian-int} is the exchange-interaction part of Eq.~\eqref{eq:KJG-Hamiltonian}.  The Zeeman term is included in the quadratic mean-field Hamiltonian and therefore modifies the bare susceptibility, but, being a one-body term, it does not generate a four-fermion RPA vertex.
Expressing $\mathcal{H}_{\mathrm{int}}$ in terms of a two-body interaction of Abrikosov fermions, one obtains
\begin{align}
    \label{eq:Hamiltonian-AF-int-real-space}
    \mathcal{H}_{\mathrm{int}}=\frac{1}{2}\frac{1}{4}\sum_{i,j}\sum_{\alpha,\alpha^{\prime}}\sum_{\mu,\nu,\rho,\lambda}J_{ij}^{\alpha\alpha^{\prime}}\sigma^{\alpha}_{\mu\nu}\sigma^{\alpha^{\prime}}_{\rho\lambda}f_{i\mu}^{\dagger}f_{j\rho}^{\dagger}f_{j\lambda}f_{i\nu},
\end{align}
Applying a Fourier transformation, this becomes
\begin{align}
    \label{eq:Hamiltonian_AF-int-momentum}
    \mathcal{H}_{\mathrm{int}}=\frac{1}{2}\frac{2}{N}
    \sum_{\bm{k},\bm{k}^{\prime},\bm{q}}
    \sum_{u,v}\sum_{\mu,\nu,\rho,\lambda}
    \left[V_{\bm{q}}^{\mu\rho\lambda\nu}\right]_{uv}
    f_{\bm{k}+\bm{q},u,\mu}^{\dagger}
    f_{\bm{k}^{\prime}-\bm{q},v,\rho}^{\dagger}
    f_{\bm{k}^{\prime},v,\lambda}
    f_{\bm{k},u,\nu},
\end{align}
Here $u,v\in\{\mathrm{A},\mathrm{B}\}$ label the two sublattices.  For fixed spin indices, the corresponding $2\times2$ sublattice block is
\begin{align}
    \label{eq:definition-vertex}
    \left[V_{\bm{q}}^{\mu\rho\lambda\nu}\right]_{uv}
    =\frac{1}{4}\frac{2}{N}
    \sum_{i\in u,j\in v}\sum_{\alpha,\alpha^{\prime}}
    J_{ij}^{\alpha\alpha^{\prime}}
    \sigma^{\alpha}_{\mu\nu}\sigma^{\alpha^{\prime}}_{\rho\lambda}
    e^{-i\bm{q}\cdot\left(\bm{r}_{i}-\bm{r}_{j}\right)}.
\end{align}
The full vertex matrix uses the same channel--sublattice indices as the susceptibilities:
\begin{align}
    \label{eq:vertex-channel-elements}
    \left[\bm V_{\bm q}\right]_{(a,u),(b,v)}
    =\left[V_{\bm q}^{\mu\rho\lambda\nu}\right]_{uv}.
\end{align}
Here $a$ specifies $(\mu,\nu)$ and $b$ specifies $(\rho,\lambda)$, with the correspondence $+\leftrightarrow(\uparrow,\downarrow)$, $-\leftrightarrow(\downarrow,\uparrow)$, $\uparrow\leftrightarrow(\uparrow,\uparrow)$, and $\downarrow\leftrightarrow(\downarrow,\downarrow)$. Four channels and two sublattices give eight values for each matrix index. The Cartesian indices $\alpha,\alpha^{\prime}$ in Eq.~\eqref{eq:definition-vertex} are summed over and are not matrix labels. The explicit $8\times8$ block form is given in Appendix~\ref{app:Detailed calculation of the random-phase approximation}.
For a directed A-to-B bond, $\bm r_j-\bm r_i=\bm\delta_\gamma$, so the vertex phase is $e^{i\bm q\cdot\bm\delta_\gamma}$. This is the same site-position convention used for the mean-field Hamiltonian and the bare susceptibility.
The matrices $\bm{\chi}^{(0)}(\bm{q},i\omega_{n})$ and $\bm{V}_{\bm{q}}$ define the Dyson equation shown diagrammatically in Fig.~\ref{fig:RPA_diagram.pdf}(c) and written as
\begin{align}
    \label{eq:Dyson-eq}
    \bm{\chi}_{\text{RPA}}(\bm{q},i\omega_{n})
    =
    \bm{\chi}^{(0)}(\bm{q},i\omega_{n})
    -
    \bm{\chi}^{(0)}(\bm{q},i\omega_{n})
    \bm{V}_{\bm{q}}\,
    \bm{\chi}_{\text{RPA}}(\bm{q},i\omega_{n}).
\end{align}
The minus sign in the second term is the perturbative sign associated with one insertion of the interaction vertex.
Its solution is
\begin{align}
    \label{eq:sol-Dyson-eq}
    \bm{\chi}_{\text{RPA}}(\bm{q},i\omega_{n})
    =
    \left[
    \bm{1}_{8}
    +
    \bm{\chi}^{(0)}(\bm{q},i\omega_{n})\bm{V}_{\bm{q}}
    \right]^{-1}
    \bm{\chi}^{(0)}(\bm{q},i\omega_{n}),
\end{align}
where $\bm{1}_{8}$ is the $8\times8$ identity matrix.
For the two-sublattice response formulated here, the cell-normalized retarded RPA susceptibility is defined as
\begin{align}
    \label{eq:retarded-susceptibility-convention}
    \chi_{\text{RPA}}^{\alpha\alpha^{\prime},R}(\bm{q},t)
    &=
    i\theta(t)\frac{2}{N}\sum_{i,j}
    e^{-i\bm q\cdot(\bm r_i-\bm r_j)}
    \left\langle
    \left[\delta S_i^{\alpha}(t),\delta S_j^{\alpha^{\prime}}(0)\right]
    \right\rangle_{\text{RPA}},
\end{align}
so that it is obtained from the Matsubara susceptibility by
\begin{align}
    \label{eq:analytic-continuation-rpa}
    \bm{\chi}_{\text{RPA}}^{R}(\bm{q},\omega)
    &=
    \bm{\chi}_{\text{RPA}}(\bm{q},i\omega_{n}\to \omega+i\delta).
\end{align}
The Cartesian components $\chi^{\alpha\alpha^{\prime},R}_{\text{RPA}}(\bm{q},\omega)$ are obtained after this matrix resummation by taking the appropriate linear combinations of the eight-channel susceptibility and summing over sublattice indices without an additional position-dependent phase, as described in Appendix~\ref{app:Detailed calculation of the random-phase approximation}.
For the diagonal spin components used in this work, the fluctuation-dissipation theorem gives the interaction-renormalized dynamical structure factor in the form
\begin{align}
    \label{eq:Dynamical-structure-factor-RPA}
    S^{\alpha\alpha}_{\text{RPA}}(\bm{q},\omega)
    &=
    \frac{1}{2}\frac{1}{\pi}\frac{1}{1-e^{-\beta\omega}}
    \mathrm{Im}\chi_{\text{RPA}}^{\alpha\alpha,R}(\bm{q},\omega).
\end{align}
The prefactor $1/2$ accounts for the normalization convention of the two-sublattice formulation.  The channel susceptibility used in the RPA is defined with $2/N$, whereas Eq.~\eqref{eq:definition-dynamical-spin-structure-factor} uses $1/N$.

\section{Results}
\label{sec:results}

We present the dynamical spin structure factor obtained using Abrikosov-fermion mean-field theory combined with RPA.
The quantity shown below is the trace over the diagonal spin components,
\begin{align}
    S(\bm{q},\omega)=\sum_{\alpha=x,y,z}S_{\text{RPA}}^{\alpha\alpha}(\bm{q},\omega).
\end{align}
The momentum path is $\text{K}^{\prime}-\Gamma-\text{M}-\text{K}-\Gamma^{\prime}-\text{M}^{\prime}-\text{K}^{\prime}$ [Fig.~\ref{fig:honeycomb_lattice.pdf}(b)].
The numerical conditions for all plots of $S(\bm q,\omega)$ presented below are summarized in Appendix~\ref{app:numerical-details}.

\subsection{Benchmark for pure Kitaev model}
\label{sec:Benchmark for pure Kitaev model}

\begin{figure*}[t]
  \centering
      \includegraphics[width=2\columnwidth,clip]{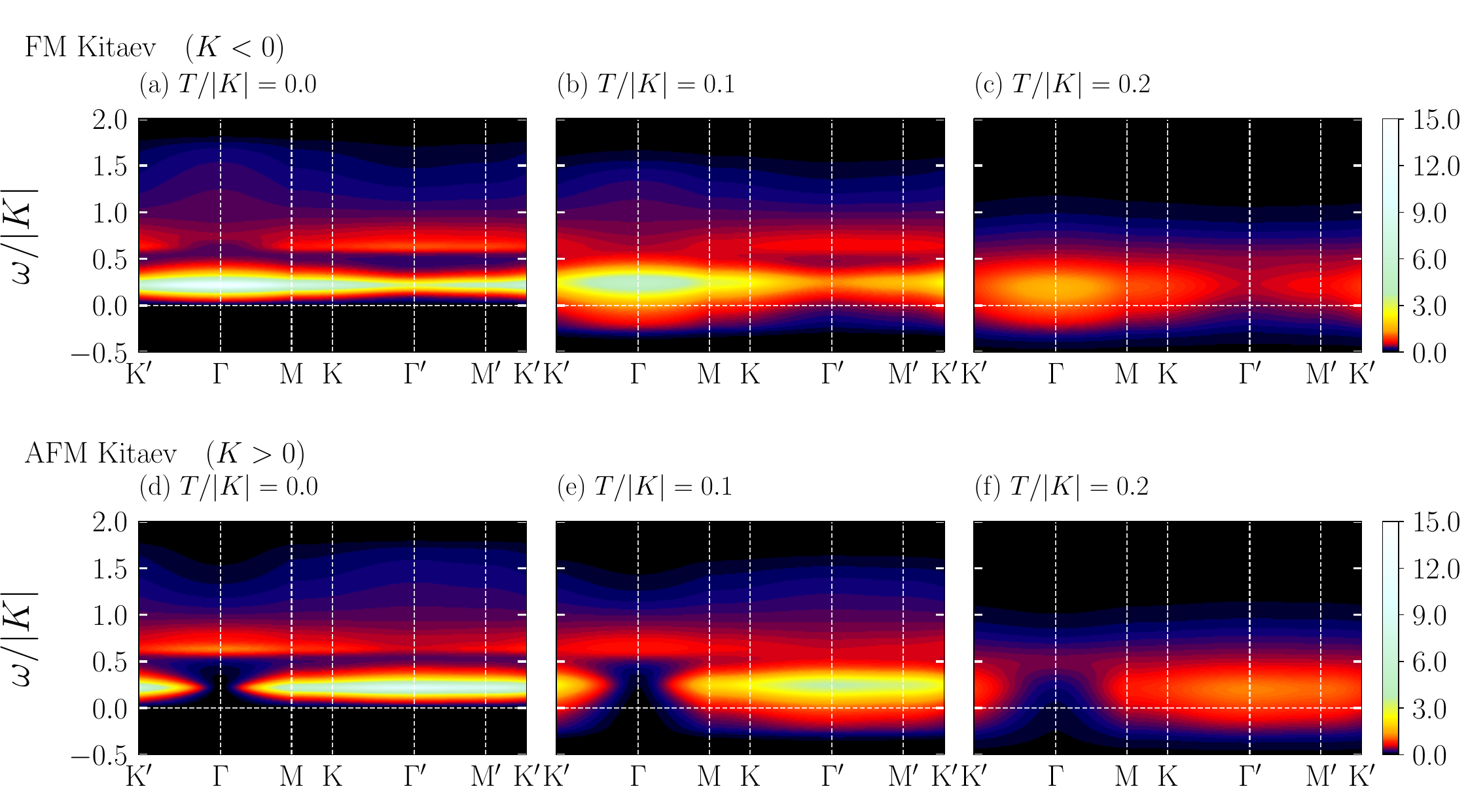}
      \caption{
      Temperature evolution of the dynamical spin structure factor $S(\bm{q},\omega)$ for the pure Kitaev model, calculated using the present parton-RPA approach.
      The upper panels [(a)--(c)] show the results for the FM Kitaev model ($K<0$), while the lower panels [(d)--(f)] show those for the AFM Kitaev model ($K>0$).
      From left to right, the panels correspond to $T/|K|=0.0$, $T/|K|=0.1$, and $T/|K|=0.2$.
      }
      \label{fig:pure_Kitaev.pdf}
\end{figure*}

We first present the results for the pure Kitaev model without non-Kitaev interactions.
Figure~\ref{fig:pure_Kitaev.pdf} shows the temperature evolution of the dynamical structure factor calculated using the present parton-RPA approach.
Figures~\ref{fig:pure_Kitaev.pdf}(a) and~\ref{fig:pure_Kitaev.pdf}(d), Figs.~\ref{fig:pure_Kitaev.pdf}(b) and~\ref{fig:pure_Kitaev.pdf}(e), and Figs.~\ref{fig:pure_Kitaev.pdf}(c) and~\ref{fig:pure_Kitaev.pdf}(f) show the data at $T/|K|=0.0$, $0.1$, and $0.2$, respectively.
Figures~\ref{fig:pure_Kitaev.pdf}(a)--\ref{fig:pure_Kitaev.pdf}(c) correspond to the FM Kitaev model $(K<0)$, while Figs.~\ref{fig:pure_Kitaev.pdf}(d)--\ref{fig:pure_Kitaev.pdf}(f) correspond to the AFM Kitaev model $(K>0)$.
All six panels contain a continuum extending over the full momentum path.
At $T/|K|=0.0$, the continuum is clearly visible up to approximately $\omega/|K|=1.5$ for both signs of $K$, while much weaker residual weight extends to approximately $\omega/|K|=1.8$.
The strongest component lies below $\omega/|K|\simeq0.4$, while a weaker, nearly horizontal band is visible around $\omega/|K|=0.6$--$0.8$ and is followed by a broad high-energy tail.

For the FM Kitaev model in Fig.~\ref{fig:pure_Kitaev.pdf}(a), the low-energy band has its largest intensity over the momentum region containing the $\Gamma$ point and remains visible along the adjacent $\mathrm{K}^{\prime}$--$\Gamma$ and $\Gamma$--$\mathrm{M}$ segments.
Its intensity is lower in the region containing the $\Gamma^{\prime}$ point, although finite spectral weight is maintained throughout the displayed path.
At $T/|K|=0.1$ [Fig.~\ref{fig:pure_Kitaev.pdf}(b)], the strongest response broadens into a low-energy region whose maximum remains at a small positive energy, while the high-energy tail is reduced.
At $T/|K|=0.2$ [Fig.~\ref{fig:pure_Kitaev.pdf}(c)], most of the visible weight is confined to a still broader low-energy strip and the high-energy tail is further suppressed.
The enhancement around the $\Gamma$ point remains visible at both finite temperatures, while its contrast relative to the rest of the momentum path decreases.

For the AFM Kitaev model at $T/|K|=0.0$ [Fig.~\ref{fig:pure_Kitaev.pdf}(d)], the low-energy band is strongly suppressed in a narrow momentum region around the $\Gamma$ point.
Away from this region, the band extends across the remaining path, with its strongest weight distributed over the side containing the $\Gamma^{\prime}$ and $\mathrm{M}^{\prime}$ points.
The suppression around the $\Gamma$ point remains visible at $T/|K|=0.1$ and $0.2$ [Figs.~\ref{fig:pure_Kitaev.pdf}(e) and~\ref{fig:pure_Kitaev.pdf}(f)], but the boundary of the suppressed region becomes smoother as the low-energy feature broadens.
As in the FM case, the clearly visible high-energy extent contracts continuously with increasing temperature.

For both signs of $K$, increasing temperature suppresses the high-energy tail, concentrates the displayed weight into a narrower energy interval, and smooths the momentum dependence of the low-energy band.
The contrast between the FM pattern centered at the $\Gamma$ point and the AFM pattern suppressed at the $\Gamma$ point remains present over the three temperatures, while the overall momentum-space modulation becomes weaker at the highest temperature.

The results in the pure Kitaev model provide a benchmark against the Majorana CDMFT+CTQMC calculation of Ref.~\cite{Yoshitake-Nasu-Motome-2016}.
The present calculation does not explicitly sample $\mathbb{Z}_2$ flux configurations. The use of parton RPA is motivated by its success at the pure Kitaev point in Ref.~\cite{Rao-Moessner-Knolle-2025}, where close agreement with the benchmark indicated that local flux effects were effectively captured.
In the present parton-RPA spectra shown in Fig.~\ref{fig:pure_Kitaev.pdf}, the low-energy response is strongest around the $\Gamma$ point in the FM Kitaev model, whereas it is suppressed at the $\Gamma$ point and enhanced toward the zone boundary in the AFM Kitaev model.
With increasing temperature, the present parton-RPA spectra also show a gradual redistribution of the dominant low-energy spectral weight toward lower energies. Both the contrasting FM and AFM momentum dependences and this thermal redistribution are qualitatively consistent with previous finite-temperature calculations~\cite{Yoshitake-Nasu-Motome-2016,Nasu-Motome-2021}.
To compare these qualitative trends with Ref.~\cite{Yoshitake-Nasu-Motome-2016}, Appendix~\ref{app:Comparison with prior work on the temperature evolution of the pure Kitaev model} presents the same spectra along the momentum path used in that study.

\subsection{Extended Kitaev models}
\label{sec:Extended Kitaev model}

We next consider the effects of a Heisenberg interaction, an off-diagonal $\Gamma$ interaction, and self-consistent magnetic order.

\subsubsection{Kitaev--Heisenberg model}
\label{sec:results-Kitaev-Heisenberg}

\begin{figure*}[t]
  \centering
      \includegraphics[width=2\columnwidth,clip]{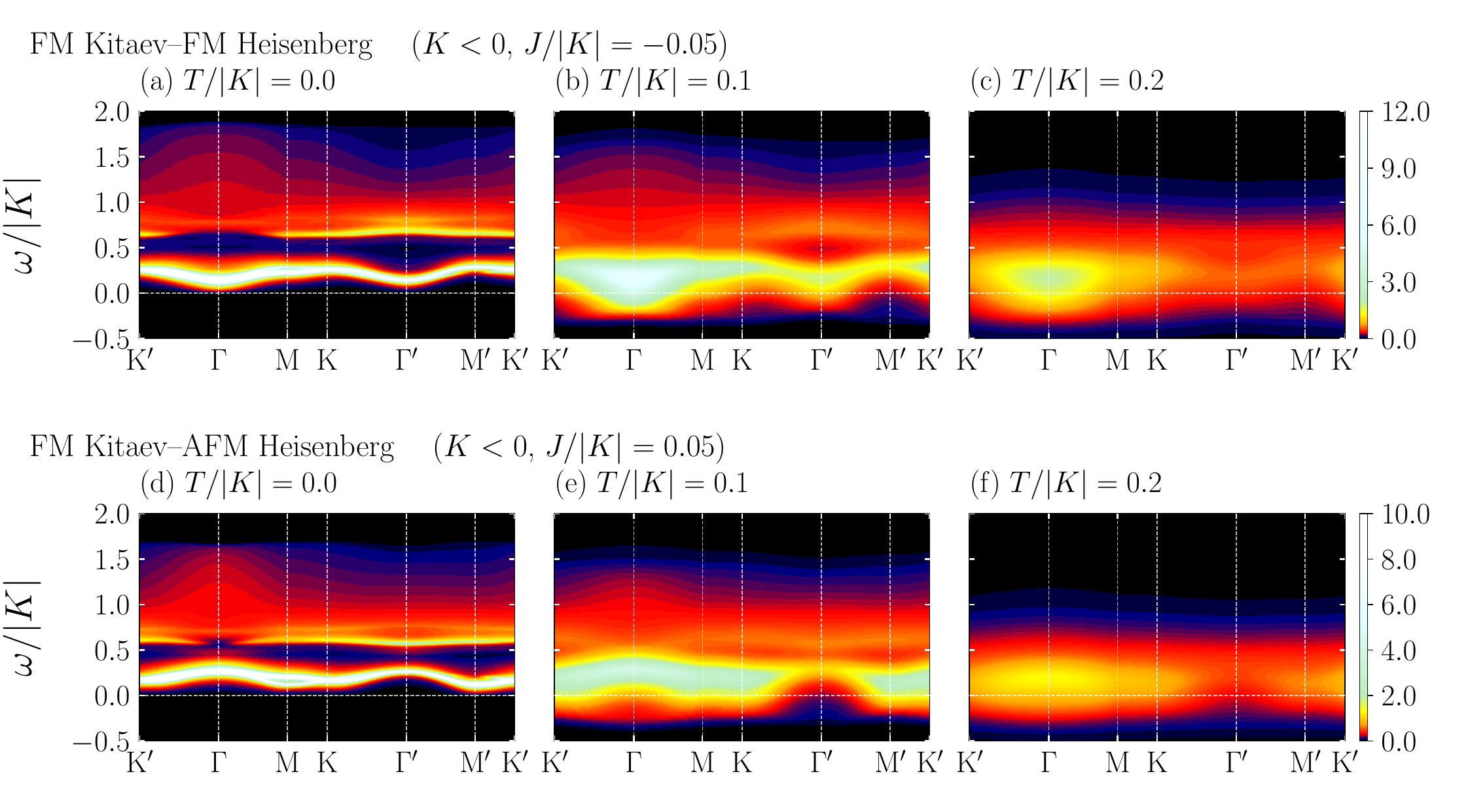}
      \caption{
      Temperature evolution of the liquid-saddle response $S(\bm{q},\omega)$ for the Kitaev--Heisenberg model, calculated using the present parton-RPA approach.
      The upper and lower rows correspond to the FM Kitaev--FM Heisenberg ($K<0$, $J/|K|=-0.05$) and FM Kitaev--AFM Heisenberg ($K<0$, $J/|K|=0.05$) models, respectively.
      From left to right, the panels correspond to $T/|K|=0.0$, $T/|K|=0.1$, and $T/|K|=0.2$.
      }
      \label{fig:Heisenberg_dssf.pdf}
\end{figure*}

Figure~\ref{fig:Heisenberg_dssf.pdf} shows the liquid-saddle response for $J/|K|=-0.05$ in the upper row and for $J/|K|=0.05$ in the lower row.
At all three temperatures, visible spectral weight extends over the full momentum path.
At $T/|K|=0.0$, several narrow low-energy peaks below $\omega/|K|\simeq0.5$ are superposed on a broader continuum.
The upper edge of the visible continuum reaches approximately $\omega/|K|=1.9$ for $J<0$ and $1.7$ for $J>0$.

For $J<0$ [Fig.~\ref{fig:Heisenberg_dssf.pdf}(a)], the strongest low-energy peak follows a shallow U-shaped momentum dependence along the $\mathrm{K}^{\prime}$--$\Gamma$--$\mathrm{M}$ segment, with a minimum around the $\Gamma$ point and higher energies toward $\mathrm{K}^{\prime}$ and $\mathrm{M}$.
A related modulation is repeated along the portion of the path containing the $\Gamma^{\prime}$ and $\mathrm{M}^{\prime}$ points.
The dispersive peaks intersect or approach one another near several high-symmetry points, producing alternating high- and low-intensity regions below $\omega/|K|\simeq0.4$.
For $J>0$ [Fig.~\ref{fig:Heisenberg_dssf.pdf}(d)], the momentum dependence is reversed: the low-energy peak reaches a higher energy around the $\Gamma$ point and lower energies toward the ends of the $\mathrm{K}^{\prime}$--$\Gamma$--$\mathrm{M}$ segment.
This inverted-U-shaped momentum dependence is repeated over the extended momentum path.

At $T/|K|=0.1$ [Figs.~\ref{fig:Heisenberg_dssf.pdf}(b) and~\ref{fig:Heisenberg_dssf.pdf}(e)], the separate low-energy peaks merge into a broader low-energy strip extending through $\omega=0$.
Residual momentum modulation remains visible, including the opposite momentum dependences around the $\Gamma$ point for the two signs of $J$.
At $T/|K|=0.2$ [Figs.~\ref{fig:Heisenberg_dssf.pdf}(c) and~\ref{fig:Heisenberg_dssf.pdf}(f)], this modulation is further reduced and the high-intensity region becomes nearly horizontal over most of the path.
Across each row, the visible upper edge decreases monotonically with temperature, reaching approximately $\omega/|K|=1.2$ for $J<0$ and $0.8$ for $J>0$ at $T/|K|=0.2$.
The high-energy intensity is simultaneously reduced, and the remaining weight is concentrated in the low-energy strip.

\subsubsection{\texorpdfstring{Kitaev--$\Gamma$ model}{Kitaev-Gamma model}}
\label{sec:results-Kitaev-Gamma}

\begin{figure}[t]
      \centering
      \includegraphics[width=\columnwidth,clip]{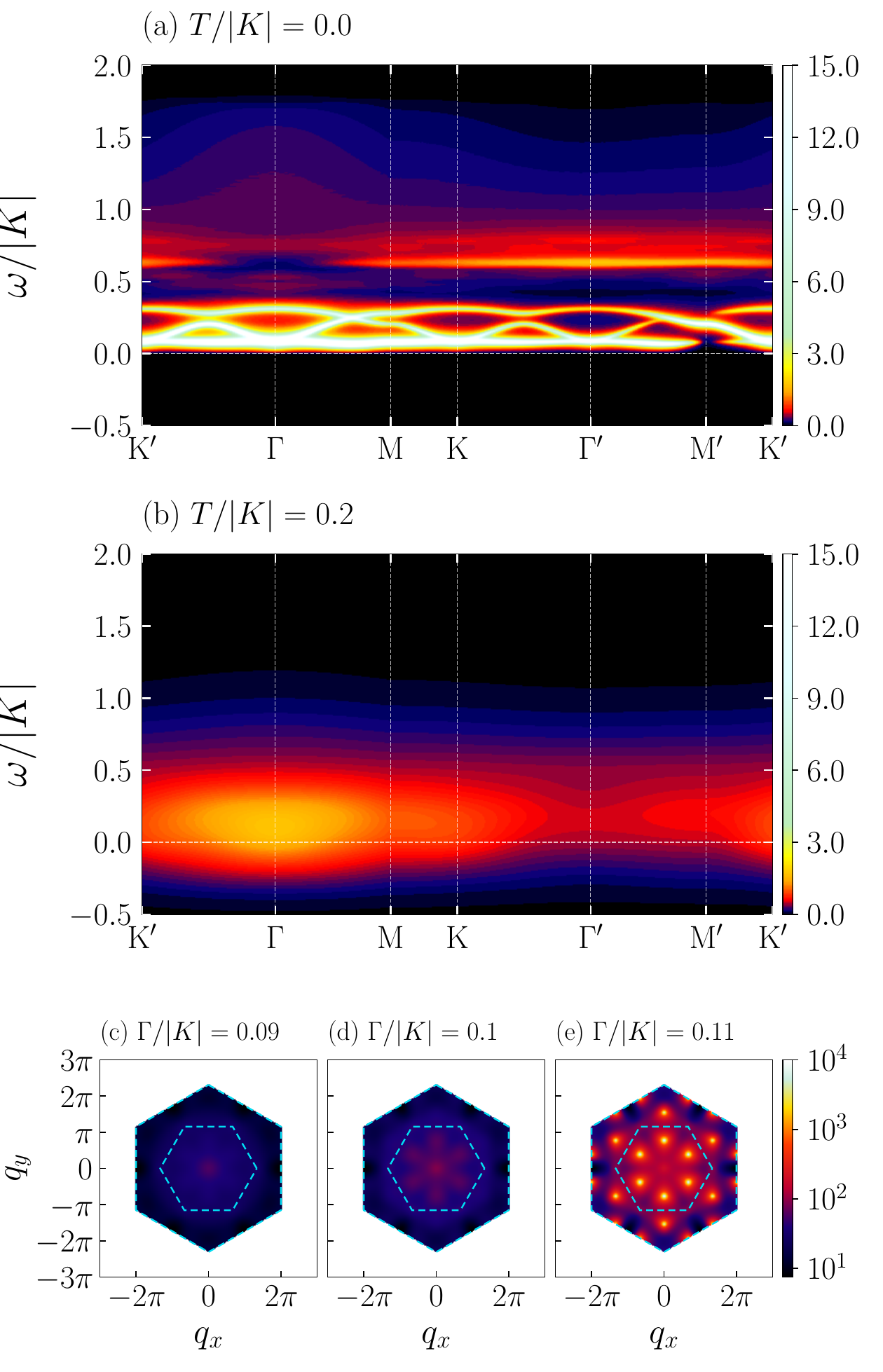}
      \caption{
      Dynamical and static responses of the Kitaev--$\Gamma$ model with FM Kitaev interaction ($K<0$), calculated using the present parton-RPA approach.
      Panels (a) and (b) show the dynamical spin structure factor $S(\bm{q},\omega)$ at $\Gamma/|K|=0.1$ for $T/|K|=0.0$ and $0.2$, respectively.
      Panels (c)--(e) show the zero-frequency static response $\chi_{\mathrm{static}}(\bm{q})$ per site at $T/|K|=0.0$ for $\Gamma/|K|=0.09$, $0.1$, and $0.11$, respectively.
      The color scale in panels (c)--(e) is logarithmic, and the cyan dashed lines indicate the boundaries of the first and extended Brillouin zones.
      }
      \label{fig:Ferro_Kitaev_Anti_Gamma.pdf}
\end{figure}

We calculate the dynamical and zero-frequency static responses of the Kitaev--$\Gamma$ model with $K<0$, $J=0$, and $\Gamma>0$.
Figures~\ref{fig:Ferro_Kitaev_Anti_Gamma.pdf}(a) and~\ref{fig:Ferro_Kitaev_Anti_Gamma.pdf}(b) show the dynamical spin structure factor $S(\bm{q},\omega)$ at $\Gamma/|K|=0.1$ for $T/|K|=0.0$ and $0.2$, respectively.
Both panels show a continuum over the entire momentum path.
At $T/|K|=0.0$ [Fig.~\ref{fig:Ferro_Kitaev_Anti_Gamma.pdf}(a)], the strongest low-energy response consists of multiple narrow, dispersive peaks below $\omega/|K|\simeq0.4$.
These peaks form a repeated pattern of crossings and local maxima along the extended high-symmetry path.
A weaker band remains visible around $\omega/|K|=0.6$--$0.8$, and diffuse weight extends to approximately $\omega/|K|=1.8$.
At $T/|K|=0.2$ [Fig.~\ref{fig:Ferro_Kitaev_Anti_Gamma.pdf}(b)], the separate low-energy peaks are no longer resolved and are replaced by a broad low-energy strip whose strongest weight lies at a small positive energy.
The strip is most intense around the $\Gamma$ point, is weaker around the $\Gamma^{\prime}$ point, and has only a shallow momentum dependence over the remaining path.
The high-energy extent is reduced, and the intensity above $\omega/|K|\simeq0.5$ is strongly suppressed relative to the zero-temperature panel.

For the static results in Figs.~\ref{fig:Ferro_Kitaev_Anti_Gamma.pdf}(c)--\ref{fig:Ferro_Kitaev_Anti_Gamma.pdf}(e), we plot the per-site zero-frequency trace
\begin{align*}
    \chi_{\mathrm{static}}(\bm{q})\equiv
    \frac{1}{2}
    \sum_{\alpha=x,y,z}
    \mathrm{Re}\,
    \chi_{\mathrm{RPA}}^{\alpha\alpha,R}
    (\bm{q},\omega=0).
\end{align*}
The factor $1/2$ converts the two-sublattice unit-cell normalization to a per-site response, and $\chi_{\mathrm{RPA}}^{\alpha\alpha,R}$ denotes the sublattice-summed Cartesian susceptibility defined in Eqs.~\eqref{eq:chi-RPA-cartesian}--\eqref{eq:chi-RPA-cartesian-zz}.
Figures~\ref{fig:Ferro_Kitaev_Anti_Gamma.pdf}(c)--\ref{fig:Ferro_Kitaev_Anti_Gamma.pdf}(e) show $\chi_{\mathrm{static}}(\bm{q})$ for $\Gamma/|K|=0.09$, $0.1$, and $0.11$, respectively.
The finite-$\bm q$ maxima are located near the sixfold symmetry star of the period-three wave vector $\bm Q=2\bm q_{\mathrm M}/3$, where $\bm q_{\mathrm M}$ denotes the momentum of a high-symmetry $\mathrm{M}$ point in the first Brillouin zone.
For $\Gamma/|K|=0.09$ [Fig.~\ref{fig:Ferro_Kitaev_Anti_Gamma.pdf}(c)], the response is broadly distributed and contains weak, spatially extended maxima around these wave vectors.
At $\Gamma/|K|=0.1$ [Fig.~\ref{fig:Ferro_Kitaev_Anti_Gamma.pdf}(d)], the same pattern becomes more intense and the individual maxima become more clearly separated from the surrounding background.
At $\Gamma/|K|=0.11$ [Fig.~\ref{fig:Ferro_Kitaev_Anti_Gamma.pdf}(e)], the response is concentrated into narrow peaks near these six symmetry-related positions in the first Brillouin zone, together with their reciprocal-lattice repetitions in the surrounding zones.
On the logarithmic color scale, the peak values in Fig.~\ref{fig:Ferro_Kitaev_Anti_Gamma.pdf}(e) approach the upper end of the displayed range, while the regions between the peaks remain substantially weaker.

The softening wave vector can be compared directly with the magnetic tendencies found in previous studies of the ferromagnetic Kitaev--$\Gamma$ model.
The same $2\bm q_{\mathrm M}/3$ ordering wave vector is the dominant Bragg vector of the six-site, period-three branches obtained in large-cluster classical simulated-annealing and energy-minimization studies of the $K<0$, $\Gamma>0$ model family at zero or finite $[111]$ field~\cite{Chern-Kaneko-Lee-Kim-2020,Rayyan-Luo-Kee-2021,Stavropoulos-Yang-Rousochatzakis-Perkins-2024}.
For the spin-$1/2$ model in a field, a thermodynamic-limit iTPS calculation likewise stabilizes the corresponding six-site state~\cite{Lee-Kaneko-Chern-Okubo-Yamaji-Kawashima-Kim-2020}, while zero-field 24-site exact diagonalization finds a compatible threefold-degenerate sector but does not resolve whether its order is six-site or spiral~\cite{Yamada-Suzuki-Suga-2020}.
The enhancement of the static response near $\bm Q=2\bm q_{\mathrm M}/3$ suggests magnetic softening at the wave vector associated with these period-three ordering tendencies.
More broadly, recent studies of the $K$--$J$--$\Gamma$ model using large-scale classical optimization and PFFRG have clarified its rich phase diagram, which includes competing noncollinear multiple-$Q$ and incommensurate phases as well as extended highly frustrated regions~\cite{Fukui-Motome-2026}.

\subsubsection{Proximate spin-liquid regime}
\label{sec:results-proximate-liquid}

\begin{figure*}[t]
  \centering
      \includegraphics[width=2\columnwidth,clip]{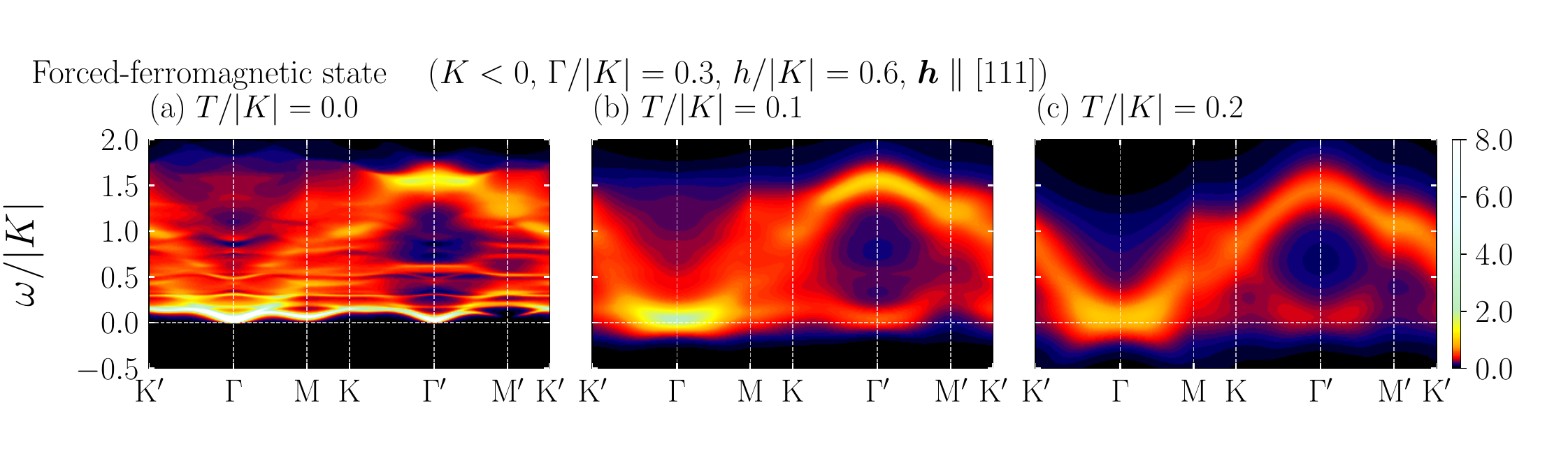}
      \caption{
      Temperature evolution of the dynamical spin structure factor $S(\bm{q},\omega)$ for a representative forced-ferromagnetic saddle of the FM Kitaev--positive-$\Gamma$ model in a $[111]$ field, with $\Gamma/|K|=0.3$ and $h/|K|=0.6$.
      From left to right, panels (a)--(c) correspond to $T/|K|=0.0$, $0.1$, and $0.2$.
      }
      \label{fig:Gamma_field_dssf.pdf}
\end{figure*}

The choice of the high-field spatially uniform branch is motivated by recent experiments on $\alpha$-RuCl$_3$.
For in-plane fields up to $24$~T, Ref.~\cite{Imamura-2026-high-field} reported that thermodynamic signatures associated with the Kitaev-spin-liquid regime persist even when the magnetization reaches approximately $90\%$ of its saturation value, while the response remains distinct from the response of a trivial spin-polarized state.
Although that experimental field geometry differs from the one used here, it motivates examining whether fractionalization-related dynamical features can remain visible deep in a nearly polarized, spatially uniform regime.

We use the spin-space $[111]$ direction, corresponding to the out-of-plane $c$-axis direction in the idealized honeycomb geometry.
Because all three field components are nonzero, a weak field in this direction gaps the Majorana cones and produces the non-Abelian chiral spin-liquid phase, while its continuation to intermediate and high fields has been widely studied in connection with field-induced phases, dynamical response, and the approach to polarization~\cite{Kitaev-2006,Janssen-Vojta-2019,Zhu-Kimchi-Sheng-Fu-2018,Gohlke-Moessner-Pollmann-2018,Hickey-Trebst-2019,Gordon-Catuneanu-Sorensen-Kee-2019,Kaib-Winter-Valenti-2019,Jiang-Devereaux-Jiang-2019,Lee-Kaneko-Chern-Okubo-Yamaji-Kawashima-Kim-2020,Zhang-Halasz-Batista-2022}.
We therefore use the widely studied $[111]$ direction to examine the forced-ferromagnetic saddle in the high-field regime.  The higher-field spatially uniform branch is consistent with the field evolution observed experimentally~\cite{Sears-2015,Kubota-2015,Banerjee-2016,Imamura-2026-high-field}.  Figure~\ref{fig:Gamma_field_dssf.pdf} shows the temperature evolution of the dynamical spin structure factor for this representative forced-ferromagnetic saddle.  More general Kitaev--$\Gamma$ and material Hamiltonians can support additional commensurate, enlarged-unit-cell, incommensurate, and field-induced states~\cite{Rau-Lee-Kee-2014,Maksimov-Chernyshev-2020,Lee-Kaneko-Chern-Okubo-Yamaji-Kawashima-Kim-2020,Rousochatzakis-Perkins-Luo-Kee-2024,Moller-Maksimov-Jiang-White-Valenti-Chernyshev-2025,Rao-Moessner-Knolle-2026}.

In the spatially uniform solutions considered here, full self-consistency drives the intersite fields $C_{\gamma}$ and $Q_{\gamma}^{\alpha\beta}$ to zero, leaving local field-split levels with flat parton dispersions.
To retain dispersive correlations inherited from the proximate Kitaev spin-liquid regime, we follow Ref.~\cite{Rao-Moessner-Knolle-2026} and first determine these intersite fields and the Lagrange multipliers $\lambda^\alpha$ in a liquid solution with $m_i^\alpha=0$ at each temperature.
Keeping these quantities fixed, we then solve self-consistently only for the onsite moments $m_i^\alpha$ in Eq.~\eqref{eq:general-majorana-decoupling} at the same temperature, with $\Gamma/|K|=0.3$ and $h/|K|=0.6$.
The liquid-state calculation is performed at the lower magnetic field at which the optimized nonuniform classical spin configuration becomes spatially uniform, as detailed in Appendix~\ref{app:classical-field-selection}.
The onsite moments are determined in the primitive two-sublattice unit cell and satisfy $\bm m_{\mathrm A}=\bm m_{\mathrm B}$, with the uniform moment aligned along $[111]$ at $T/|K|=0.0$.

At $T/|K|=0.0$, sharp low-energy structures are superposed on a broad continuum.
As the temperature increases, these structures broaden and spectral weight accumulates at low energies.
At the same time, the diffuse high-energy background is progressively suppressed, and the visible continuum contracts toward lower energies over much of the momentum path, although a broad dispersive feature remains around the $\Gamma^{\prime}$ point.
A broad continuum remains visible throughout the momentum path up to $T/|K|=0.2$.

\section{Discussion}
\label{sec:Discussion}

\begin{figure*}[t]
  \centering
      \includegraphics[width=2\columnwidth,clip]{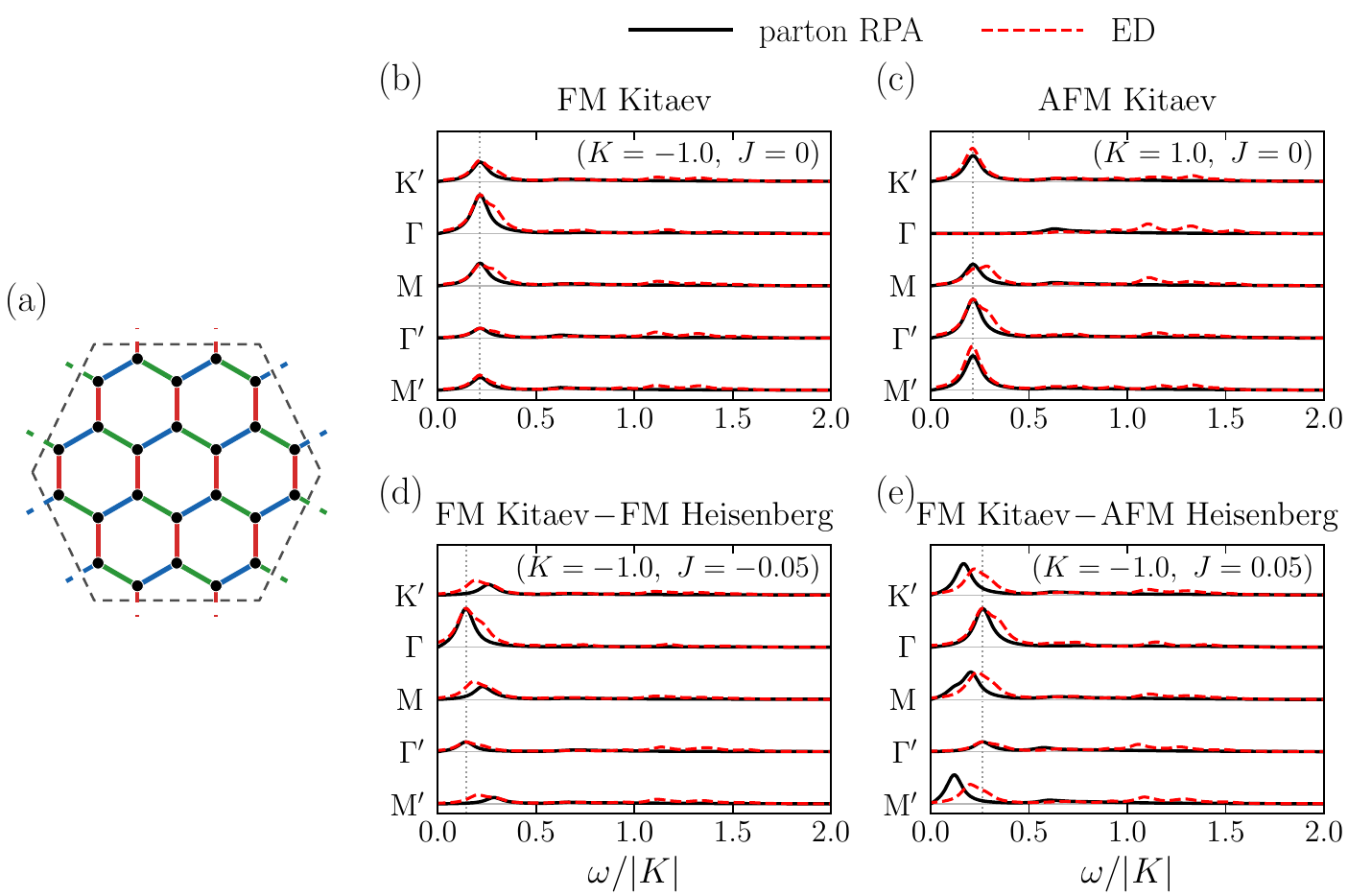}
      \caption{
      (a) $24$-site periodic cluster used for ED.
      (b)--(e) Zero-temperature $S(\bm{q},\omega)$ at the labeled high-symmetry points for the indicated models, comparing parton RPA (black solid) and ED (red dashed).
      Both calculations use $\delta/|K|=0.05$.
      The spectra are normalized by their maximum at $\bm q_{\rm ref}$, with $\bm q_{\rm ref}=\Gamma$ in panels (b), (d), and (e) and $\Gamma^{\prime}$ in panel (c); only the ED curves are shifted in energy as described in the text.
      }
      \label{fig:RPA_ED.pdf}
\end{figure*}

\begin{figure*}[t]
  \centering
      \includegraphics[width=2\columnwidth,clip]{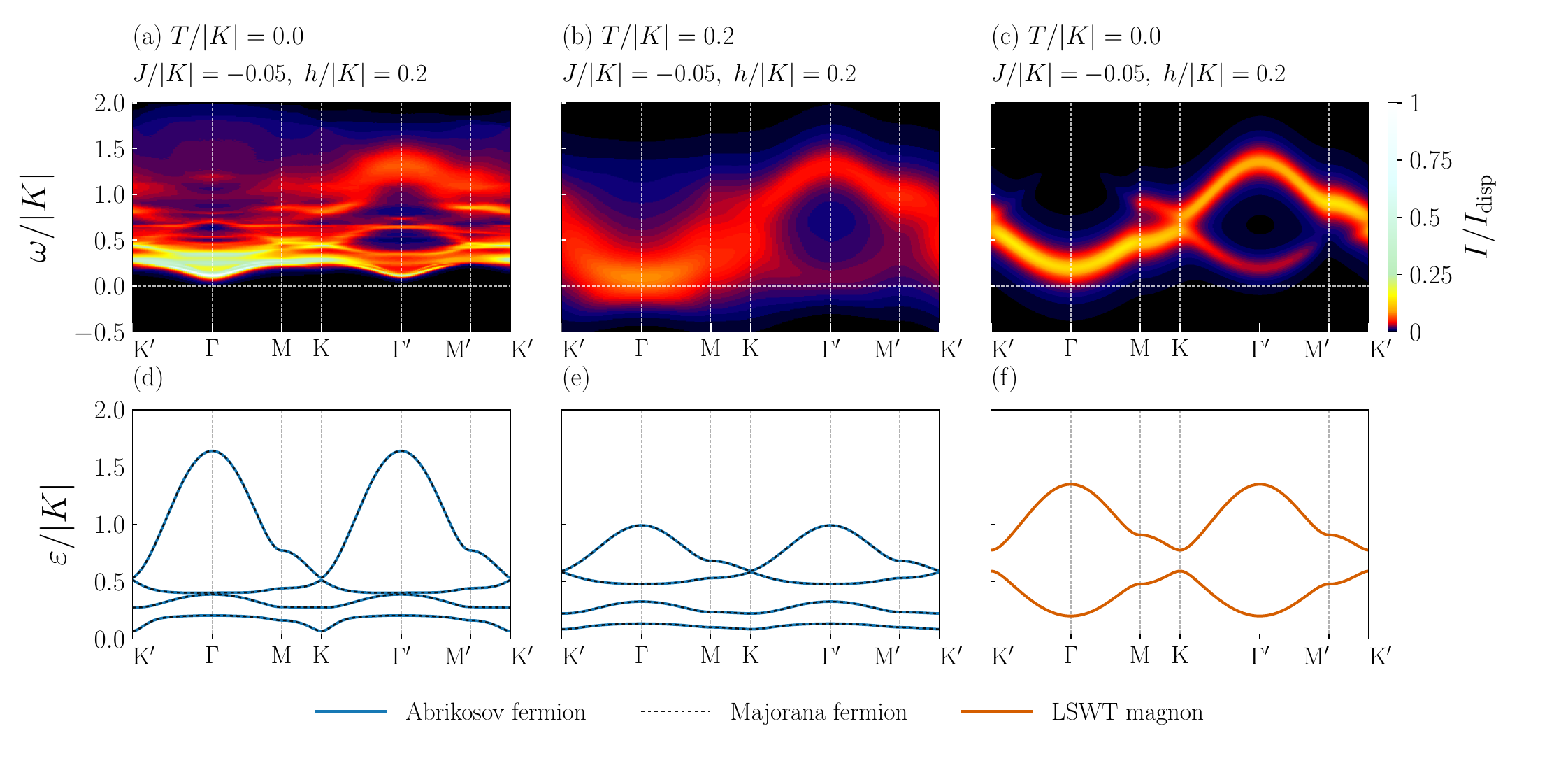}
      \caption{
      Comparison of the dynamical structure factors and the corresponding parton or magnon dispersions for ferromagnetic Kitaev coupling ($K<0$), $J/|K|=-0.05$, and $h/|K|=0.2$ along $[111]$.
      (a), (b) Parton-RPA results for the selected uniform-ferromagnetic saddle at $T/|K|=0.0$ and $0.2$, respectively.
      (c) LSWT result for the $[111]$ ferromagnetic state at the same $J$, $h$, and temperature as in panel (a).
      In panels (a)--(c), intensities are divided by the common display scale $I_{\mathrm{disp}}=12$.
      (d), (e) Positive parton bands underlying panels (a) and (b), respectively.  The solid blue and dotted black curves denote the Abrikosov- and Majorana-fermion dispersions, which coincide within numerical precision.
      (f) One-magnon bands underlying panel (c).
      }
      \label{fig:Heisenberg_magnon.pdf}
\end{figure*}

\begin{figure*}[t]
  \centering
      \includegraphics[width=2\columnwidth,clip]{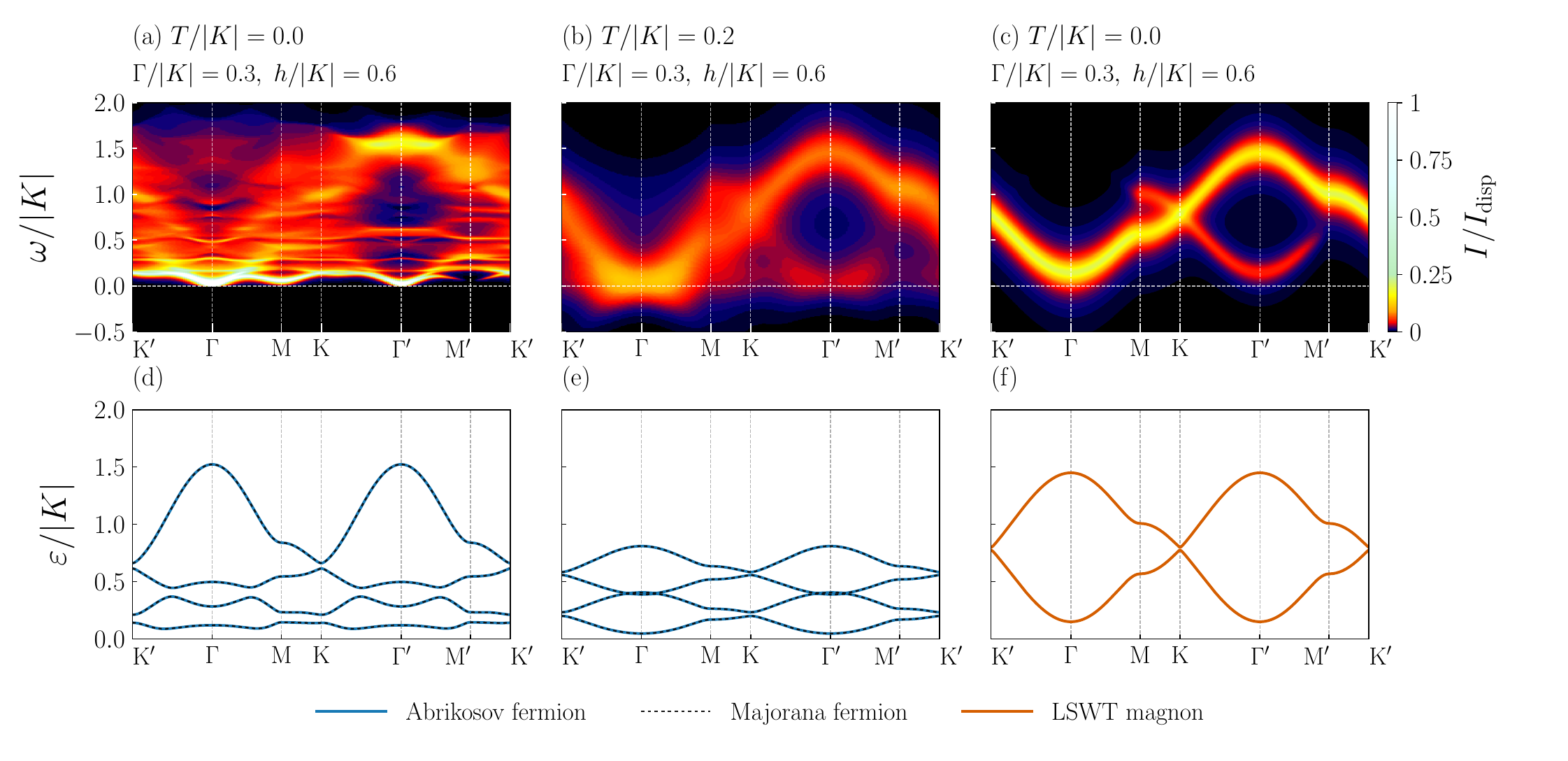}
      \caption{
      Field-induced comparison of the dynamical structure factors and the corresponding parton or magnon dispersions for ferromagnetic Kitaev coupling ($K<0$), $\Gamma/|K|=0.3$, and $h/|K|=0.6$ along $[111]$.
      (a), (b) Parton-RPA results for the selected uniform-ferromagnetic saddle at $T/|K|=0.0$ and $0.2$, respectively.
      (c) LSWT result for the $[111]$ ferromagnetic state at the same couplings, field, and temperature as in panel (a).
      In panels (a)--(c), intensities are divided by the common display scale $I_{\mathrm{disp}}=8$.
      (d), (e) Positive parton bands underlying panels (a) and (b), respectively.  The solid blue and dotted black curves denote the Abrikosov- and Majorana-fermion dispersions, which coincide within numerical precision.
      (f) One-magnon bands underlying panel (c).
      }
      \label{fig:Gamma_magnon.pdf}
\end{figure*}

A common feature of the finite-temperature spectra in Figs.~\ref{fig:pure_Kitaev.pdf},~\ref{fig:Heisenberg_dssf.pdf},~\ref{fig:Ferro_Kitaev_Anti_Gamma.pdf}, and~\ref{fig:Gamma_field_dssf.pdf} is that the visible high-energy extent of the broad continuum contracts as the temperature increases.
In fractionalized systems, broad continua commonly arise when a local spin operator couples to multiparticle states, as exemplified by the two- and multispinon response of the spin-$1/2$ Heisenberg chain~\cite{Muller-Thomas-Beck-Bonner-1981,Caux-Hagemans-2006} and by the Majorana-fermion and flux response of the Kitaev spin liquid~\cite{Baskaran-Mandal-2007,Knolle-2014,Knolle-2015,Hermanns-Kimchi-2018}.
In the present parton formulation, this connection is explicit.  The spin operator is bilinear in fermions, and the bare susceptibility is a two-quasiparticle bubble whose poles are fixed by differences between BdG eigenenergies, as shown in Eq.~\eqref{eq:chi0-channel-pole-form}.
The energy range of the bare continuum, and consequently much of the broad spectral weight retained after the RPA resummation, therefore follows the parton bandwidth.
The intersite channels $C_{\gamma}$ and $Q_{\gamma}^{\alpha\beta}$ enter the quadratic Majorana Hamiltonian and map onto the hopping and pairing amplitudes of the Abrikosov-fermion BdG Hamiltonian, as detailed in Appendix~\ref{app:Connection between Majorana-fermion and Abrikosov-fermion formulations}.
As documented in Appendix~\ref{app:numerical-details}, the magnitudes of the dominant bond amplitudes decrease smoothly with temperature for the representative liquid saddles.
Within the present theory, this temperature-induced reduction narrows the parton bands and compresses the associated two-parton phase space, thereby driving the contraction of the broad continuum toward lower energies.  The thermal occupation factors and the RPA redistribution further control the detailed spectral pattern.
Pronounced finite-temperature redistribution of the Kitaev continuum has also been established by Majorana CDMFT and unbiased QMC+CTQMC calculations~\cite{Yoshitake-Nasu-Motome-2016,Yoshitake-Nasu-2017,Yoshitake-Nasu-Motome-2017-QMC}, together with the thermal-fractionalization and flux-proliferation crossovers found in finite-temperature studies of the Kitaev model~\cite{Nasu-Udagawa-Motome-2014,Nasu-Udagawa-Motome-2015,Motome-Nasu-2020}.

The local-moment normalization requires separate care because the present framework combining mean-field theory and RPA is not a conserving approximation and the final RPA susceptibility does not automatically satisfy the exact local-moment sum rule.  For each saddle and temperature, we rescale the bare Wick susceptibility to satisfy the fluctuation local-moment sum rule, as detailed in Appendix~\ref{app:local-moment-normalization}.  Because the rescaled bubble also enters the RPA denominator, this prescription can change peak energies and instability conditions, not only spectral intensities.  At this fixed normalization, our analysis focuses on momentum dependence, relative peak positions and intensities, qualitative spectral shapes, and the redistribution of spectral weight, rather than on the absolute integrated intensity or gap.

Figure~\ref{fig:RPA_ED.pdf} compares the zero-temperature parton-RPA and exact-diagonalization (ED) spectra of the pure Kitaev and Kitaev--Heisenberg models at representative high-symmetry points.
The ED method is described in Appendix~\ref{app:Zero-temperature-exact-diagonalization-calculation-of-the-dynamical-spin-structure-factor}.
For each model, we align the ED reference peak with its parton-RPA counterpart by plotting all ED traces against $\omega-\Delta\omega$, while leaving the parton-RPA energy axis unchanged.
We define $\Delta\omega=\omega_{\mathrm{peak}}^{\rm ED}(\bm q_{\rm ref})-\omega_{\mathrm{peak}}^{\rm RPA}(\bm q_{\rm ref})$, where $\omega_{\mathrm{peak}}^{\rm ED}(\bm q_{\rm ref})$ and $\omega_{\mathrm{peak}}^{\rm RPA}(\bm q_{\rm ref})$ are the ED and RPA reference-peak energies, respectively.
The reference momentum is $\bm q_{\rm ref}=\Gamma$ for the FM Kitaev and Kitaev--Heisenberg panels and $\bm q_{\rm ref}=\Gamma^{\prime}$ for the AFM Kitaev panel.
For each method, all traces in a panel are normalized by the maximum intensity at $\bm q_{\rm ref}$ within $0\leq\omega/|K|\leq2$. This comparison tests relative peak positions, intensities, and line shapes, not the absolute gap.

Figure~\ref{fig:RPA_ED.pdf} shows several common spectral trends.
In the pure FM Kitaev case [Fig.~\ref{fig:RPA_ED.pdf}(b)], the dominant peaks at the $\mathrm{K}^{\prime}$, $\Gamma$, $\mathrm{M}$, and $\Gamma^{\prime}$ points are nearly aligned in energy, reflecting the flat low-energy structure also seen in Fig.~\ref{fig:pure_Kitaev.pdf}(a), and the two calculations exhibit the same qualitative intensity hierarchy.
In the AFM Kitaev case [Fig.~\ref{fig:RPA_ED.pdf}(c)], the strong suppression at the $\Gamma$ point and the enhancement toward the $\Gamma^{\prime}$ point are likewise captured.
Once a FM Heisenberg interaction is introduced [Fig.~\ref{fig:RPA_ED.pdf}(d)], the parton-RPA spectrum shows that the peaks at $\mathrm{K}^{\prime}$ and $\mathrm{M}$ move to higher energies relative to the dominant peak at the $\Gamma$ point. This behavior is consistent with the U-shaped momentum dependence of the low-energy peak along the $\mathrm{K}^{\prime}$-$\Gamma$-$\mathrm{M}$ path in Fig.~\ref{fig:Heisenberg_dssf.pdf}(a) and is also seen in the ED spectra.
For an AFM Heisenberg interaction [Fig.~\ref{fig:RPA_ED.pdf}(e)], the momentum dependence of the peak energy is reversed. The peak at the $\Gamma$ point lies at higher energy than the $\mathrm{K}^{\prime}$ and $\mathrm{M}$ peaks, consistently with the inverted-U-shaped momentum dependence of the low-energy peak in Fig.~\ref{fig:Heisenberg_dssf.pdf}(d).
The ED data likewise show a U-shaped momentum dependence of the low-energy peak energy for $J<0$ and an inverted-U-shaped dependence for $J>0$ along the $\mathrm{K}^{\prime}$--$\Gamma$--$\mathrm{M}$ path.
Residual differences in low-energy weights and detailed line shapes, including those near $\mathrm{M}^{\prime}$, can arise from both finite-size effects in ED and the parton-RPA approximation.
Taken together, the qualitative consistency with CDMFT+CTQMC for the pure model and with ED in the presence of Heisenberg interactions indicates that parton RPA provides a useful description of the characteristic temperature evolution and perturbation-dependent redistribution of spectral weight.

We finally turn to the relation between the proximate-spin-liquid response and conventional magnons.
The leading LSWT response consists of one-magnon poles, whereas parton RPA resums the two-fermion response generated by the bilinear spin operator. Broad high-energy weight alone does not identify fractionalized excitations, since it also occurs in classical Kitaev dynamics~\cite{Samarakoon-Banerjee-2017}.

Figure~\ref{fig:Heisenberg_magnon.pdf} compares the parton-RPA and LSWT responses and illustrates the temperature evolution of the parton response for the uniform-ferromagnetic branch of the Kitaev--Heisenberg model at $J/|K|=-0.05$ and $h/|K|=0.2$ along $[111]$.
At $T/|K|=0.0$, we compare the selected uniform-ferromagnetic parton-RPA response [Fig.~\ref{fig:Heisenberg_magnon.pdf}(a)] with LSWT [Fig.~\ref{fig:Heisenberg_magnon.pdf}(c)] at the same couplings, field, and temperature. Both spectra exhibit a low-energy peak whose energy is minimized near the $\Gamma$ point along $\mathrm{K}^{\prime}$--$\Gamma$--$\mathrm{M}$ and a higher-energy feature whose peak energy is maximized near the $\Gamma^{\prime}$ point along $\mathrm{K}$--$\Gamma^{\prime}$--$\mathrm{M}^{\prime}$. In LSWT, these features follow the lower and upper magnon branches, respectively, whereas the higher-energy feature in parton RPA is broadened within the continuum.
The lower row demonstrates that the Abrikosov- and Majorana-fermion formulations have identical one-parton dispersions [Figs.~\ref{fig:Heisenberg_magnon.pdf}(d) and~\ref{fig:Heisenberg_magnon.pdf}(e)] and contrasts them with the one-magnon bands [Fig.~\ref{fig:Heisenberg_magnon.pdf}(f)].
Upon heating to $T/|K|=0.2$, the parton bands narrow [Figs.~\ref{fig:Heisenberg_magnon.pdf}(d) and~\ref{fig:Heisenberg_magnon.pdf}(e)]. In the response [Fig.~\ref{fig:Heisenberg_magnon.pdf}(b)], the fine peak structure becomes less pronounced with the larger broadening used at this temperature, while broad low-energy weight remains concentrated near the $\Gamma$ point and the higher-energy dispersive feature retains its maximum near the $\Gamma^{\prime}$ point.

Figure~\ref{fig:Gamma_magnon.pdf} compares the zero-temperature parton-RPA and LSWT responses and illustrates the temperature evolution of the parton response for the uniform branch of the Kitaev--$\Gamma$ model at $\Gamma/|K|=0.3$ and $h/|K|=0.6$.
At the same couplings, field, and temperature [Figs.~\ref{fig:Gamma_magnon.pdf}(a) and~\ref{fig:Gamma_magnon.pdf}(c)], the magnonlike similarity lies primarily in the high-energy structure: the broad parton-RPA feature reaches its highest energy near the $\Gamma^{\prime}$ point and moves to lower energies toward $\mathrm{K}$ and $\mathrm{M}^{\prime}$, resembling the upper magnon branch in LSWT.
Figure~\ref{fig:Gamma_magnon.pdf}(b) shows the subsequent temperature evolution within the parton approach. Upon heating to $T/|K|=0.2$, the parton bands narrow [Figs.~\ref{fig:Gamma_magnon.pdf}(d) and~\ref{fig:Gamma_magnon.pdf}(e)] and the high-energy continuum contracts. With the larger broadening used at this temperature, the fine peak structure becomes less pronounced, while the high-energy maximum near the $\Gamma^{\prime}$ point remains visible.

In both models, the smearing of fine peak structure at elevated temperature makes the correspondence with the LSWT magnon dispersion more apparent in the dynamical response, particularly for the high-energy feature near the $\Gamma^{\prime}$ point. These similarities in momentum dependence, together with the remaining two-parton continuum, are consistent with a spinon-to-magnon crossover at the level of the dynamical response. Related evolution from an incoherent fractionalized response to dispersive magnon bands has been reported in finite-temperature CTQMC calculations for the FM Kitaev model in a $[111]$ field~\cite{Yoshitake-Nasu-Kato-Motome-2020}, in field- and temperature-dependent calculations for a material-motivated model~\cite{Winter-Riedl-Kaib-2018}, and in semiclassical dynamical calculations of field-polarized Kitaev magnets~\cite{Zhang-Wilke-Kim-2023}.

To discuss the spinon-to-magnon conversion more rigorously, future work should follow a continuous evolution of self-consistent states and address the microscopic formation of spin-one excitations. This requires treating the local constraint and dynamical gauge-field fluctuations beyond the present approximation to resolve confinement and parton bound-state formation.

\section{Summary}
\label{sec:Summary}

We have developed a finite-temperature Abrikosov-fermion mean-field theory combined with RPA for extended Kitaev models by using the exact canonical mapping between the quadratic Majorana and Abrikosov-fermion Hamiltonians and constructing the corresponding matrix response theory for Heisenberg and $\Gamma$ interactions.
This framework treats liquid and selected ordered saddles within a common formalism while retaining identical Majorana- and Abrikosov-fermion dispersions.  In the pure Kitaev limit, it qualitatively captures the contrasting ferromagnetic and antiferromagnetic momentum structures and the downward thermal redistribution of the continuum found in CTQMC calculations.  The results identify the temperature-induced reduction of the intersite bond amplitudes as the mean-field mechanism that narrows the parton bands, compresses the two-particle phase space, and contracts the continuum toward lower energies.
For ferromagnetic Kitaev coupling, the low-energy peak has its minimum energy at the $\Gamma$ point for $J<0$ and its maximum energy there for $J>0$, in qualitative agreement with zero-temperature exact diagonalization.
In the ferromagnetic Kitaev--$\Gamma$ model, the static response develops finite-momentum softening, indicating an instability toward the period-three magnetic order found by independent numerical methods.
Finally, the finite-field comparisons with LSWT in the Kitaev--Heisenberg and Kitaev--$\Gamma$ models show that pronounced magnonlike spectral features can appear while a broad two-parton continuum remains present.
This coexistence produces spectral features consistent with a spinon-to-magnon crossover and supports the use of finite-temperature parton RPA to connect characteristic fractionalized dynamics near the Kitaev limit with collective spectral structures in proximate ordered regimes.

The present formulation opens several directions for future work.  It can be extended to material-motivated Hamiltonians containing $\Gamma^{\prime}$, longer-range interactions, and additional anisotropic interactions, and to frustrated lattices beyond the honeycomb geometry where Abrikosov-fermion parton descriptions are already widely used.  Calculations over continuous temperature and field ranges, together with extensions to the response functions appropriate to the coupling operators and selection rules of each probe, would enable more direct comparisons with inelastic neutron scattering, Raman, THz, and RIXS measurements in proximate Kitaev materials and other frustrated magnets.  Self-consistent treatments of competing ordered saddles and collective instabilities could further connect the evolution of dynamical spectra to field- and temperature-driven phase boundaries.  Beyond the present static parton mean-field framework combined with RPA, incorporating the local constraint, vertex and self-energy feedback, and dynamical gauge-field fluctuations in a conserving framework would provide a route to the flux dynamics, parton bound-state formation, and confinement processes that underlie the microscopic evolution from fractionalized continua to magnons.  Combining such developments with ED, tensor-network methods, and sign-problem-free numerical benchmarks where available should clarify absolute energy scales and critical behavior and broaden the use of finite-temperature parton response theory as a quantitative bridge between microscopic spin models and experiment.

\begin{acknowledgments}
  The authors thank A.~Ono for fruitful discussions.
  We also thank J.~Knolle for valuable advice.
  D.S. is grateful to P.~Rao for kindly providing the numerical code used in Ref.~\cite{Rao-Moessner-Knolle-2025}, as well as for insightful comments and useful advice.
  D.S. also thanks A.~Rosch, U.~F.~P.~Seifert, and A.~P.~Joy for valuable discussions.
  Parts of the numerical calculations were performed on the supercomputing systems at ISSP, The University of Tokyo.
  This work was supported by Grant-in-Aid for Scientific Research from
  JSPS, KAKENHI Grant Nos.~JP23H01129, JP23H04865, JP24K00563, JP26H00624, JP26H02230.
  D.S. acknowledges support from GP-Spin at Tohoku University.
\end{acknowledgments}

\appendix

\section{Connection between Majorana fermion and Abrikosov fermion formulations}
\label{app:Connection between Majorana-fermion and Abrikosov-fermion formulations}
These two representations are related by Eqs.~\eqref{eq:MF-AF-relation-up} and~\eqref{eq:MF-AF-relation-down}.
For an oriented bond $(i,j)$, we use the Abrikosov-fermion bond operators
\begin{align}
    \chi_{ij}
    &=\frac{1}{2}\sum_{\mu,\nu}f_{i\mu}^{\dagger}\sigma^{0}_{\mu\nu}f_{j\nu},
    &
    \eta_{ij}
    &=-\frac{i}{2}\sum_{\mu,\nu}f_{i\mu}\sigma^{y}_{\mu\nu}f_{j\nu},\\
    E_{ij}^{\alpha}
    &=\frac{1}{2}\sum_{\mu,\nu}f_{i\mu}^{\dagger}\sigma_{\mu\nu}^{\alpha}f_{j\nu},
    &
    D_{ij}^{\alpha}
    &=\frac{i}{2}\sum_{\mu,\nu}f_{i\mu}\left(\sigma^{y}\sigma^{\alpha}\right)_{\mu\nu}f_{j\nu}.
\end{align}
Here, $\sigma^{0}$ is the $2\times2$ identity matrix.
The corresponding unitary transformations are
\begin{align}
    \begin{pmatrix}
    f_{i\uparrow}\\f_{i\downarrow}\\f_{i\uparrow}^{\dagger}\\f_{i\downarrow}^{\dagger}
    \end{pmatrix}
    &=\frac{1}{\sqrt{2}}\begin{pmatrix}
        1&0&0&i\\
        0&i&-1&0\\
        1&0&0&-i\\
        0&-i&-1&0
    \end{pmatrix}
    \begin{pmatrix}
        c_{i}\\b_{i}^{x}\\b_{i}^{y}\\b_{i}^{z}
    \end{pmatrix},
    \label{eq:MF-AF-relation-matrix-c-to-f}\\
\begin{pmatrix}
    c_{i}\\b_{i}^{x}\\b_{i}^{y}\\b_{i}^{z}
\end{pmatrix}
&=\frac{1}{\sqrt{2}}\begin{pmatrix}
        1&0&1&0\\
        0&-i&0&i\\
        0&-1&0&-1\\
        -i&0&i&0
    \end{pmatrix}
    \begin{pmatrix}
    f_{i\uparrow}\\f_{i\downarrow}\\f_{i\uparrow}^{\dagger}\\f_{i\downarrow}^{\dagger}
    \end{pmatrix}.
    \label{eq:MF-AF-relation-matrix-f-to-c}
\end{align}
Using the channel truncation of Eq.~\eqref{eq:general-majorana-decoupling}, we transform the retained bond mean fields $C_{\gamma}$ and $Q_{\gamma}^{\alpha\beta}$ through Eq.~\eqref{eq:MF-AF-relation-matrix-c-to-f}. The brackets below denote the corresponding Abrikosov mean-field parameters:
\begin{align}
    \langle\chi_{ij}\rangle&=-\frac{i}{4}\left(C_{\gamma}+Q_{\gamma}^{zz}+Q_{\gamma}^{xx}+Q_{\gamma}^{yy}+iQ_{\gamma}^{xy}-iQ_{\gamma}^{yx}\right),\\
    \langle\eta_{ij}\rangle&=-\frac{i}{4}\left(Q_{\gamma}^{zx}+iQ_{\gamma}^{zy}-Q_{\gamma}^{xz}-iQ_{\gamma}^{yz}\right),\\
    \langle E_{ij}^{x}\rangle&=-\frac{i}{4}\left(Q_{\gamma}^{zx}+iQ_{\gamma}^{zy}+Q_{\gamma}^{xz}-iQ_{\gamma}^{yz}\right),\\
    \langle E_{ij}^{y}\rangle&=-\frac{i}{4}\left(-iQ_{\gamma}^{zx}+Q_{\gamma}^{zy}+iQ_{\gamma}^{xz}+Q_{\gamma}^{yz}\right),\\
    \langle E_{ij}^{z}\rangle&=-\frac{i}{4}\left(C_{\gamma}+Q_{\gamma}^{zz}-Q_{\gamma}^{xx}-Q_{\gamma}^{yy}-iQ_{\gamma}^{xy}+iQ_{\gamma}^{yx}\right),\\
    \langle D_{ij}^{x}\rangle&=-\frac{i}{4}\left(C_{\gamma}-Q_{\gamma}^{zz}+Q_{\gamma}^{xx}-Q_{\gamma}^{yy}+iQ_{\gamma}^{xy}+iQ_{\gamma}^{yx}\right),\\
    \langle D_{ij}^{y}\rangle&=\frac{1}{4}\left(C_{\gamma}-Q_{\gamma}^{zz}-Q_{\gamma}^{xx}+Q_{\gamma}^{yy}-iQ_{\gamma}^{xy}-iQ_{\gamma}^{yx}\right),\\
    \langle D_{ij}^{z}\rangle&=-\frac{i}{4}\left(Q_{\gamma}^{zx}+iQ_{\gamma}^{zy}+Q_{\gamma}^{xz}+iQ_{\gamma}^{yz}\right).
\end{align}
These relations map the Majorana mean-field solution onto the Abrikosov-fermion BdG Hamiltonian used in the numerical calculation.
We therefore use the bond fields
$\chi_{ij}$, $\eta_{ij}$, $E_{ij}^{\alpha}$, and $D_{ij}^{\alpha}$ directly, rather than introducing a separate SU($2$) gauge-matrix ansatz for the quadratic Hamiltonian.

We retain the Majorana saddle-point amplitudes $C_{\gamma}$ and $Q_{\gamma}^{\alpha\beta}$ to show which self-consistent fields enter each interaction.
The bond orientation is from the A to the B sublattice, as specified below Eq.~\eqref{eq:Kitaev-Hamiltonian-Majorana-mean-field}.
In Eqs.~\eqref{eq:abrikosov-mf-kitaev-explicit}--\eqref{eq:abrikosov-mf-gamma-explicit},
$\mathcal{H}_{X}^{\mathrm{MF}}$ denotes the quadratic contribution generated by the intersite fields $C_{\gamma}$ and $Q_{\gamma}^{\alpha\beta}$ for $X=K,J,\Gamma$.
The saddle-point constants and the magnetic-channel terms for an ordered state are fixed by Eq.~\eqref{eq:general-majorana-decoupling} and are not repeated below.

\begin{widetext}
For the Kitaev interaction, the quadratic contribution is
\begin{align}
    \mathcal{H}_{K}^{\mathrm{MF}}
    &=-\frac{iK}{2}\sum_{\gamma=x,y,z}\sum_{\langle i,j\rangle_{\gamma}}
    Q_{\gamma}^{\gamma\gamma}
    \left(D_{ij}^{x}-iD_{ij}^{y}-\chi_{ij}^{\dagger}-E_{ij}^{z\dagger}
    +\chi_{ij}+E_{ij}^{z}-D_{ij}^{x\dagger}-iD_{ij}^{y\dagger}\right)
    \notag\\
    &\quad-\frac{iK}{2}\sum_{\langle i,j\rangle_{x}}C_{x}
    \left(D_{ij}^{x}+iD_{ij}^{y}-\chi_{ij}^{\dagger}+E_{ij}^{z\dagger}
    +\chi_{ij}-E_{ij}^{z}-D_{ij}^{x\dagger}+iD_{ij}^{y\dagger}\right)
    \notag\\
    &\quad+\frac{iK}{2}\sum_{\langle i,j\rangle_{y}}C_{y}
    \left(D_{ij}^{x}+iD_{ij}^{y}+\chi_{ij}^{\dagger}-E_{ij}^{z\dagger}
    -\chi_{ij}+E_{ij}^{z}-D_{ij}^{x\dagger}+iD_{ij}^{y\dagger}\right)
    \notag\\
    &\quad+\frac{iK}{2}\sum_{\langle i,j\rangle_{z}}C_{z}
    \left(D_{ij}^{x}-iD_{ij}^{y}+\chi_{ij}^{\dagger}+E_{ij}^{z\dagger}
    -\chi_{ij}-E_{ij}^{z}-D_{ij}^{x\dagger}-iD_{ij}^{y\dagger}\right).
    \label{eq:abrikosov-mf-kitaev-explicit}
\end{align}
For the Heisenberg interaction, it is
\begin{align}
    \mathcal{H}_{J}^{\mathrm{MF}}
    &=\frac{iJ}{2}\sum_{\gamma=x,y,z}\sum_{\langle i,j\rangle_{\gamma}}C_{\gamma}
    \left(D_{ij}^{x}-iD_{ij}^{y}+3\chi_{ij}^{\dagger}-E_{ij}^{z\dagger}
    -3\chi_{ij}+E_{ij}^{z}-D_{ij}^{x\dagger}-iD_{ij}^{y\dagger}\right)
    \notag\\
    &\quad-\frac{iJ}{2}\sum_{\gamma=x,y,z}\sum_{\langle i,j\rangle_{\gamma}}
    \sum_{\alpha=x,y,z}Q_{\gamma}^{\alpha\alpha}
    \left(D_{ij}^{x}-iD_{ij}^{y}-\chi_{ij}^{\dagger}-E_{ij}^{z\dagger}
    +\chi_{ij}+E_{ij}^{z}-D_{ij}^{x\dagger}-iD_{ij}^{y\dagger}\right).
    \label{eq:abrikosov-mf-heisenberg-explicit}
\end{align}
For the off-diagonal $\Gamma$ interaction, the corresponding contribution is
\begin{align}
    \mathcal{H}_{\Gamma}^{\mathrm{MF}}
    &=-\frac{i\Gamma}{2}\sum_{\gamma=x,y,z}\sum_{\langle i,j\rangle_{\gamma}}
    \left(Q_{\gamma}^{\alpha_{\gamma}\beta_{\gamma}}
    +Q_{\gamma}^{\beta_{\gamma}\alpha_{\gamma}}\right)
    \left(D_{ij}^{x}-iD_{ij}^{y}-\chi_{ij}^{\dagger}-E_{ij}^{z\dagger}
    +\chi_{ij}+E_{ij}^{z}-D_{ij}^{x\dagger}-iD_{ij}^{y\dagger}\right)
    \notag\\
    &\quad-\Gamma\sum_{\langle i,j\rangle_{x}}C_{x}
    \left(D_{ij}^{z}-iE_{ij}^{y\dagger}+iE_{ij}^{y}+D_{ij}^{z\dagger}\right)
    \notag\\
    &\quad-i\Gamma\sum_{\langle i,j\rangle_{y}}C_{y}
    \left(D_{ij}^{z}-E_{ij}^{x\dagger}+E_{ij}^{x}-D_{ij}^{z\dagger}\right)
    \notag\\
    &\quad-\Gamma\sum_{\langle i,j\rangle_{z}}C_{z}
    \left(D_{ij}^{x}+iD_{ij}^{y}+D_{ij}^{x\dagger}-iD_{ij}^{y\dagger}\right).
    \label{eq:abrikosov-mf-gamma-explicit}
\end{align}
\end{widetext}
\section{Local-moment normalization of the Wick bubble}
\label{app:local-moment-normalization}

The unrescaled Wick susceptibility $\widetilde{\bm{\chi}}^{(0)}$ evaluated at the quadratic Abrikosov-fermion saddle includes both normal and anomalous contractions.
With the fluctuation definition in Eq.~\eqref{eq:definition-dynamical-spin-structure-factor}, the physical momentum--frequency zeroth moment reduces to
\begin{align}
    \label{eq:local-moment-sum-rule}
    \mathcal{M}_0^{\mathrm{phys}}
    \equiv
    \frac{1}{N}\sum_i\sum_{\alpha=x,y,z}
    \left\langle\left(\delta S_i^\alpha\right)^2\right\rangle
    =\frac{1}{N}\sum_i
    \left(\frac{3}{4}-\lvert\bm m_i\rvert^2\right).
\end{align}
For a nonmagnetic physical spin-$1/2$ state, this gives $\mathcal{M}_0^{\mathrm{phys}}=S(S+1)=3/4$.
Because the single-occupancy constraint is imposed only on average, the Gaussian saddle also contains empty and doubly occupied local states. The resulting local-moment deficit follows from the Abrikosov-fermion identity
\begin{align}
    \label{eq:abrikosov-local-moment-identity}
    \sum_{\alpha=x,y,z}\left(S_i^\alpha\right)^2
    =
    \frac{3}{4}
    \left(n_i-2n_{i\uparrow}n_{i\downarrow}\right).
\end{align}
For the nonmagnetic liquid saddle,
$\langle n_{i\uparrow}\rangle=\langle n_{i\downarrow}\rangle=1/2$ and the onsite spin-mixing and pairing averages vanish.  Wick's theorem then gives
\begin{align}
    \bigl\langle n_{i\uparrow}n_{i\downarrow}\bigr\rangle_0
    &=
    \bigl\langle n_{i\uparrow}\bigr\rangle_0
    \bigl\langle n_{i\downarrow}\bigr\rangle_0
    +\bigl|\bigl\langle f_{i\uparrow}f_{i\downarrow}\bigr\rangle_0\bigr|^2
    -\bigl|\bigl\langle f_{i\uparrow}^{\dagger}f_{i\downarrow}\bigr\rangle_0\bigr|^2
    =\frac{1}{4},
    \label{eq:wick-double-occupancy}\\
    \sum_{\alpha=x,y,z}
    \left\langle\left(S_i^\alpha\right)^2\right\rangle_0
    &=\frac{3}{8}
    =\frac{1}{2}S(S+1),
    \qquad S=\frac{1}{2}.
    \label{eq:wick-local-moment}
\end{align}
Equivalently, the unprojected saddle has probabilities
$P_0=P_2=1/4$ and $P_1=1/2$ for empty, doubly occupied, and singly occupied local states, respectively, whereas the exact constraint retains only $P_1=1$.
We verified Eq.~\eqref{eq:wick-local-moment} directly using the BdG covariance matrices. For both signs of the pure Kitaev coupling and for $T/|K|=0.0$, $0.1$, and $0.2$, we obtain
$\langle n_{i\uparrow}\rangle=\langle n_{i\downarrow}\rangle=1/2$,
$\langle n_{i\uparrow}n_{i\downarrow}\rangle=1/4$, and
$\sum_\alpha\langle(S_i^\alpha)^2\rangle=3/8$ to machine precision.

For a general saddle, the unrescaled $\delta S$ moment is
\begin{align}
    \widetilde{\mathcal M}_{0}
    &=
    \frac{1}{N}\sum_i\sum_{\alpha=x,y,z}
    \left[
    \left\langle\left(S_i^\alpha\right)^2\right\rangle_0
    -\left\langle S_i^\alpha\right\rangle_0^2
    \right]
    \notag\\
    &=\frac{1}{N}\sum_i\sum_{\alpha=x,y,z}
    \left\langle\left(\delta S_i^\alpha\right)^2\right\rangle_0.
    \label{eq:wick-fluctuation-local-moment}
\end{align}
Keeping the saddle-point magnetization $\bm m_i=\langle\bm S_i\rangle_0$ fixed, we determine the normalization factor of the bare fluctuation bubble from the zeroth-moment sum rule:
\begin{align}
    Z
    &=
    \frac{\mathcal M_{0}^{\mathrm{phys}}}
    {\widetilde{\mathcal M}_{0}},
    \qquad
    \bm{\chi}^{(0)}
    =Z\widetilde{\bm{\chi}}^{(0)}.
    \label{eq:sum-rule-rescaled-bubble}
\end{align}
Here $\widetilde{\bm{\chi}}^{(0)}$ is the unrescaled Wick matrix of $\delta S$ and $\bm{\chi}^{(0)}$ is the normalized matrix entering the RPA resummation.  For the nonmagnetic liquid saddles, Eqs.~\eqref{eq:wick-local-moment} and~\eqref{eq:wick-fluctuation-local-moment} give $\widetilde{\mathcal M}_{0}=3/8$ and $\mathcal M_{0}^{\mathrm{phys}}=3/4$, so Eq.~\eqref{eq:sum-rule-rescaled-bubble} reduces to the factor $Z=2$ used for the liquid calculations.  In magnetized saddles, $Z$ is instead evaluated directly from the local BdG covariance matrix for each parameter set and temperature.

\newpage
\section{Detailed calculation of the random phase approximation}
\label{app:Detailed calculation of the random-phase approximation}

The two-body interaction terms entering the RPA vertex are
\begin{align}
    \mathcal{H}_{\mathrm{pert}}
    &=\mathcal{H}_{\text{K}}+\mathcal{H}_{J}+\mathcal{H}_{\Gamma},
    \label{eq:perturb-Hamiltonian}\\
    \mathcal{H}_{\text{K}}
    &=K \sum_{\gamma=x,y,z} \sum_{\langle i,j\rangle_{\gamma}}S_{i}^{\gamma}S_{j}^{\gamma},
    \label{eq:perturb-Hamiltonian-K}\\
    \mathcal{H}_{J}
    &=J\sum_{\langle i,j\rangle}\bm{S}_{i}\cdot\bm{S}_{j},
    \label{eq:perturb-Hamiltonian-J}\\
    \mathcal{H}_{\Gamma}
    &=\Gamma\sum_{\gamma=x,y,z}\sum_{\langle i,j\rangle_{\gamma}}
    \left(S_{i}^{\alpha_{\gamma}}S_{j}^{\beta_{\gamma}}
    +S_{i}^{\beta_{\gamma}}S_{j}^{\alpha_{\gamma}}\right).
    \label{eq:perturb-Hamiltonian-Gamma}
\end{align}
Here $(\alpha_{\gamma},\beta_{\gamma})$ denotes the two spin components different from $\gamma$.
We use $(\alpha_{x},\beta_{x})=(y,z)$, $(\alpha_{y},\beta_{y})=(z,x)$, and $(\alpha_{z},\beta_{z})=(x,y)$.
Thus the off-diagonal $\Gamma$ interaction is counted once on each $\gamma$ bond.
The Zeeman term, if present,
\begin{align}
    \mathcal{H}_{h}
    =
    -\sum_{i,\alpha}h^{\alpha}S_{i}^{\alpha}
    =
    -\frac{1}{2}\sum_{i,\alpha}\sum_{\mu,\nu}h^{\alpha}
    f_{i\mu}^{\dagger}\sigma^{\alpha}_{\mu\nu}f_{i\nu},
\end{align}
is a one-body, quadratic term in the Abrikosov fermions.
It is therefore included in the mean-field Hamiltonian that determines the unrescaled Wick susceptibility $\widetilde{\bm{\chi}}^{(0)}$, but it does not generate a four-point RPA vertex.
In the Abrikosov fermion representation, $\mathcal{H}_{\mathrm{pert}}$ has the following three contributions.
\begin{widetext}
For the Kitaev interaction, we obtain
\begin{align}
    &\mathcal{H}_{\text{K}}=\frac{K}{4}\sum_{\langle i,j\rangle_{x}}\left(f_{i\uparrow}^{\dagger}f_{j\uparrow}^{\dagger}f_{j\downarrow}f_{i\downarrow}+f_{i\uparrow}^{\dagger}f_{j\downarrow}^{\dagger}f_{j\uparrow}f_{i\downarrow}+f_{i\downarrow}^{\dagger}f_{j\uparrow}^{\dagger}f_{j\downarrow}f_{i\uparrow}+f_{i\downarrow}^{\dagger}f_{j\downarrow}^{\dagger}f_{j\uparrow}f_{i\uparrow}\right)\nonumber\\
    &-\frac{K}{4}\sum_{\langle i,j\rangle_{y}}\left(f_{i\uparrow}^{\dagger}f_{j\uparrow}^{\dagger}f_{j\downarrow}f_{i\downarrow}-f_{i\uparrow}^{\dagger}f_{j\downarrow}^{\dagger}f_{j\uparrow}f_{i\downarrow}-f_{i\downarrow}^{\dagger}f_{j\uparrow}^{\dagger}f_{j\downarrow}f_{i\uparrow}+f_{i\downarrow}^{\dagger}f_{j\downarrow}^{\dagger}f_{j\uparrow}f_{i\uparrow}\right)\nonumber\\
    &+\frac{K}{4}\sum_{\langle i,j\rangle_{z}}\left(f_{i\uparrow}^{\dagger}f_{j\uparrow}^{\dagger}f_{j\uparrow}f_{i\uparrow}-f_{i\uparrow}^{\dagger}f_{j\downarrow}^{\dagger}f_{j\downarrow}f_{i\uparrow}-f_{i\downarrow}^{\dagger}f_{j\uparrow}^{\dagger}f_{j\uparrow}f_{i\downarrow}+f_{i\downarrow}^{\dagger}f_{j\downarrow}^{\dagger}f_{j\downarrow}f_{i\downarrow}\right).
\end{align}

For the Heisenberg interaction, we obtain
\begin{align}
    &\mathcal{H}_{J}=\frac{J}{2}\sum_{\langle i,j\rangle}\left(f_{i\uparrow}^{\dagger}f_{j\downarrow}^{\dagger}f_{j\uparrow}f_{i\downarrow}+f_{i\downarrow}^{\dagger}f_{j\uparrow}^{\dagger}f_{j\downarrow}f_{i\uparrow}\right)+\frac{J}{4}\sum_{\langle i,j\rangle}\left(f_{i\uparrow}^{\dagger}f_{j\uparrow}^{\dagger}f_{j\uparrow}f_{i\uparrow}-f_{i\uparrow}^{\dagger}f_{j\downarrow}^{\dagger}f_{j\downarrow}f_{i\uparrow}-f_{i\downarrow}^{\dagger}f_{j\uparrow}^{\dagger}f_{j\uparrow}f_{i\downarrow}+f_{i\downarrow}^{\dagger}f_{j\downarrow}^{\dagger}f_{j\downarrow}f_{i\downarrow}\right).
\end{align}

For the $\Gamma$ interaction, we obtain
\begin{align}
    &\mathcal{H}_{\Gamma}=-i\frac{\Gamma}{4}\sum_{\langle i,j\rangle_{x}}\left(
    f_{i\uparrow}^{\dagger}f_{j\uparrow}^{\dagger}f_{j\uparrow}f_{i\downarrow}
    -f_{i\uparrow}^{\dagger}f_{j\downarrow}^{\dagger}f_{j\downarrow}f_{i\downarrow}
    -f_{i\downarrow}^{\dagger}f_{j\uparrow}^{\dagger}f_{j\uparrow}f_{i\uparrow}
    +f_{i\downarrow}^{\dagger}f_{j\downarrow}^{\dagger}f_{j\downarrow}f_{i\uparrow}
    +f_{i\uparrow}^{\dagger}f_{j\uparrow}^{\dagger}f_{j\downarrow}f_{i\uparrow}
    -f_{i\uparrow}^{\dagger}f_{j\downarrow}^{\dagger}f_{j\uparrow}f_{i\uparrow}
    -f_{i\downarrow}^{\dagger}f_{j\uparrow}^{\dagger}f_{j\downarrow}f_{i\downarrow}
    +f_{i\downarrow}^{\dagger}f_{j\downarrow}^{\dagger}f_{j\uparrow}f_{i\downarrow}
    \right)
    \nonumber\\
    &+\frac{\Gamma}{4}\sum_{\langle i,j\rangle_{y}}\left(
    f_{i\uparrow}^{\dagger}f_{j\uparrow}^{\dagger}f_{j\downarrow}f_{i\uparrow}
    +f_{i\uparrow}^{\dagger}f_{j\downarrow}^{\dagger}f_{j\uparrow}f_{i\uparrow}
    -f_{i\downarrow}^{\dagger}f_{j\uparrow}^{\dagger}f_{j\downarrow}f_{i\downarrow}
    -f_{i\downarrow}^{\dagger}f_{j\downarrow}^{\dagger}f_{j\uparrow}f_{i\downarrow}
    +f_{i\uparrow}^{\dagger}f_{j\uparrow}^{\dagger}f_{j\uparrow}f_{i\downarrow}
    -f_{i\uparrow}^{\dagger}f_{j\downarrow}^{\dagger}f_{j\downarrow}f_{i\downarrow}
    +f_{i\downarrow}^{\dagger}f_{j\uparrow}^{\dagger}f_{j\uparrow}f_{i\uparrow}
    -f_{i\downarrow}^{\dagger}f_{j\downarrow}^{\dagger}f_{j\downarrow}f_{i\uparrow}
    \right)
    \nonumber\\
    &-i\frac{\Gamma}{2}\sum_{\langle i,j\rangle_{z}}\left(f_{i\uparrow}^{\dagger}f_{j\uparrow}^{\dagger}f_{j\downarrow}f_{i\downarrow}-f_{i\downarrow}^{\dagger}f_{j\downarrow}^{\dagger}f_{j\uparrow}f_{i\uparrow}\right).
\end{align}
\end{widetext}

For completeness, the sublattice-resolved vertex block is
\begin{align}
    \left[V^{\mu\rho\lambda\nu}_{\bm{q}}\right]_{uv}
    =\frac{1}{4}\frac{2}{N}\sum_{i\in u,j\in v}\sum_{\alpha,\alpha^{\prime}}
    J_{ij}^{\alpha\alpha^{\prime}}e^{-i\bm{q}\cdot(\bm{r}_{i}-\bm{r}_{j})}
    \sigma^{\alpha}_{\mu\nu}\sigma^{\alpha^{\prime}}_{\rho\lambda}.
\end{align}
\clearpage
For example, the $(u,v)=(\text{A},\text{B})$ component of $V_{\bm{q}}^{\uparrow\downarrow\uparrow\downarrow}$ is
\begin{widetext}
\begin{align}
    \left(V_{\bm{q}}^{\uparrow\downarrow\uparrow\downarrow}\right)_{\text{A}\text{B}}
    =\frac{2}{N}\sum_{i\in\text{A}}\left[\frac{K}{4}e^{-i\bm{q}\cdot\left(\bm{r}_{i}-\left(\bm{r}_{i}+\bm{\delta}_{x}\right)\right)}+\frac{K}{4}e^{-i\bm{q}\cdot\left(\bm{r}_{i}-\left(\bm{r}_{i}+\bm{\delta}_{y}\right)\right)}+\sum_{\gamma=x,y,z}\frac{J}{2}e^{-i\bm{q}\cdot\left(\bm{r}_{i}-\left(\bm{r}_{i}+\bm{\delta}_{\gamma}\right)\right)}\right]
    =\frac{K}{4}\left(e^{i\bm{q}\cdot\bm{\delta}_{x}}+e^{i\bm{q}\cdot\bm{\delta}_{y}}\right)+\frac{J}{2}\sum_{\gamma=x,y,z}e^{i\bm{q}\cdot\bm{\delta}_{\gamma}}.
\end{align}
\end{widetext}
Here, $\bm{\delta}_{\gamma}$ is the directed A-to-B bond vector defined in Sec.~\ref{sec:Model}.
The four-point vertex functions are then represented by the following matrices.
\begin{widetext}
\begin{align}
    &V_{\bm{q}}^{\uparrow\uparrow\downarrow\downarrow}
    =\begin{pmatrix}
        0&\dfrac{K}{4}\left(e^{i\bm{q}\cdot\bm{\delta}_{x}}-e^{i\bm{q}\cdot\bm{\delta}_{y}}\right)-i\dfrac{\Gamma}{2} e^{i\bm{q}\cdot\bm{\delta}_{z}}\\
        \dfrac{K}{4}\left(e^{-i\bm{q}\cdot\bm{\delta}_{x}}-e^{-i\bm{q}\cdot\bm{\delta}_{y}}\right)-i\dfrac{\Gamma}{2} e^{-i\bm{q}\cdot\bm{\delta}_{z}}&0
    \end{pmatrix},\label{eq:vertex-uudd}\\
    &V_{\bm{q}}^{\uparrow\downarrow\uparrow\downarrow}
    =V_{\bm{q}}^{\downarrow\uparrow\downarrow\uparrow}
    =\begin{pmatrix}
        0&\dfrac{K}{4}\left(e^{i\bm{q}\cdot\bm{\delta}_{x}}+e^{i\bm{q}\cdot\bm{\delta}_{y}}\right)+\dfrac{J}{2}\displaystyle\sum_{\gamma=x,y,z}e^{i\bm{q}\cdot\bm{\delta}_{\gamma}}\\
        \dfrac{K}{4}\left(e^{-i\bm{q}\cdot\bm{\delta}_{x}}+e^{-i\bm{q}\cdot\bm{\delta}_{y}}\right)+\dfrac{J}{2}\displaystyle\sum_{\gamma=x,y,z}e^{-i\bm{q}\cdot\bm{\delta}_{\gamma}}&0
    \end{pmatrix},\label{eq:vertex-udud}\\
    &V_{\bm{q}}^{\downarrow\downarrow\uparrow\uparrow}
    =\begin{pmatrix}
        0&\dfrac{K}{4}\left(e^{i\bm{q}\cdot\bm{\delta}_{x}}-e^{i\bm{q}\cdot\bm{\delta}_{y}}\right)+i\dfrac{\Gamma}{2} e^{i\bm{q}\cdot\bm{\delta}_{z}}\\
        \dfrac{K}{4}\left(e^{-i\bm{q}\cdot\bm{\delta}_{x}}-e^{-i\bm{q}\cdot\bm{\delta}_{y}}\right)+i\dfrac{\Gamma}{2} e^{-i\bm{q}\cdot\bm{\delta}_{z}}&0
    \end{pmatrix},\label{eq:vertex-dduu}\\
    &V_{\bm{q}}^{\uparrow\uparrow\uparrow\uparrow}
    =V_{\bm{q}}^{\downarrow\downarrow\downarrow\downarrow}
    =\begin{pmatrix}
        0&\dfrac{K}{4}e^{i\bm{q}\cdot\bm{\delta}_{z}}+\dfrac{J}{4}\displaystyle\sum_{\gamma=x,y,z}e^{i\bm{q}\cdot\bm{\delta}_{\gamma}}\\
        \dfrac{K}{4}e^{-i\bm{q}\cdot\bm{\delta}_{z}}+\dfrac{J}{4}\displaystyle\sum_{\gamma=x,y,z}e^{-i\bm{q}\cdot\bm{\delta}_{\gamma}}&0
    \end{pmatrix},\label{eq:vertex-uuuu}\\
    &V_{\bm{q}}^{\uparrow\downarrow\downarrow\uparrow}
    =V_{\bm{q}}^{\downarrow\uparrow\uparrow\downarrow}
    =\begin{pmatrix}
        0&-\dfrac{K}{4}e^{i\bm{q}\cdot\bm{\delta}_{z}}-\dfrac{J}{4}\displaystyle\sum_{\gamma=x,y,z}e^{i\bm{q}\cdot\bm{\delta}_{\gamma}}\\
        -\dfrac{K}{4}e^{-i\bm{q}\cdot\bm{\delta}_{z}}-\dfrac{J}{4}\displaystyle\sum_{\gamma=x,y,z}e^{-i\bm{q}\cdot\bm{\delta}_{\gamma}}&0
    \end{pmatrix}\label{eq:vertex-uddu}.
\end{align}
The remaining $\Gamma$-induced spin-nonconserving vertex blocks are
\begin{align}
    &V_{\bm{q}}^{\uparrow\uparrow\uparrow\downarrow}
    =V_{\bm{q}}^{\uparrow\uparrow\downarrow\uparrow}
    =
    \begin{pmatrix}
        0&-i\dfrac{\Gamma}{4}e^{i\bm{q}\cdot\bm{\delta}_{x}}+\dfrac{\Gamma}{4}e^{i\bm{q}\cdot\bm{\delta}_{y}}\\
        -i\dfrac{\Gamma}{4}e^{-i\bm{q}\cdot\bm{\delta}_{x}}+\dfrac{\Gamma}{4}e^{-i\bm{q}\cdot\bm{\delta}_{y}}&0
    \end{pmatrix},
    \label{eq:vertex-gamma-mixing}\\
    &V_{\bm{q}}^{\uparrow\downarrow\downarrow\downarrow}
    =V_{\bm{q}}^{\downarrow\uparrow\downarrow\downarrow}
    =
    \begin{pmatrix}
        0&i\dfrac{\Gamma}{4}e^{i\bm{q}\cdot\bm{\delta}_{x}}-\dfrac{\Gamma}{4}e^{i\bm{q}\cdot\bm{\delta}_{y}}\\
        i\dfrac{\Gamma}{4}e^{-i\bm{q}\cdot\bm{\delta}_{x}}-\dfrac{\Gamma}{4}e^{-i\bm{q}\cdot\bm{\delta}_{y}}&0
    \end{pmatrix},
    \label{eq:vertex-gamma-mixing-uddd}\\
    &V_{\bm{q}}^{\downarrow\uparrow\uparrow\uparrow}
    =V_{\bm{q}}^{\uparrow\downarrow\uparrow\uparrow}
    =
    \begin{pmatrix}
        0&i\dfrac{\Gamma}{4}e^{i\bm{q}\cdot\bm{\delta}_{x}}+\dfrac{\Gamma}{4}e^{i\bm{q}\cdot\bm{\delta}_{y}}\\
        i\dfrac{\Gamma}{4}e^{-i\bm{q}\cdot\bm{\delta}_{x}}+\dfrac{\Gamma}{4}e^{-i\bm{q}\cdot\bm{\delta}_{y}}&0
    \end{pmatrix},
    \label{eq:vertex-gamma-mixing-duuu}\\
    &V_{\bm{q}}^{\downarrow\downarrow\downarrow\uparrow}
    =V_{\bm{q}}^{\downarrow\downarrow\uparrow\downarrow}
    =
    \begin{pmatrix}
        0&-i\dfrac{\Gamma}{4}e^{i\bm{q}\cdot\bm{\delta}_{x}}-\dfrac{\Gamma}{4}e^{i\bm{q}\cdot\bm{\delta}_{y}}\\
        -i\dfrac{\Gamma}{4}e^{-i\bm{q}\cdot\bm{\delta}_{x}}-\dfrac{\Gamma}{4}e^{-i\bm{q}\cdot\bm{\delta}_{y}}&0
    \end{pmatrix}.
    \label{eq:vertex-gamma-mixing-dddu}
\end{align}
These $2\times2$ sublattice matrices form the block elements of the full $8\times8$ vertex matrix in the channel basis
$\{+,-,\uparrow,\downarrow\}\otimes\{\mathrm{A},\mathrm{B}\}$.
The resulting matrix is
\begin{align}
    \label{eq:vertex-8x8}
    \bm{V}_{\bm{q}}=
    \begin{pmatrix}
        V_{\bm{q}}^{\uparrow\uparrow\downarrow\downarrow}
        &V_{\bm{q}}^{\uparrow\downarrow\uparrow\downarrow}
        &V_{\bm{q}}^{\uparrow\uparrow\uparrow\downarrow}
        &V_{\bm{q}}^{\uparrow\downarrow\downarrow\downarrow}\\
        V_{\bm{q}}^{\downarrow\uparrow\downarrow\uparrow}
        &V_{\bm{q}}^{\downarrow\downarrow\uparrow\uparrow}
        &V_{\bm{q}}^{\downarrow\uparrow\uparrow\uparrow}
        &V_{\bm{q}}^{\downarrow\downarrow\downarrow\uparrow}\\
        V_{\bm{q}}^{\uparrow\uparrow\downarrow\uparrow}
        &V_{\bm{q}}^{\uparrow\downarrow\uparrow\uparrow}
        &V_{\bm{q}}^{\uparrow\uparrow\uparrow\uparrow}
        &V_{\bm{q}}^{\uparrow\downarrow\downarrow\uparrow}\\
        V_{\bm{q}}^{\downarrow\uparrow\downarrow\downarrow}
        &V_{\bm{q}}^{\downarrow\downarrow\uparrow\downarrow}
        &V_{\bm{q}}^{\downarrow\uparrow\uparrow\downarrow}
        &V_{\bm{q}}^{\downarrow\downarrow\downarrow\downarrow}
    \end{pmatrix}.
\end{align}

The unrescaled Matsubara Wick susceptibility entering the RPA is defined directly from the spin fluctuations.  In sublattice space it is
\begin{align}
    \label{eq:definition-matsubara-channel-susceptibility}
    \widetilde{\chi}_{u,v}^{\alpha\alpha^{\prime}(0)}(\bm{q},i\omega_{n})
    &=
    \frac{2}{N}\sum_{i\in u,j\in v}
    e^{-i\bm{q}\cdot\left(\bm{r}_{i}-\bm{r}_{j}\right)}
    \int_{0}^{\beta}d\tau\,e^{i\omega_{n}\tau}
    \left\langle \mathcal{T}_{\tau}\delta S_{i}^{\alpha}(\tau)
    \delta S_{j}^{\alpha^{\prime}}(0)\right\rangle_{0}
    \notag\\
    &=
    \frac{2}{N}\sum_{i\in u,j\in v}
    e^{-i\bm{q}\cdot\left(\bm{r}_{i}-\bm{r}_{j}\right)}
    \int_{0}^{\beta}d\tau\,e^{i\omega_{n}\tau}
    \frac{1}{4}
    \sum_{\mu,\nu,\rho,\lambda}
    \sigma^{\alpha}_{\mu\nu}\sigma^{\alpha^{\prime}}_{\rho\lambda}
    \left[
    \left\langle
    \mathcal{T}_{\tau}\bar{f}_{i\mu}(\tau)f_{i\nu}(\tau)
    \bar{f}_{j\rho}(0)f_{j\lambda}(0)
    \right\rangle_{0}
    -\left\langle\bar{f}_{i\mu}f_{i\nu}\right\rangle_0
    \left\langle\bar{f}_{j\rho}f_{j\lambda}\right\rangle_0
    \right].
\end{align}
Here, \(\omega_{n}=2\pi nT\) is a bosonic Matsubara frequency, and the subscript 0 indicates that the expectation value is taken with the quadratic mean-field Hamiltonian.
The subtraction in the second line follows directly from the $\delta S$ operators in the first line and is applied to every component of the eight-channel basis, including density channels with nonzero expectation values.
Equivalently, after the internal fermionic Matsubara-frequency summation is performed, each channel component at $i\omega_n\ne0$ has the pole representation
\begin{align}
    \label{eq:chi0-channel-pole-form}
    \widetilde{\chi}^{ab(0)}_{u,v}(\bm{q},i\omega_{n})
    =
    \frac{2}{N}\sum_{\bm{k}}^{\mathrm{BZ}}\sum_{\eta,\eta^{\prime}=1}^{8}
    \mathcal{W}^{ab}_{uv\eta\eta^{\prime}}(\bm{k},\bm{q})
    \frac{
    g(\varepsilon_{\bm{k},\eta})-g(\varepsilon_{\bm{k}+\bm{q},\eta^{\prime}})
    }
    {i\omega_{n}+\varepsilon_{\bm{k},\eta}-\varepsilon_{\bm{k}+\bm{q},\eta^{\prime}}},
\end{align}
where \(a,b\in\{+,-,\uparrow,\downarrow\}\) and $\eta,\eta^{\prime}=1,\ldots,8$ label the signed BdG branches.
To specify the weight explicitly, let $P_{\mathrm A}=\operatorname{diag}(1,0)$ and $P_{\mathrm B}=\operatorname{diag}(0,1)$ be the sublattice projectors, and define $\sigma^{\pm}=(\sigma^x\pm i\sigma^y)/2$, $\sigma^{\uparrow}=(\bm 1_2+\sigma^z)/2$, and $\sigma^{\downarrow}=(\bm 1_2-\sigma^z)/2$. In the Nambu ordering of Eq.~\eqref{eq:definition-Fk}, the channel matrices are
\begin{align}
    \label{eq:nambu-channel-matrices}
    \mathcal S_u^a=
    \begin{pmatrix}
        \sigma^a\otimes P_u & 0\\
        0 & -(\sigma^a)^T\otimes P_u
    \end{pmatrix}.
\end{align}
They satisfy $S_{(l,u)}^a=\mathcal F_l^\dagger\mathcal S_u^a\mathcal F_l/2+\operatorname{Tr}(\sigma^a)/2$; the constant term in the density channels cancels in the connected correlation. The weight in Eq.~\eqref{eq:chi0-channel-pole-form} is
\begin{align}
    \label{eq:explicit-bubble-weight}
    \mathcal W^{ab}_{uv\eta\eta^{\prime}}(\bm k,\bm q)
    =-\frac{1}{2}
    \left[T_{\bm k}^{\dagger}\mathcal S_u^a T_{\bm k+\bm q}\right]_{\eta\eta^{\prime}}
    \left[T_{\bm k+\bm q}^{\dagger}\mathcal S_v^b T_{\bm k}\right]_{\eta^{\prime}\eta}.
\end{align}
The minus sign follows from the fermionic Wick contraction with the susceptibility convention in Eq.~\eqref{eq:definition-matsubara-channel-susceptibility}, and the factor $1/2$ removes the Nambu double counting when all eight signed branches are summed. The full Nambu matrices include both normal and anomalous contractions.
We use the eight-component channel space
$\{+,-,\uparrow,\downarrow\}\otimes\{\mathrm{A},\mathrm{B}\}$.
Here, $S_{i}^{\uparrow}=f_{i\uparrow}^{\dagger}f_{i\uparrow}$ and $S_{i}^{\downarrow}=f_{i\downarrow}^{\dagger}f_{i\downarrow}$.
The unrescaled Wick bubble is collected into the $8\times8$ matrix
\begin{align}
    \label{eq:chi0-8x8}
    \widetilde{\bm{\chi}}^{(0)}(\bm{q},i\omega_{n})=
    \begin{pmatrix}
        \widetilde{\bm{\chi}}^{++(0)}&\widetilde{\bm{\chi}}^{+-(0)}&\widetilde{\bm{\chi}}^{+\uparrow(0)}&\widetilde{\bm{\chi}}^{+\downarrow(0)}\\
        \widetilde{\bm{\chi}}^{-+(0)}&\widetilde{\bm{\chi}}^{--(0)}&\widetilde{\bm{\chi}}^{-\uparrow(0)}&\widetilde{\bm{\chi}}^{-\downarrow(0)}\\
        \widetilde{\bm{\chi}}^{\uparrow +(0)}&\widetilde{\bm{\chi}}^{\uparrow -(0)}&\widetilde{\bm{\chi}}^{\uparrow\uparrow(0)}&\widetilde{\bm{\chi}}^{\uparrow\downarrow(0)}\\
        \widetilde{\bm{\chi}}^{\downarrow +(0)}&\widetilde{\bm{\chi}}^{\downarrow -(0)}&\widetilde{\bm{\chi}}^{\downarrow\uparrow(0)}&\widetilde{\bm{\chi}}^{\downarrow\downarrow(0)}
    \end{pmatrix}_{(\bm{q},i\omega_{n})}.
\end{align}
Each entry in Eq.~\eqref{eq:chi0-8x8} is a $2\times2$ matrix in the sublattice indices $(u,v)$.
Thus, $\widetilde{\bm{\chi}}^{(0)}$ denotes the unrescaled Wick susceptibility of $\delta S$, and $\bm{\chi}^{(0)}=Z\widetilde{\bm{\chi}}^{(0)}$ denotes the normalized fluctuation matrix entering the RPA, following Appendix~\ref{app:local-moment-normalization}.
\end{widetext}

By connecting $\bm{\chi}^{(0)}$ with the full vertex matrix in Eq.~\eqref{eq:vertex-8x8}, we obtain the Dyson equation.
Suppressing the common argument $(\bm q,i\omega_n)$ in Eqs.~\eqref{eq:Dyson_eq_8x8} and~\eqref{eq:sol-Dyson_eq_8x8},
\begin{align}
    \bm{\chi}_{\text{RPA}}
    &=\bm{\chi}^{(0)}-\bm{\chi}^{(0)}\bm{V}_{\bm q}\,\bm{\chi}_{\text{RPA}}.
    \label{eq:Dyson_eq_8x8}
\end{align}
Solving this $8\times8$ matrix equation gives
\begin{align}
    \bm{\chi}_{\text{RPA}}
    &=\left[\bm{1}_{8}+\bm{\chi}^{(0)}\bm{V}_{\bm q}\right]^{-1}\bm{\chi}^{(0)}.
    \label{eq:sol-Dyson_eq_8x8}
\end{align}
With the convention in Eq.~\eqref{eq:retarded-susceptibility-convention}, the retarded susceptibility used in the dynamical structure factor is obtained after this matrix resummation by
\begin{align}
    \label{eq:analytic-continuation-rpa-appendix}
    \bm{\chi}_{\text{RPA}}^{R}(\bm{q},\omega)
    =
    \bm{\chi}_{\text{RPA}}(\bm{q},i\omega_{n}\to\omega+i\delta).
\end{align}
Only after this $8\times8$ resummation and analytic continuation do we transform back to the Cartesian spin basis.
The diagonal components used in the dynamical structure factor are
written below with $s,s'\in\{+1,-1\}$ labeling the $+,-$ channels and
$\eta_{\uparrow}=-\eta_{\downarrow}=1$:
\begin{align}
    \chi_{\text{RPA}}^{xx,R}
    &=\frac{1}{4}\sum_{u,v}\sum_{s,s'}[\bm{\chi}_{\text{RPA}}^{R,ss'}]_{uv},
    \label{eq:chi-RPA-cartesian}\\
    \chi_{\text{RPA}}^{yy,R}
    &=-\frac{1}{4}\sum_{u,v}\sum_{s,s'}ss'[\bm{\chi}_{\text{RPA}}^{R,ss'}]_{uv},
    \label{eq:chi-RPA-cartesian-yy}\\
    \chi_{\text{RPA}}^{zz,R}
    &=\frac{1}{4}\sum_{u,v}\sum_{\sigma,\sigma'}
    \eta_{\sigma}\eta_{\sigma'}[\bm{\chi}_{\text{RPA}}^{R,\sigma\sigma'}]_{uv}.
    \label{eq:chi-RPA-cartesian-zz}
\end{align}
Applying the fluctuation-dissipation theorem, the final dynamical structure factor $S^{\alpha\alpha}_{\text{RPA}}(\bm{q},\omega)$ is obtained as
\begin{align}
    \label{eq:Sqw-RPA}
    S^{\alpha\alpha}_{\text{RPA}}(\bm{q},\omega)
    =\frac{1}{2}\frac{1}{\pi}
    \frac{\mathrm{Im}\chi^{\alpha\alpha,R}_{\text{RPA}}(\bm{q},\omega)}
    {1-e^{-\beta\omega}}.
\end{align}
The prefactor $1/2$ is required to return to the normalization of the dynamical structure factor defined in Eq.~\eqref{eq:definition-dynamical-spin-structure-factor}.
In the present two-sublattice formulation, the channel susceptibility is introduced with the prefactor $2/N$ in Eq.~\eqref{eq:definition-matsubara-channel-susceptibility}, while Eqs.~\eqref{eq:chi-RPA-cartesian}--\eqref{eq:chi-RPA-cartesian-zz} sum the sublattice-resolved susceptibility over $(u,v)$.
The extra factor $1/2$ therefore compensates for this convention.

\section{Majorana-fermion perturbation theory and its relation to the present formulation}
\label{app:majorana-perturbation-theory}

\subsection{Majorana-fermion RPA}

The Majorana response theory follows Ref.~\cite{Rao-Moessner-Knolle-2025} and uses the representation and normalization defined in Eqs.~\eqref{eq:spin-Majorana-representation} and~\eqref{eq:commutation-relation-Majorana}.
The magnetic field is part of the quadratic mean-field Hamiltonian and does not generate a four-Majorana RPA vertex.
The bare susceptibility is the same quantity defined in Eq.~\eqref{eq:definition-matsubara-channel-susceptibility}, now evaluated by inserting $S_i^\alpha=ic_i b_i^\alpha$.
Its Wick decomposition contains the direct $cc\,bb$ contraction and the crossed $cb\,cb$ contraction with the relative minus sign required by fermionic anticommutation.
The crossed contribution vanishes when the $c$ and $b^\alpha$ sectors are decoupled, as at a zero-field liquid saddle, but it must be retained in a magnetic field or at a magnetically ordered saddle.
After the internal Matsubara-frequency sum and the continuation $i\omega_n\rightarrow\omega+i\delta$, the resulting bubble contains the pair poles $\omega\mp(\varepsilon_{\bm k+\bm q,\eta}+\varepsilon_{-\bm k,\eta^{\prime}})+i\delta$ at zero temperature.  Here $\varepsilon_{\bm k,\eta}$ is the positive Majorana-quasiparticle energy defined in Eq.~\eqref{eq:eigenvalue-equation-majorana-particle}. At finite temperature, transitions involving thermally excited quasiparticles also give poles at energy differences, with the corresponding Fermi occupation factors.

The interaction kernel is fixed by the same $J_{ij}^{\alpha\alpha'}$ introduced in Eq.~\eqref{eq:Hamiltonian-int} and the same Fourier transform that defines $\bm V_{\bm q}$ in Eq.~\eqref{eq:definition-vertex}.
Here it is represented in the Cartesian spin--sublattice channels $(\alpha,u)$ instead of the Abrikosov channels used in Appendix~\ref{app:Detailed calculation of the random-phase approximation}.
Thus the two matrices have different dimensions, but contain the same physical exchange couplings and bond phases.
In this subsection, $\bm\chi^{(0)}$ denotes the bubble obtained directly from the Majorana Wick contractions.
The RPA series then gives~\cite{Rao-Moessner-Knolle-2025}
\begin{align}
    \bm\chi_{\mathrm{RPA}}(\bm q,i\omega_n)
    =
    \left[
    \bm 1+\bm\chi^{(0)}(\bm q,i\omega_n)
    \bm V_{\bm q}
    \right]^{-1}
    \bm\chi^{(0)}(\bm q,i\omega_n).
    \label{eq:majorana-rpa-equation}
\end{align}
Finally, the physical Cartesian susceptibility is obtained by contracting the sublattice indices,
\begin{align}
    \chi_{\mathrm{RPA}}^{\alpha\beta}(\bm q,i\omega_n)
    =\sum_{u,v}
    [\bm\chi_{\mathrm{RPA}}(\bm q,i\omega_n)]_{u\alpha,v\beta}.
    \label{eq:majorana-sublattice-contraction}
\end{align}
This is the same sublattice contraction used in Eqs.~\eqref{eq:chi-RPA-cartesian}--\eqref{eq:chi-RPA-cartesian-zz}.

\subsection{Relation to the present Abrikosov-fermion formulation}

The canonical transformation in Eqs.~\eqref{eq:MF-AF-relation-matrix-c-to-f} and~\eqref{eq:MF-AF-relation-matrix-f-to-c} maps the quadratic Majorana mean-field Hamiltonian to the Abrikosov BdG Hamiltonian and therefore reproduces its quasiparticle bands.
The equality of the quadratic Hamiltonians does not by itself imply equality of unprojected spin correlation functions, however, because the response also depends on the operator inserted at the two ends of the bubble.

Substituting Eq.~\eqref{eq:MF-AF-relation-matrix-f-to-c} into $S_i^\alpha=ic_i b_i^\alpha$ gives, in the enlarged Fock space,
\begin{align}
    ic_i b_i^x
    ={}&\frac{1}{2}\left(
    f_{i\uparrow}^\dagger f_{i\downarrow}
    +f_{i\downarrow}^\dagger f_{i\uparrow}
    \right)
    +\frac{1}{2}\left(
    f_{i\uparrow}f_{i\downarrow}
    -f_{i\uparrow}^\dagger f_{i\downarrow}^\dagger
    \right),
    \notag\\
    ic_i b_i^y
    ={}&-\frac{i}{2}\left(
    f_{i\uparrow}^\dagger f_{i\downarrow}
    -f_{i\downarrow}^\dagger f_{i\uparrow}
    \right)
    -\frac{i}{2}\left(
    f_{i\uparrow}f_{i\downarrow}
    +f_{i\uparrow}^\dagger f_{i\downarrow}^\dagger
    \right),
    \notag\\
    ic_i b_i^z
    ={}&\frac{1}{2}\left(n_{i\uparrow}-n_{i\downarrow}\right)
    +\frac{1}{2}\left(n_{i\uparrow}+n_{i\downarrow}-1\right).
    \label{eq:majorana-spin-in-abrikosov-full-space}
\end{align}
The first term in each line is the Abrikosov spin operator in Eq.~\eqref{eq:Abrikosov fermion representation}.
The remaining terms are local gauge-constraint-sector bilinears.
They vanish after exact projection because $n_{i\uparrow}+n_{i\downarrow}=1$ and the on-site pair-annihilation and pair-creation operators annihilate the singly occupied physical subspace.
Thus, $ic_i b_i^\alpha$ and $f_i^\dagger\sigma^\alpha f_i/2$ represent exactly the same physical spin after projection, but they are not identical operators on the full enlarged Fock space.

The RPA vertices in the two constructions originate from the same microscopic exchange interaction in Eq.~\eqref{eq:KJG-Hamiltonian} and the same bond form factors.
Restricting this interaction to the physical spin sector and using $S_i^+=f_{i\uparrow}^\dagger f_{i\downarrow}$, $S_i^-=f_{i\downarrow}^\dagger f_{i\uparrow}$, and $S_i^z=(n_{i\uparrow}-n_{i\downarrow})/2$ gives precisely the $8\times8$ vertex in Eq.~\eqref{eq:vertex-8x8}.
In particular, the factors $K/4$, $J/2$, $\Gamma/4$, and $\Gamma/2$ in Eqs.~\eqref{eq:vertex-uudd}--\eqref{eq:vertex-gamma-mixing-dddu} introduce no additional interaction parameter.
This equivalence in the physical spin sector should be distinguished from a canonical transformation of the unprojected four-fermion theory, which would also retain the additional local bilinears in Eq.~\eqref{eq:majorana-spin-in-abrikosov-full-space}.

At the exact projected level, the two representations therefore give the same susceptibility.
At Gaussian mean-field level, however, both formulations impose their local constraints only on average.
The Majorana Wick bubble, when expressed in Abrikosov variables, contains not only the correlation of the first terms in Eq.~\eqref{eq:majorana-spin-in-abrikosov-full-space}, but also mixed correlations with the constraint-sector terms and correlations between the constraint-sector terms themselves.
These additional contributions are not required to vanish in an unprojected mean-field state, although particular symmetries may eliminate some of them.
The bare susceptibility used in the present Abrikosov formulation instead inserts only $f_i^\dagger\sigma^\alpha f_i/2$.
Consequently, the two mean-field bubbles can differ even when their quadratic quasiparticle Hamiltonians are canonically equivalent.
This difference is a mean-field artifact of the incomplete local projection rather than evidence that either parton representation is intrinsically more accurate.
Because the RPA response depends nonlinearly on the bare susceptibility through its matrix denominator, this difference in the bubble propagates to, and can be enhanced in, the final RPA spectrum.

Momentum-dependent factors enter at the same stages in the two representations.
With the site-position Fourier transforms in Eqs.~\eqref{eq:fourier-transformation-majorana} and~\eqref{eq:fourier-transformation}, both quadratic Hamiltonians contain the directed-bond phases $e^{i\bm k\cdot\bm\delta_\gamma}$.
No additional bond form factor is inserted in the sublattice-resolved bare bubble.
The exchange-bond form factors enter $\bm V_{\bm q}$, represented explicitly for the present Abrikosov channels in Eq.~\eqref{eq:vertex-8x8}, and the final Cartesian response is obtained by the direct sublattice sum in Eq.~\eqref{eq:majorana-sublattice-contraction}.
The site-position phases are already included in the sublattice-resolved susceptibility of Eq.~\eqref{eq:definition-matsubara-channel-susceptibility}, so no further sublattice phase multiplies this sum.
The sum-rule normalization of the Abrikosov Wick bubble described in Appendix~\ref{app:local-moment-normalization} is separate from the representation change discussed above.

\section{Numerical parameters for the \texorpdfstring{dynamical structure factor}{dynamical structure factor}}
\label{app:numerical-details}

\begin{figure}[t]
  \centering
  \includegraphics[width=\columnwidth,clip]{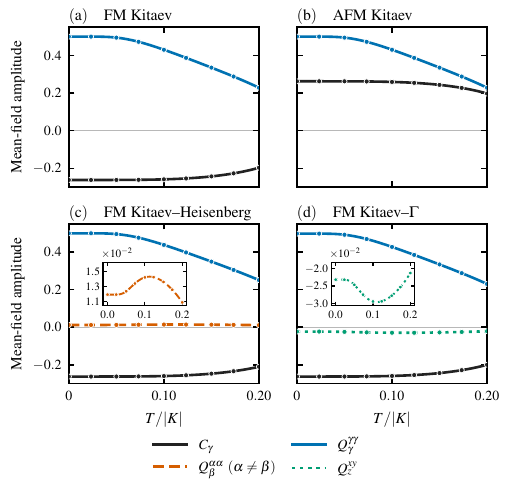}
  \caption{
  Temperature evolution of representative self-consistent liquid-saddle mean fields used in the main-text calculations.
  Panels (a) and (b) show the FM and AFM pure Kitaev models, respectively; panel (c) shows the FM Kitaev--Heisenberg model with $J/|K|=-0.05$; and panel (d) shows the FM Kitaev--$\Gamma$ model with $\Gamma/|K|=0.1$.
  All calculations have $h=0$ and $m_i^\alpha=0$.
  The amplitudes follow the convention of Eqs.~\eqref{eq:mean-field-channel-Kitaev-Majorana-C} and~\eqref{eq:mean-field-channel-Kitaev-Majorana-Q}.
  In panel (c), the orange curve denotes the symmetry-equivalent off-bond diagonal channels $Q_{\beta}^{\alpha\alpha}$ with $\alpha\ne\beta$; in panel (d), the green curve denotes $Q_z^{xy}$ and its symmetry-related partners.
  The curves connect 64 self-consistent temperature points, and nine approximately equally spaced points are marked by filled circles.
  }
  \label{fig:mean_fields_temperature.pdf}
\end{figure}

Figure~\ref{fig:mean_fields_temperature.pdf} shows the temperature evolution of the representative self-consistent liquid-saddle mean fields used in the main-text calculations.
The pure-Kitaev values $C_\gamma\simeq\mp0.2624$ and $Q_\gamma^{\gamma\gamma}=1/2$ agree with Eqs.~\eqref{eq:mean-field-value-of-Kitaev-model-C} and~\eqref{eq:mean-field-value-of-Kitaev-model-Q}.

All quantities are measured in units of $|K|$.
At finite temperature, the same $T$ is used in the mean-field and unrescaled Wick-susceptibility calculations.
The analytic continuation $i\omega_n\to\omega+i\delta$ introduces the energy broadening $\delta$.
The internal Brillouin-zone momentum sum in the bare Wick susceptibility is evaluated on a uniform $100\times100$ mesh.
Unless stated otherwise, we use $\delta/|K|=0.01$ at $T/|K|=0.0$ and set $\delta/|K|=T/|K|$ at finite temperature.
For the pure-Kitaev spectra at $T/|K|=0.0$ in Fig.~\ref{fig:pure_Kitaev.pdf} and in Appendix~\ref{app:Comparison with prior work on the temperature evolution of the pure Kitaev model}, we instead use $\delta/|K|=0.05$.
For the zero-temperature comparison in Figs.~\ref{fig:RPA_ED.pdf}(b)--\ref{fig:RPA_ED.pdf}(e), both parton RPA and ED use $\delta/|K|=0.05$.
For the LSWT spectra in Figs.~\ref{fig:Heisenberg_magnon.pdf}(c) and~\ref{fig:Gamma_magnon.pdf}(c), we use $\delta/|K|=0.1$.
All parton-RPA spectra are shown on their directly calculated energy axes without an energy shift.
The temperature-dependent broadening is introduced phenomenologically to represent unresolved finite-temperature damping that is not generated explicitly within the present static parton mean-field framework combined with RPA.  Accordingly, linewidths and the degree of peak merging are not interpreted quantitatively.

\begin{figure*}[t]
  \centering
      \includegraphics[width=2\columnwidth,clip]{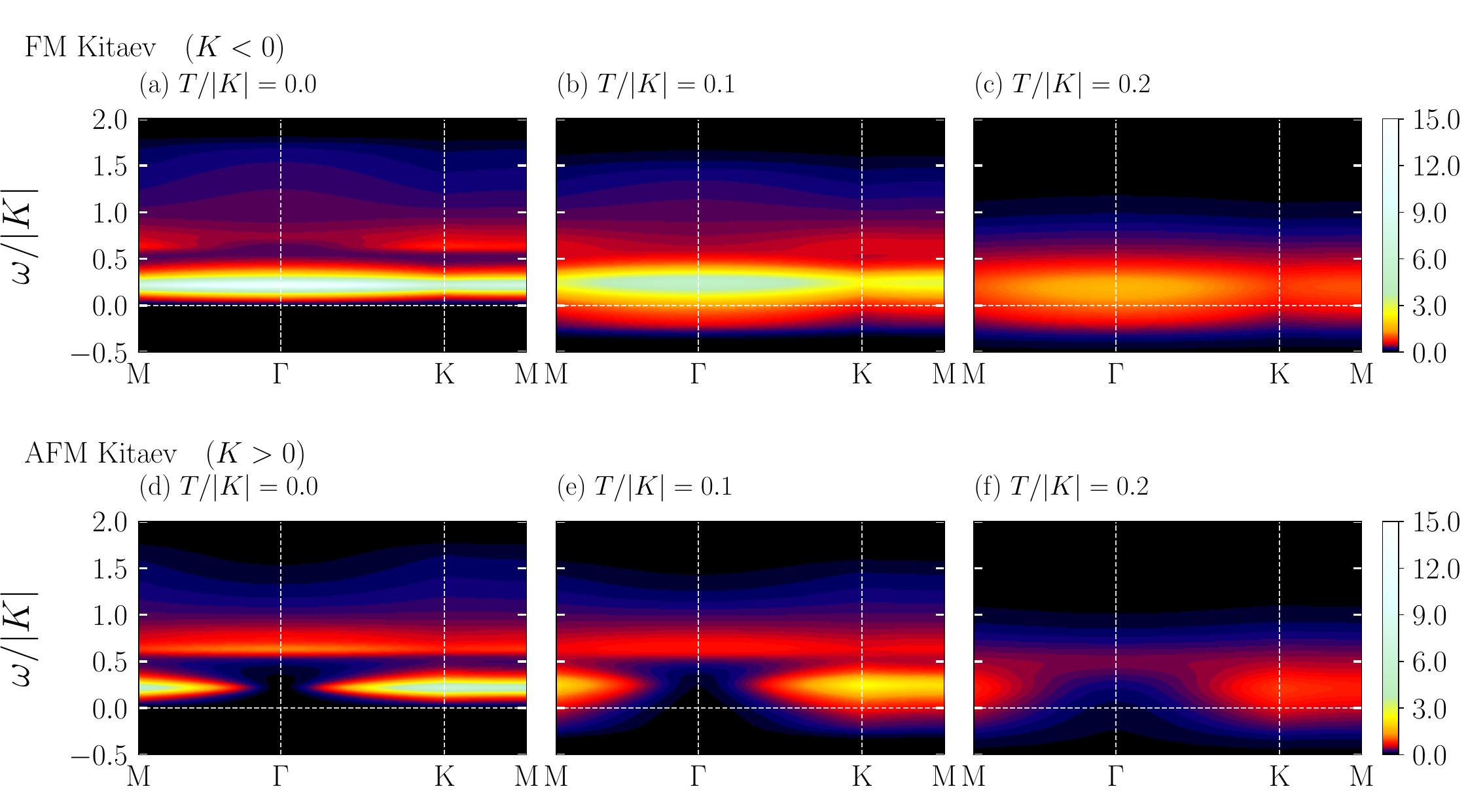}
      \caption{
      Temperature evolution of the dynamical spin structure factor $S(\bm{q},\omega)$ for the pure Kitaev model, obtained using the present parton-RPA approach, along the momentum path $\mathrm{M}$-$\Gamma$-$\mathrm{K}$-$\mathrm{M}$ used in Ref.~\cite{Yoshitake-Nasu-Motome-2016}.
      The upper panels [(a)--(c)] show the FM Kitaev model ($K<0$), while the lower panels [(d)--(f)] show the AFM Kitaev model ($K>0$).
      From left to right, the panels correspond to $T/|K|=0.0$, $0.1$, and $0.2$.
      }
      \label{fig:pure_Kitaev_benchmark.pdf}
\end{figure*}

For the forced-ferromagnetic Kitaev--$\Gamma$ spectra in Fig.~\ref{fig:Gamma_field_dssf.pdf}, Table~\ref{tab:forced-ferro-reference-fields} lists the reference mean fields and the normalization factor $Z$ determined separately at each temperature.
\begin{table}[t]
  \caption{Frozen reference fields and normalization factors used for the forced-ferromagnetic spectra in Fig.~\ref{fig:Gamma_field_dssf.pdf}.  At each temperature, a separate reference saddle is converged self-consistently at $\Gamma/|K|=0.3$ and $h_{\rm ref}/|K|=0.371$; its intersite sector and Lagrange multiplier are then held fixed in the calculation at $h/|K|=0.6$.  For $\{\alpha,\beta,\gamma\}=\{x,y,z\}$, the $C_3$-symmetric entries listed below are $C=C_\gamma$, $Q_\gamma^{\gamma\gamma}$, and $Q_\gamma^{\alpha\beta}=Q_\gamma^{\beta\alpha}$.  The two $Q$ entries determine the coefficient $KQ_\gamma^{\gamma\gamma}+\Gamma(Q_\gamma^{\alpha\beta}+Q_\gamma^{\beta\alpha})$ entering the quadratic Hamiltonian.  Here $\lambda=\lambda_i^\alpha$ and $Z$ is defined in Eq.~\eqref{eq:sum-rule-rescaled-bubble}.}
  \label{tab:forced-ferro-reference-fields}
  \centering
  
  \begin{ruledtabular}
  \begin{tabular}{lccc}
    & \multicolumn{3}{c}{$T/|K|$} \\
    Quantity & $0.0$ & $0.1$ & $0.2$ \\
    \colrule
    $C$                              & $-0.2373$ & $-0.2157$ & $-0.1380$ \\
    $Q_\gamma^{\gamma\gamma}$       & $ 0.4161$ & $ 0.3016$ & $ 0.1397$ \\
    $Q_\gamma^{\alpha\beta}$       & $-0.0483$ & $-0.0461$ & $-0.0324$ \\
    $\lambda/|K|$                    & $ 0.1090$ & $ 0.1158$ & $ 0.1308$ \\
    $Z$                              & $ 1.616$  & $ 1.521$  & $ 1.550$  \\
  \end{tabular}
  \end{ruledtabular}
  
\end{table}

For the Kitaev--Heisenberg results in Figs.~\ref{fig:Heisenberg_magnon.pdf}(a) and~\ref{fig:Heisenberg_magnon.pdf}(b), the intersite fields and Lagrange multiplier are first converged separately at each temperature at $J/|K|=-0.05$ and $h/|K|=0.2$ with $m_i^\alpha=0$. They are then held fixed while the uniform onsite moment is determined at the same couplings, field, and temperature. The normalization factors are $Z=1.858$ and $1.845$ for the displayed $T/|K|=0.0$ and $0.2$ results, respectively.

\section{Comparison with prior work on the temperature evolution of the pure Kitaev model}
\label{app:Comparison with prior work on the temperature evolution of the pure Kitaev model}

Figure~\ref{fig:pure_Kitaev_benchmark.pdf} shows the calculated spectra along the momentum path used in Ref.~\cite{Yoshitake-Nasu-Motome-2016}.  The spectra exhibit the same characteristic sign-dependent momentum structure and downward redistribution of spectral weight reported in that study.  Because the temperature points and normalization conventions are not matched one-to-one, we do not interpret this figure as a point-by-point quantitative benchmark.
\FloatBarrier

\section{\texorpdfstring{Classical-energy selection in the FM Kitaev--positive-$\Gamma$ model under a $[111]$ field}{Classical-energy selection in the FM Kitaev-positive-Gamma model under a [111] field}}
\label{app:classical-field-selection}

\begin{figure}[t]
  \centering
  \includegraphics[width=0.82\columnwidth,clip]{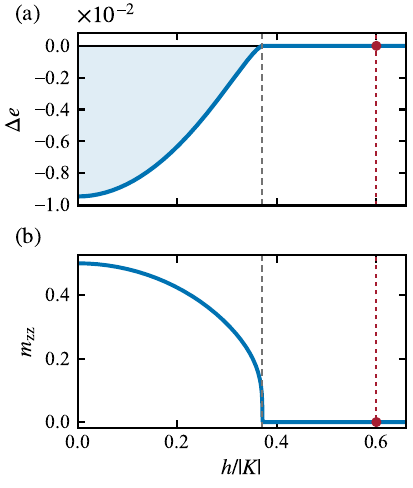}
  \caption{
  Classical field sweep at $K=-|K|$, $\Gamma/|K|=0.3$, and $S=1/2$ for a $[111]$ field.
  Panel (a) shows the dimensionless energy difference $\Delta e$, defined as the minimum energy per site in the two-stripe variational space minus that in the uniform sector, divided by $|K|$. Panel (b) shows the staggered zigzag component defined in Eq.~\eqref{eq:classical-zigzag-component}.
  The red dotted lines mark $h/|K|=0.6$, and the gray dashed lines mark the collapse field $h_c/|K|\simeq0.371$.
  Above $h_c$, the optimized two-stripe configuration is itself uniform, so the energy difference in panel (a) vanishes.
  }
  \label{fig:classical_field_selection.pdf}
\end{figure}

Figure~\ref{fig:classical_field_selection.pdf} shows the classical field sweep used to determine the collapse field $h_c$ and verify that the selected high-field configuration is spatially uniform.
We set $J=0$, $K=-|K|$, and $\Gamma/|K|=0.3$ in Eq.~\eqref{eq:KJG-Hamiltonian}, take $\bm h=h\bm e_c$, and replace each spin by a classical vector $\bm S_i=S\bm n_i$ with $S=1/2$ and $|\bm n_i|=1$.
We compare a spatially uniform configuration, $\bm n_i=\bm n_{\rm F}$, with the $2\times1$ two-stripe space used for the zigzag state, $\bm n_{\rm A}=\bm n_{\rm B}=\bm n_0$ and $\bm n_{\rm C}=\bm n_{\rm D}=\bm n_1$.
All spin directions are varied independently to minimize the classical energy at each field.
Thus, ``forced ferromagnetic'' denotes a spatially uniform state and does not impose either a fixed moment direction or saturation.
The two-stripe space contains this uniform limit when $\bm n_0=\bm n_1$.

We define the dimensionless energy per site of a configuration $X$ as $e_X\equiv E_X/(N|K|)$.
The uniform-configuration energy density is
\begin{align}
e_{\rm Ferro}
&=\frac{S^2}{2|K|}\left[K-\Gamma+3\Gamma u^2\right]-\frac{h}{|K|}Su,
\quad
u=\bm n_{\rm F}\cdot\bm e_c,
\label{eq:classical-uniform-energy}
\end{align}
and is minimized over $-1\leq u\leq1$.
For the two-stripe space, we evaluate Eq.~\eqref{eq:KJG-Hamiltonian} in the $2\times1$ cell and minimize it over the four angular variables specifying $\bm n_0$ and $\bm n_1$.
We monitor the staggered component
\begin{align}
m_{\rm zz}=\frac{S}{2}|\bm n_0-\bm n_1|,
\label{eq:classical-zigzag-component}
\end{align}
which is finite in the canted zigzag state and vanishes in the uniform limit.

The two optimized stripe directions approach each other continuously as the field increases, and $m_{\rm zz}$ vanishes at $h_c/|K|\simeq0.371$.
At $h/|K|=0.6$, the optimized $2\times1$ state is uniform, with $m_{\rm zz}<10^{-7}$ and a dimensionless energy per site of approximately $-0.350$.

The zero energy difference above $h_c$ reflects the identity of the optimized configurations, rather than a degeneracy between distinct ordered states.
These results identify the state at $h/|K|=0.6$ as spatially uniform and motivate the forced-ferromagnetic saddle used in the main text.
At each temperature, the liquid-state intersite fields and Lagrange multipliers are first converged at $h_{\rm ref}=h_c\simeq0.371|K|$ with $m_i^\alpha=0$. They are then held fixed while only the onsite moments $m_i^\alpha$ are determined self-consistently at $h/|K|=0.6$ and the same temperature.

This comparison is deliberately restricted to the uniform and two-stripe sectors used to determine the collapse field.
It is not an exhaustive classical phase diagram and does not exclude lower-energy states with longer magnetic periods.
Moreover, the classical calculation fixes the local spin length, whereas the ordered Majorana saddle determines the moment magnitude and direction self-consistently on top of the frozen liquid channels.
The classical sweep is therefore used only to select the spatial ordering pattern, not to replace the quantum saddle or establish a global phase boundary.

\section{Zero-temperature exact-diagonalization calculation of the dynamical spin structure factor}
\label{app:Zero-temperature-exact-diagonalization-calculation-of-the-dynamical-spin-structure-factor}

The ED comparisons are performed at zero field, where $\bm m_i=0$ and hence $\delta S_i^\mu=S_i^\mu$.
The quantity shown in Figs.~\ref{fig:RPA_ED.pdf}(b)--\ref{fig:RPA_ED.pdf}(e) is given by
\begin{align}
\label{eq:ed_lehmann_sqw}
S(\bm{q},\omega)
&=\sum_{\mu=x,y,z}S^{\mu\mu}(\bm{q},\omega)\notag\\
&=
\sum_{\mu=x,y,z}\sum_{n}
\left|
\langle n | \delta S_{-\bm{q}}^\mu |0\rangle
\right|^2
\delta\left(\omega-(E_n-E_0)\right),
\end{align}
where $|0\rangle$ and $|n\rangle$ are the ground state and excited eigenstates of Hamiltonian $\mathcal{H}$ with eigenvalues $E_0$ and $E_n$, respectively.
Here,
\begin{align}
\label{eq:ed_spin_operator_q}
\delta S_{\bm{q}}^{\mu}
=
\frac{1}{\sqrt{N}}
\sum_{i}
e^{-i\bm{q}\cdot\bm{r}_{i}}
\delta S_{i}^{\mu}.
\end{align}
With this Fourier convention, the phase in Eq.~\eqref{eq:definition-dynamical-spin-structure-factor} requires the Lehmann matrix element of $\delta S_{-\bm q}^{\mu}$ in Eq.~\eqref{eq:ed_lehmann_sqw}.
We evaluate Eq.~\eqref{eq:ed_lehmann_sqw} through the resolvent
\begin{align}
\label{eq:ed_resolvent_sqw}
S^{\mu\mu}(\bm{q},\omega)
&=
-\frac{1}{\pi}\mathrm{Im}
\left\langle 0 \left|
\left(\delta S_{-\bm{q}}^{\mu}\right)^\dagger
\frac{1}{\omega+i\delta+E_{0}-\mathcal{H}}
\delta S_{-\bm{q}}^{\mu}
\right|0\right\rangle.
\end{align}
For a finite cluster, the spectrum consists of discrete poles, and the finite $\delta$ broadens each pole into a Lorentzian of width $\delta$.
Introducing the normalized Krylov vector and the Lanczos recursion,
\begin{align}
|\phi_0\rangle
&=
\frac{\delta S_{-\bm{q}}^\mu|0\rangle}
{\sqrt{\langle 0|(\delta S_{-\bm{q}}^\mu)^\dagger \delta S_{-\bm{q}}^\mu|0\rangle}},
\label{eq:ed-initial-krylov-vector}\\
\mathcal{H}|\phi_{m}\rangle
&=
b_{m+1}|\phi_{m+1}\rangle + a_m|\phi_m\rangle + b_m|\phi_{m-1}\rangle,
\label{eq:ed_lanczos_recursion}
\end{align}
with $a_{m}=\langle\phi_{m}|\mathcal{H}|\phi_{m}\rangle$ and $b_{0}=0$, one obtains a tridiagonal representation of $\mathcal{H}$ in the Krylov subspace generated from $\delta S_{-\bm{q}}^{\mu}|0\rangle$.
Defining $z=\omega+i\delta+E_0$, the corresponding resolvent matrix element is written as the continued fraction
\begin{align}
\label{eq:ed_continued_fraction}
&\left\langle\phi_{0}\left|
\frac{1}{\omega+i\delta+E_{0}-\mathcal{H}}
\right|\phi_{0}\right\rangle\notag\\
&=
\cfrac{1}{z-a_{0}
-\cfrac{b_{1}^{2}}{z-a_{1}
-\cfrac{b_{2}^{2}}{z-a_{2}-\ddots}}},
\end{align}
and therefore
\begin{align}
S^{\mu\mu}(\bm{q},\omega)
&=
-\frac{1}{\pi}\,\mathrm{Im}\,
\Biggl[
\langle 0|(\delta S_{-\bm{q}}^\mu)^\dagger \delta S_{-\bm{q}}^\mu|0\rangle\notag\\
&\quad\times
\left\langle \phi_0 \left|
\frac{1}{\omega+i\delta+E_{0}-\mathcal{H}}
\right|\phi_0\right\rangle
\Biggr].
\label{eq:ed_final_sqw}
\end{align}
\bibliography{./refs}

\end{document}